# Energy or Mass and Interaction.

## Gustavo R. González-Martín[a]

## Departamento de Física, Universidad Simón Bolívar,

## Apartado 89000, Caracas 1080-A, Venezuela.

A review. Problems: 1-Many empirical parameters and large dimension number; 2-Gravitation and Electrodynamics are challenged by dark matter and energy. Energy and nonlinear electrodynamics are fundamental in a unified nonlinear interaction. **Nuclear energy appears as nonlinear SU(2) magnetic energy**. Gravitation and electromagnetism are unified giving **Einstein's equation and a geometric energy momentum tensor**. A solution energy in the newtonian limit gives the **gravitational constant $G$**. Outside of this limit $G$ is variable. May be interpreted as **dark matter or energy**. In vacuum, **known gravitational solutions** are obtained. **Electromagnetism is an SU(2)$_Q$ subgroup**. A U(1) limit gives **Maxwell's equations**. Geometric fields determine a **generalized Dirac equation** and are the **germ of quantum physics**. Planck's $h$ and of Einstein's $c$ are given by the potential and the metric. Excitations have **quanta of charge, flux and spin** determining the FQHE. There are only three stable 1/2 spin fermions. **Mass is a form of energy**. The rest energies of the fermions give the **proton/electron mass ratio**. Potential excitations have energies equal to the **weak boson masses** allowing a geometric interpretation of **Weinberg's angle**. SU(2)$_Q$ gives **the anomalous magnetic moments** of proton, electron, neutron and generates **nuclear range attractive potentials** strong enough to produce the **binding energies of the deuteron and other nuclides**. **Lepton and meson masses** are due to topological excitations. The geometric mass spectrum is satisfactory. The **proton has a triple structure**. The **alpha constant** is a geometric number.

[a] Webpage http://prof.usb.ve/ggonzalm

# 1. Energy.

Our physical notions of energy, space, mass, force and time are local manifestations of a nonlinear physical geometry. We are able to sense and experiment this geometry which appears to be generated by the evolution of matter currents. In many present day geometric physical theories it is implicitly assumed that matter is linearly made of particles related to points, strings, membranes, etc. Nevertheless, this assumption may not be sufficient nor necessary. Instead, it appears necessary to assume that the action of physical matter currents $J$ nonlinearly determines a physical geometry which reacts back on the current. The reaction is locally experienced by matter as the action of an interaction potential $A$ which may be represented by a geometric connection $\omega$ associated to an interaction group [1] sufficient to account for the physical energy effects. Energy, or the action-reaction (or interaction) $J.A$, is the fundamental dynamical concept. The group generates infinitesimal excitations of the geometry which are representations of the group and behave as physical particles. We call this group the structure group of the physical theory.

Starting from space and time we shall inquire the properties of this geometry. In order to describe relativistic motion we need a space-time with 4 parameters or coordinates. Measurements require standard units which introduce a metric. Relativity allows us to say that space-time is locally an orthonormal space with a invariant metric under the Lorentz group SO(3,1). The use of the Lorentz group as structure group of a curved space-time leads to gravitation. We should require, at least, that the structure group also represent electrodynamics [1].

It is known that the structure group of electromagnetism is U(1). Since this group is complex we felt, in a first attempt, that in order to approach unification it was desirable to work with the spin group SL(2,C), rather than the Lorentz group itself SO(3,1). A gravitation theory related to SL(2,C) was discussed by Carmelli [2]. The simplest way to enlarge the group, apparently, was to use the group U(1) ⊗ SL(2,C) which is the group that preserves the metric associated to a tetrad induced from a spinor base.

It was known to Infeld and Van der Waerden [3, 4, 5], when using this group, that there appeared arbitrary fields which admitted interpretation as electromagnetic potentials because they obeyed the necessary field equations. To admit this interpretation we further required that the electrodynamic Lorentz force equation be a consequence of the field equations. Otherwise the equation of motion, necessarily implied by the nonlinear theory, contradicts the experimentally well-established motion of charged particles and the theory should be rejected.

This attempt [6], using U(1) ⊗ SL(2,C) as the group, led to a negative result, because the equations of motion depend on the commutators of the gravitational and electromagnetic parts which commute. This means that a charged particle would follow the same path followed by a neutral particle. This proves that it is not possible, without contradictions, to consider that the U(1) part represents electromagnetism as suggested by Infeld and Van der Waerden. This also means that to obtain the correct motion we must enlarge the chosen group in such a way that the electromagnetic generators do not commute with the gravitational ones. It is not true that any structure group that contains SL(2,C) ⊗ U(1) as a subgroup gives a unified theory without contradicting the electrodynamic equations of motion. The correct classical motion is a fundamental requirement of a unified theory.

In addition to this commutation physical problem there are two mathematical problems due to the SO(3,1) group. The second problem is that orthogonal groups are double valued due to their quadratic metric. The third problem is the definition of the square root of unit vectors associated to the metric signature.

Clifford algebras were developed to solve the third problem for general orthonormal spaces $R^{m,n}$. The geometric elements in the algebras include real numbers R, complex numbers C, anticommuting quaternions H, Clifford operators $Cl(3,1)...Cl(m,n)$. The groups generated by $Cl(m,n)$ solve the three problems. They are also useful in taking the square root of operators and have a richer mathematical structure which determines a higher predictive power.

Therefore, we should use the maximal group of Clifford algebra transformations which preserve the metric structure of the associated orthonormal tetradimesional space. The construction of the theory is accomplished [7, 8, 9] by taking the group of correlated spinor automorphisms of the space-time geometric algebra as the structure group $G$ of the theory. This group is essentially SL(4,R). This construction appears to be sufficient to obtain a structure group which describes all known physical energy interactions and corresponds to Hilbert's sixth problem [10].

## 1.1. Extension of Relativity.

Associated to any orthonormal flat space there is a Clifford geometric algebra [11]. There are inclusion mappings $\kappa$ of the orthonormal space into the algebra, mapping orthonormal vector bases to orthonormal sets of the algebra. The different images of a base determine a subspace of the algebra. The geometrical reason for the introduction of these algebras is to obtain geometric objects whose square is the negative of the scalar product of a vector $x$ with itself,

$$\left(\kappa\left(x\right)\right)^2 = -x.x\,I = -g\left(x,x\right)I \quad . \tag{1.1.1}$$

In a sense, this is a generalization of the introduction of imaginary numbers for the real line. These algebras are useful in defining square roots of operators.

For tridimensional euclidian space, the even Clifford subalgebra also has the structure of the Lie algebra of the SU(2) group, 2 to 1 homomorphic to the rotation group. SU(2) transformations by $2\pi$ and $4\pi$ are different but associated to a rotation by $2\pi$. Furthermore, it is known that a rotation by $4\pi$ is not geometrically equivalent to a rotation by $2\pi$ when its orientation entanglement relation with its surrounding is considered [12]. To preserve this geometric difference in a space-time subspace we must require the use of, at least, the even geometric subalgebra for the treatment of a relativistic space-time.

When the complete algebra is defined for Minkowski space-time, the observer orthonormal tetrads are mapped to orthonormal sets of the algebra. Now the number of possible orthonormal sets in the algebra is much larger than the possible orthonormal tetrads in space-time. There are operations, within the algebra, that transform all possible orthonormal sets among each other. These are the inner automorphisms of the algebra. Geometrically this means that the algebra space contains many copies of the orthonormal Minkowski space. A relativistic observer may be imbedded in the algebra in many equivalent ways. It may be said that a normal space-time observer is algebraically "blind". Usually the algebra is restricted to its even part, when the symmetry is extended from the Lorentz group (automorphisms of space-time) to the corresponding spin group SL(2,C) [13], (automorphisms of the even subalgebra). In this manner a fixed copy of Minkowski space is chosen within the geometric algebra. This copy remains invariant under the spin group.

The situation is similar to the imbedding of a three dimensional observer carrying a spatial triad into tetradimensional space-time. This imbedding is not unique, depending on the relative state of motion of the observer. There are many spatial tridimensional spaces in space-time, defining the concept of  simultaneity which is different for observers with different constant velocities. These possible physical observers may all be transformed into each other by the group of automorphisms of space-time, the Lorentz group.

This similarity allows the extension of the principle of relativity [1] by taking as structure group the group of correlated automorphisms of the space-time geometric algebra spinors instead of the group of automorphisms of space-time itself or only the automorphisms of the even subalgebra spinors. A relativistic observer carrying a space-time tetrad is imbedded in the geometric algebra space in a nonunique way, depending on some bias related to the orientation of a tetradimensional space-time subspace of the sixteen dimensional algebra. As Dirac once pointed out, we should let the geometrical structure itself lead to its physical meaning.

We may conceive complete observers which are not algebraically "blind". These observers should be associated to different but equivalent orthonormal sets in the algebra. Transformations among complete observers should produce algebra automorphisms, preserving the algebraic structure. This is the same situation of special relativity  for space-time observers and Lorentz transformations.

In particular the inner automorphisms of the algebra are of the form

$$a' = g^{-1} a g   ,    \tag{1.1.2}$$

where $g$ is an element of the largest subspace contained in the algebra that constitutes a group. This action corresponds to the adjoint group acting on the algebra.

For the Minkowski orthonormal space, denoted by R$^{3,1}$ , the Clifford algebra $Cl(3,1)$=R$_{3,1}$ is $\mathcal{Q}(2)$, where $\mathcal{Q}$ may be called the ring of pseudoquaternions [9] and the corresponding group is GL(2,$\mathcal{Q}$). The adjoint of the center of this group, acting on the algebra, corresponds to the identity. The quotient by its normal subgroup R$^+$ is the simple group SL(2,$\mathcal{Q}$). Therefore, the simple group nontrivially transforming the complete observers among each other is SL(2,$\mathcal{Q}$). This group is precisely the group $G$ of correlated automorphisms of the spinor space associated to the geometric algebra. We may associate a spinor base to a complete observer. A transformation by $G$ of a complete observer into another produces an adjoint transformation of the algebra and, consequently, a transformation of an orthonormal set onto an equivalent set in a different Minkowski subspace of the algebra. The metric in the equivalent Minkowski spaces in the algebra is the same. These transformations preserve the scalar products of space-time vectors mapped into the algebra. The subgroup of $G$ whose $Ad(G)$, in addition, leaves invariant the original Minkowski subspace, is known as the Clifford group. The Spin subgroup of the Clifford group is used in physics, in a standard manner, to extend the relativity principle from vectors to spinors. Since there are many copies of the spin group $L$ in SL(2,$\mathcal{Q}$), in our extension we have to choose a particular copy by specifying an inclusion map $i$. Apart from choosing an element of $L$, a *standard*  vector observer, we must choose $i$, thus defining a *complete* spinor observer

This complete observer, associated to a spinor base, carries, not only space-time information but also some other internal information related to the algebra [14]. The group SL(2,$\mathcal{Q}$) of transformations of these complete observers transforms the observations made by them. The observations are relative. The special relativity principle may be extended to this situation.

Since complete observers are themselves physical sytems we state the generalized principle as follows: All observers are equivalent under structure group transformations for stating the physical laws of natural systems and are defined by spinor bases associated to the geometric algebra of space-time.

The nonuniqueness of the orthonormal set has been known in geometry for a long time. We have given physical meaning to the orthonormal sets by associating them to physical observers. This implies that the physically allowable transformations are those mapping the algebra to itself by its own operations. We also have given a relativistic meaning to these transformations.

Furthermore, we should point out that our algebra is isomorphic to the usual Dirac algebra as a vector space but not as an algebra. The algebras correspond to space-times of opposite signature. The requirement to use a timelike interval to param-etrize the timelike world line of an observer determines that the appropriate algebra is not Dirac´s algebra $R_{1,3}$ but instead the algebra $R_{3,1}$, indicated here. The main practical difference is the appearance of a second compact subgroup related to electromagnetism and charge quantization as will be seen in the following sections and chapters. The algebra $R_{3,1}$ is discussed in the appendix.

The equation of motion of matter is the integrability condition of the field equation. It may be interpreted as a generalized Dirac equation with potentials given by the generators of the structure group $SL(2,\mathcal{Q})$ or its covering group $SL(1,\Lambda\otimes Q)$. The equation for the frame using the other K ring in the group $SL(2,Q)$ does not lead to Schrödinger's equation for a particle [15] as shown in [9, chapter 3, section 3.5.2]. From this point on, we shall use the notation $SL(2,\mathcal{Q})$ or its homomorphic $SL(4,R)$ to indicate these groups or the covering group, unless otherwise explicitly stated when it is convenient to distinguish them.

In general relativity [16] the space-time manifold is permitted to have curvature, special relativity is required to be valid locally and local observer frames are introduced, depending on their positions on space-time. In this manner we get fields of orthonormal tetrads on a curved manifold. The geometry of the manifold determines the motion, introducing accelerations of inertial and gravitational nature.

Similarly in our case, in order to include accelerated systems, we let space-time have curvature and introduce local complete observers that depend on their positions. But now, these observers are represented by general spinor frames which are subject to transformations beyond relativity (Ultra relativity). In this manner we get fields of spinor bases (frames) that geometrically are local sections of a fiber bundle with a curved base space. The geometry determines the evolution of matter, but now we have, in addition to inertial and gravitational accelerations, other possible accelerations due to other fields of force represented by the additional generators. These algebraic observers are accelerated observers. In other words we now get a geometrically unified theory with extra interactions (nuclear and particle) whose properties must be investigated.

# 1.2. Energy and the Field Equation.

The group $SL(4,R)$ is known not to preserve the corresponding metric. But, if we think of general relativity as linked to general coordinate transformations changing the form of the metric, it would be in the same spirit to use such a group. Instead of coordinate transformations whose physical meaning is associated with a change of observers, we have transformations belonging to the structure group whose physical meaning should be associated with a change of spinors related to observers. Representations of this group would be linked to matter fields. If we restrict to the even part of the group, taken as a subset of the Clifford algebra, we get the group $SL_1(2,C)$, used in spinor physics. Since $SL(4,R)$ is larger (higher dimensional) than $SL(2,C)$ it gives us an opportunity to associate the extra generators with energy interactions apart from gravitation and electromagnetism. The generator that plays the role of the electromagnetic generator must be consistent with its use in other equations of physics. The physical meaning of the remaining generators should be identified.

The field equations should relate the interaction connection $\omega$ or the physically equivalent interaction potential $A$ to a geometric object representing matter. We expect that matter is represented by a current $J(m)$ function of points $m$ on a space-time manifold $M$, valued in the group Lie algebra, rather than the nongeometric stress energy tensor $T$. The simplest object of this type is a generalized curvature $\Omega$ of $\omega$ or the physical generalized Maxwell tensor $F$. This generalized tensor obeys the Bianchi identity, which we write indicating the covariant exterior derivative by $D$,

$$DF \equiv D\Omega = 0 \tag{1.2.1}$$

The next simple object is constructed using Hodge duality. In similarity with the linear Maxwell's theory, we postulate the corresponding nonlinear field equation for the curvature as the generalized geometrical electrodynamic equation,

$$D^*F \equiv D^*\Omega = k^*J \ , \tag{1.2.2}$$

where matter is represented by the current $*J$, which must be a 3-form valued in the algebra, and $k$ is a constant to be identified later. Because of the geometrical structure of the theory the source current must be a geometrical object compatible with the field equation and the geometry. The structure of $J$, of course, is given in terms of some geometric objects acted upon by the potential. The geometrical properties of the curvature and the field equations determine that $J$ obeys an integrability condition,

$$DD^*F = \left[F, {}^*F\right] = 0 \ , \tag{1.2.3}$$

$$D^*J = 0 \ . \tag{1.2.4}$$

This relation being an integrability condition on the field equations, includes all self reaction terms of the matter on itself. A physical system would be represented by matter fields and interaction fields which are solutions to this set of nonlinear simultaneous equations. There should be no worries about infinities produced by self-reaction terms. As in the EIH method in general relativity [17], when a perturbation is performed on the nonlinear equations, for example to obtain linearity of the equations, the splitting of the equations into equations of different order brings in the concepts of field produced by the source, force produced by the field and therefore the self-reaction terms. These terms, not present in the original nonlinear system, are a problem introduced by this particular method of solution. In the zeroth order a classical particle moves as a test particle without self-reaction. In the first order the field produced by the particle produces a self correction to the motion.

Enlarging the group of the potential not only unifies satisfactorily gravitation and electromagnetism [8, 9], but requires other fields [14] and it appears to give a gravitational theory that differs, in principle, from Einstein's theory and resembles Yang's theory [18]. This may be seen from the field equations of the theory, which relate the derivatives of the Ehresmann curvature to a current source $J$.

The product of the interaction potential $A$ by the matter current $J$ has units of energy or inverse length. We may naturally define a fundamental unified geometric energy $\mathfrak{M}$ associated to the geometry, which defines the concept of mass and appears to be related to the concepts of inertia,

$$\mathfrak{M}(m) = \tfrac{1}{4}\mathrm{tr}\left(J^{\mu}(m)A_{\mu}(m)\right) \; . \tag{1.2.5}$$

This unification of the concept of energy and mass leads to important physical results.

A theory of connections without any other objects is incomplete from a geometrical point of view. A connection on a principal bundle is related to the structure group and the base space of the bundle. Representations of the group provide a natural vector fiber for an associated vector bundle on which the connection may be made to act. The geometric meaning of the physical potential is related to parallel translation of the elements of the fiber at different points throughout the base space. This is, essentially, a process of comparison of elements at different events.

A vector fiber space of this type has a base and the effect of the potential is naturally defined on the base. From a geometrical point of view the potential should be complemented by a vector base. It is well known that Einstein's gravitation theory may be expressed using an orthonormal tetrad instead of the metric [19]. In this theory we have taken this idea one step further, introducing a spinor base $e$ on the fiber space of an associated vector bundle $S$, in addition to the base on the fiber of the tangent space. In other words, we work with the base of the "square root space" of the usual flat space. The potential, which represents the gravitational and electromagnetic fields, depends on a current source term. We also postulate that this source current is built from fundamental matter fields that have the geometrical interpretation of forming a base $e$ on the fiber of the associated vector bundle and defines an orthonormal subset $\kappa$ of the geometric Clifford algebra,

$$^{*}\!J = \frac{1}{3!}\,\varepsilon_{\alpha\beta\gamma\mu}\tilde{e}\kappa^{\mu}e\,dx^{\alpha}\wedge dx^{\beta}\wedge dx^{\gamma} \quad . \tag{1.2.6}$$

This base $e$, when arranged as a matrix with the vectors of the base as columns, is related to an element of the group of the principal bundle. It is natural to expect that a base field $e$ (a section in geometric language), which we shall call a frame $e$, should obey equations of motion that naturally depend on the potential field. In fact, it will be seen that a particular solution of the integrability condition, the covariant conservation equation (1.2.4) of $J$, is obtained from the equation

$$\kappa^{\mu}\nabla_{\mu}e = 0 \quad . \tag{1.2.7}$$

This equation may be interpreted as a generalized Dirac equation since the structure group is SL(4,R) or its covering group .

We should note that whenever we have an sl(2,C) potential, there is a canonical coupling of standard gravitation to spin ½ particles obtained by postulating a Dirac equation which depends on a spin frame [20, 21, 22]. Nevertheless, strictly this does not represent a real unification. Our field equation implies integrability conditions in terms of $J$. Together with the geometric structure of $J$, our conditions imply the generalized Dirac equation which, therefore, is not required to be separately postulated, as in the previously mentioned nonunified case. The theory under discussion is not a mere pasting together of canonical gravitation and canonical electromagnetism for spin ½ particles. Rather, it is the introduction of a generalized geometric structure which nontrivially modifies both canonical theories and their coupling. Actually, the nonlinear field equation for the potential and the simplest geometric structure of the current are sufficient to predict this generalized Dirac equation and provide a unified concept of energy and mass.

If we introduce a variational principle [8,9] to obtain the two fundamental equations, (1.2.2) and (1.2.4), the principle determines a third related fundamental equation,

$$\mathrm{tr}\left[4F_{\hat{\rho}\nu}F^{\mu\nu} - u_{\hat{\rho}}^{\mu}F^{\kappa\lambda}F_{\kappa\lambda}\right] = k\,\mathrm{tr}\left[4e^{-1}\iota\circ u^{(\mu}\nabla_{\hat{\rho})}e - u_{\hat{\rho}}^{\mu}e^{-1}\iota\circ u^{\nu}\nabla_{\nu}e\right] \tag{1.2.8}$$

which represents an energy balance relation. It has been shown [8, 9] that the latter equation leads to the Einstein equation of gravitation in General Relativity with a geometric tensor source $T$.

Some of the features of the theory depend only on its geometry and not on a particular field equation and may be seen directly. For example, matter must evolve as a representation of SL(4,R) instead of the Lorentz group. It follows that matter states are characterized by three quantum numbers corresponding to the discrete numbers characterizing the states of a representation of SL(4,R). One of these numbers is spin, another is associated to the electromagnetic SU(2). This gives us the opportunity to recognize the latter as the electric charge [23].

Perhaps we should realize that the idea of a quantum entered Modern Physics by the experimental determination of the discreteness of electric charge. Later atomic measurements were explained by quantum theory by assuming the quantum of spin, but quantum theory was not given the burden to quantize electric charge. The possibility of obtaining the quantum of charge as explained before, may indicate that present day quantum theory is an incomplete theory as Dirac indicated [24].

As a bonus, this theory provides a third quantum number for a matter state, which may be recognized as a quantum of magnetic flux, providing a plausible fundamental explanation to the fractional quantum Hall effect [23].

As in general relativity, the integrability conditions imply the equations of motion for a classical particle, without knowledge of the detailed form of the source $J$, if we assume that $J$ has a multipole structure. The desired classical Lorentz equations of motion were obtained [1, 7].

Nevertheless the main objective at present, is not to describe the classical motion of matter exhaustively but rather to construct the geometrical theory and to require that it be compatible with the classical motion of the sources and with modern ideas in quantum theory. In particular, it appears, as first objective, to exploit the opportunity  provided by the theory to give a geometrical interpretation to the source current in terms of fundamental geometric matter field objects. If a geometric structure is given to $J$, the first stage in the construction of the unified concept of energy is completed.

## 1.3. Inertial Effects and Mass.

The proposed nonlinear equation and its integrability condition have peculiar aspects that distinguish them from standard equations in classical physics. Normally coupled field equations and equations of motion, for example Maxwell and Lorentz equations, in presence of a current source do not provide, by themselves, a static internal solution for a source that may represent a particle or object under the influence of its own field. Use of delta functions for point particles avoid the problem rather than solve it, and may introduce self accelerated solutions [25, 26]. The choice of current density in the theory, together with the interpretation developed allows a discussion on different grounds. The frame $e$ that enters in the current represents matter. Since a measurement is always a comparison between similar objets, a measurement of $e$ entails another frame $e'$ to which its components are referred. If we choose the referential $e'$ properly we may find interesting solutions. In particular we can find a geometric background solution, which we call the substratum, whose excitations behave as particles.

The integrability condition of the nonlinear equation leads to a generalized Dirac equation for the motion of matter,  with a parameter that may be identified with a mass defined in terms of energy [7, 8, 9]. The recognition of a single concept of mass is fundamental in General Relativity and merits discussion of possible solutions of the coupled equations.

If we identify a geometrical excitation with a physical particle, the Dirac equation for a linear excitation, which now would be the linear equation for a particle, contains parameters provided by the curved (nonlinear) background solution which we shall call its  substratum solution. Some of the particle properties could be determined by a substratum geometry. In particular a mass parameter arises for the frame excitation particle from the mass-energy concept defined in terms of energy. It is clear that this parameter is not calculable from the linearized excitation equation but rather from a nonlinear substratum solution.This appears interesting, but requires a knowledge of a substratum solution to the nonlinear field equation. Thus it is necessary to find a nonlinear solution, the simpler the better, that could illustrate this ideas. It is in this context that the following solution is presented.

The nonlinear equations of the theory are applicable to an isolated physical system interacting with itself. Of course the equations must be expressed in terms of components with respect to an arbitrary reference frame. A reference frame adapted to an arbitrary observer introduces arbitrary fields which do not contain any information related to the physical system in question. The only nonarbitrary reference frame is the frame defined by the physical system itself.

Any excitation must be associated to a definite background solution or substratum solution. An arbitrary observation of an excitation property depends on both the excitation and its substratum, but the physical observer must be the same for both excitation and substratum. We may use the freedom to select the reference frame to refer the excitation to the physical frame defined by its own substratum.

We have chosen the current density 3-form $J$ to be

$$J^{\mu} = \tilde{e}\kappa^{\hat{\alpha}}u_{\hat{\alpha}}^{\mu}e \quad , \tag{1.3.1}$$

in terms of the matter spinor frame $e$ and the orthonormal space-time tetrad $u$ .

Since we selected that the substratum be referred to itself, the substratum matter local frame $e_m$, referred to $e_r$ becomes the group identity $I$. Actually this generalizes comoving coordinates (coordinates adapted to dust matter geodesics) [27]. We adopt coordinates adapted to local substratum matter frames (the only nonarbitrary frame is itself, as are the comoving coordinates). If the frame $e$ becomes the identity $I$, the comoving substratum current density becomes a constant. Comparison of an object with itself gives trivial information. For example free matter or an observer are always at rest with themselves, no velocity, no acceleration, no self forces, etc. In its own reference frame these effects actually disappear. This substratum represents inert matter. Only constant self energy terms, determined by the nonlinearity, make sense and should be the origin of the constant bare inertial mass parameter.

At the small distance $\lambda$, characteristic of excitations, the elements of the substratum, both connection and frame, appear symmetric, independent of space-time. We should remember that space-time $M$ is, mathematically, a locally symmetric space or hyperbolic manifold [28]. We recognize these as the necessary condition for the substratum to locally admit a maximal set of Killing vectors [29] that should determine the space-time symmetries of the connection (and curvature). This means that there are space-time Killing coordinates such that the connection is constant and nonzero in the small region of particle interest. A flat connection does not satisfy the field equation. The excitations may always be taken around a symmetric nonzero connection or potential.

In particular the nonlinear equation admits a local nonzero constant potential solution. This would be the potential determined by an observer at rest with the matter frame. Of course, this solution is trivial but since the potential has units of energy, mass or inverse length, this actually introduces fundamental dimensions in the theory. Furthermore, a constant nonzero solution assigns a constant mass parameter $m$ to a fundamental particle excitation and allows the calculation of mass ratios of particles by integration of $\mathfrak{m}$ on symmetric spaces [9]. The result is obtained in terms of the dimensionless coupling constant in eq. (1.2.2). Therefore, we wish to find a constant nonzero inert (trivial) solution to the nonlinear field equation which we shall call the inert substratum solution.

First we look into the left side of the field equation (1.2.2), and notice that for a constant potential form $A$ and a flat metric, the expression reduces to triple wedge products of $A$ with itself [9], which may be put in the form of a polynomial in the components of $A$. This cubic polynomial represents a self interaction of the potential field since it may also be considered as a source for the differential operator.

Rather than work with the whole group $G$ we first restrict the group to the 10 dimensional $Sp(4,\mathbb{R})$ subgroup. Furthermore we desire to look at the nongravitational part of the potential. Hence, we limit the components of the potential to the Minkowski subspace defined by the orthonormal set, which is the coset $Sp(4,\mathbb{R})/SL(2,C)$. This is possible because if the potential is odd so is the triple product giving an odd current as required. The substratum solution is

$$A_s = -\sqrt[3]{\pi\alpha/3}e^{-1}\kappa_{\hat{\alpha}}edx^{\hat{\alpha}} + e^{-1}de = -\frac{1}{4}\mathfrak{m}J + e^{-1}de = \Lambda + e^{-1}de \ . \tag{1.3.2}$$

where $\mathfrak{m}$ is a constant determining the equipartition of excitation energy. This inert potential is essentially proportional to the current, up to an automorphism. It should be noted that in the expression for $A$, the term containing the current $J$ defines a potential tensorial form $\Lambda$. Its subtraction from $A$ gives an object, $e^{-1}de$, that transforms as a potential or connection. This solution may be extended from the subgroup to the whole 15 dimensional $SL(4,R)$ group using complex coordinates on the complex coset $SL(4,R)/SL(2,C)$. We call this solution the complex inert substratum [9].

For any potential solution $A$ or connection we can always define a new potential or connection by subtracting the tensorial potential form $\Lambda$ corresponding to the substratum solution, eq. (1.3.2),

$$\widehat{A} \equiv A - \Lambda = A + \frac{\mathfrak{m}}{4}J \ \ . \tag{1.3.3}$$

In terms of the new potential defined by equation (1.3.3) the equation of motion (1.2.9), in induced representations, explicitly displays the term depending on the substratum mass required by the Dirac equation. Using the algebraic relations among the orthonormal subset $\kappa$ of the geometric Clifford algebra we get,

$$\kappa^{\mu}\nabla_{\mu}e = \kappa^{\mu}\left(\partial_{\mu}e - eA_{\mu}\right) = \kappa^{\mu}\left(\partial_{\mu}e - e\widehat{A}_{\mu} + e\left(\frac{\mathfrak{m}}{4}J_{\mu}\right)\right) = \kappa^{\mu}\widehat{\nabla}_{\mu}e + \frac{\mathfrak{m}}{4}\kappa^{\mu}\kappa_{\mu}e = \kappa^{\mu}\widehat{\nabla}_{\mu}e - \mathfrak{m}e = 0 \ , \tag{1.3.4}$$

which explicitly shows the geometric energy $\mathfrak{m}$ as the germ of the Dirac particle mass. This equation is nonlinearly coupled to the field equation (1.2.2) through the definition of $J$, eq. (1.2.6).

## 1.4. The Classical Fields.

The curvature of this geometry is a generalized curvature associated to the group $SL(4,R)$. Since it is known that the even subgroup of $SL(4,R)$ is the Spin group related to the Lorentz group, we look for a limit theory to get this reduction. When ultra relativistic effects are small, we expect that we can choose bases so that the odd part is small of order $\varepsilon$. This is accomplished mathematically by contracting the $SL(4,R)$ group with respect to its odd subspace [30]. In the contracted group this odd subspace becomes an abelian subspace. Then the $SL(4,R)$ curvature reduces to

$$F = F_{+} + O\left(\varepsilon\right) \ , \tag{1.4.1}$$

where $F_{+}$ is the curvature of the even subgroup $SL_{1}(2,C)$

The result is that in this limit the curvature reduces to its even component which splits into an $SL(2,C)$ curvature and a separate commuting $U(1)$ curvature. It is known that an $SL(2,C)$ curvature may represent gravitation [31] and a $U(1)$ curvature may represent electromagnetism [32, 33, 34].

If we take this limit $U(1)$ as representing the standard physical electromagnetism we must accept that, in the full theory, electromagnetism is related to the $SU(2)$ subgroup of $SL(4,R)$ obtained using the inner automorphisms. Similarly the $SL(2,C)$ of gravitation may be transformed into an equivalent subgroup by an automorphism. This ambiguity of the subgroups within *G* represents a symmetry of the interactions. Since the noncompact generators are equivalent to space-time boosts, their generated symmetry may be considered *external*. The *internal* symmetry is determined by the compact nonrotational SU(2) sector.

It is well known in special relativity, that motion produces a relativity of electric and magnetic fields. We find, since SL(4,R) acts on the curvature, an intrinsic relativity of the unified fields, altering the nonunified fields that are seen by an observer. Given the orthonormal set corresponding to an observer, the SL(4,R) curvature may be decomposed in terms of a base generated by the set. The quadratic terms correspond to the SL(2,C) curvature and its associated Riemann curvature seen by the observer. A field named gravitation by an observer, may appear different to another observer. These transformations disguise interactions into each other.

The algebra associates some generators to space-time and simultaneously to some interactions. This appears surprising, but on a closer look this is a natural association. In an experiment, changes due to an interaction generator are interpreted by an observer as time and distance which become parameters of change. Then, it is natural that a reorientation, a gyre of space-time within the algebra corresponds to a rearrangement of interactions. A complete observer has the capacity to sense forces not imputable to his space-time riemannian curvature. He senses them as nongravitational, nonriemannian, forces. This capacity may be interpreted as the capacity to carry some generalized charge corresponding to the nongravitational interactions. Ultra relativity is essentially interpreted as an intrinsic relativity of energy interactions.

We should separate the equations with respect to the even subalgebra or subgroup as indicated by eq. (1.4.1), because this part represents the classical fields. In other words, we have the $sl_1(2,C)$ forms, as functions of its generators $\iota$, $E$,

$$A_+ = A\iota I + \Gamma^a E_a \quad, \tag{1.4.2}$$

$$F_+ = {}^n F \iota I + {}^n R^a E_a \quad. \tag{1.4.3}$$

It should be noted that the even curvature does not just arise from the even part of the connection because it depends on the product of odd parts.

The curvature $F$ of the abelian even connection $A$ corresponds to the Maxwell curvature tensor in electrodynamics and obeys

$$D * F = d * F = k * j \equiv 4\pi\alpha * j \quad, \tag{1.4.4}$$

$$DF = dF = 0 \quad, \tag{1.4.5}$$

which are exactly the standard Maxwell's equations if we define $k$ in terms of the fine structure constant $\alpha$. The standard $j.A$ interaction is obtained back.

The curvature form $R$ of the $\Gamma$ connection corresponds to the Riemann curvature with torsion, in standard spinor formulation. They obey the equations

$$D * R = k * J_+ \quad, \tag{1.4.6}$$

$$DR = 0 \quad, \tag{1.4.7}$$

which are not Einstein's equations but represent a spinor gravitation formulation equivalent to Yang's [18] theory restricted to its SO(3,1) subgroup.

It may be claimed that the Einstein equation of the geometric unified theory is equation (1.2.8) rather than the field equation. In general this equation may be written in terms of the Einstein tensor $G$ [8, 9]

$$G_{\rho\mu} = \frac{3}{{}^n R}\Big[8\pi\, {}^j\Theta_{\rho\mu} - 8\pi\Big({}^g\Theta_{\rho\mu} + {}^t\Theta_{\rho\mu} + {}^e\Theta_{\rho\mu}\Big)\Big] \quad. \tag{1.4.8}$$

We have then a generalized Einstein equation with geometric stress energy tensors.

When we consider the external field problem, that is, space-time regions where there is no matter, the gravitational part of the field equations for $J=0$ are similar to those of Yang's gravitational theory. All vacuum solutions of Einstein's equation are solutions of these equations. In particular the Schwarzchild metric is a solution and, therefore, the newtonian motion under a $1/r$ gravitational potential is obtained as a limit of the geodesic motion under the proposed equations.

Nevertheless, in the interior field problem there are differences between the unified physical geometry and Einstein's theory. For this problem where the source $J$ is nonzero, our theory is also essentially different from Yang's theory. Since in the physical geometry the metric and the so(3,1) potential remain compatible, the base space remains pseudoriemannian with torsion avoiding the difficulties discussed by Fairchild and others for Yang's theory. This equation may be interpreted as the existence of an apparent effective stress energy tensor which includes contributions from "dark matter or dark energy".

## 1.5. Results.

The physical universe is described by matter equations associated to an evolution group. The group is obtained from geometric algebraic tranformations of relativistic space and time. The related algebra is essentially the Pauli matrices multiplied by the quaternions, as indicated in the appendix. The interaction is represented by field equations and equations of motion in terms of potential and force tensors determined by matter transformation currents. We identify the concept of energy using the potential and current fields.

A nonlinear particular solution, which may represent inertial properties, was given. The concept of  bare inertial mass is related to this solution. Microscopic physics is seen as the study of linear geometric excitations.

The results obtained indicate that gravitation and electromagnetism are unified in a nontrivial manner. There are additional generators that may represent non classical nuclear and particle interactions [9]. Strong nuclear electromagnetism is related to an SU(2)$_Q$ subgroup. If we exclusively restrict to a U(1) subgroup we obtain Maxwell's field equations. We have a generalized Einstein equation with geometric stress energy tensors and possible dark matter effects.

# 2. Quanta.

The geometrical structure of the theory, in terms of a potential describing the interaction and a frame describing matter, is determined by the field equations which relate the generalized curvature to the matter distribution. Nevertheless some aspects of the theory may be discussed by group techniques without necessarily solving the field equations.

The structure group of the theory, SL(2,$\mathcal{Q}$), the group of spinor automorphisms of the universal Clifford algebra of the tangent space at a space-time point, acts on associated spinor bundles and has generators that may represent other interactions, apart from gravitation and electromagnetism. A consequence of the theory is the necessary association of electromagnetism to an SU(2) subgroup of SL(2,$\mathcal{Q}$).

In order to give a complete picture, the frame fields should represent matter and, consequently, the frame excitations should represent particles. The potential acts on the frame excitations which may be considered linear representations of the group. It is of interest to consider the irreducible representations of this group and to discuss its physical interpretation and predictions [8]. Representations of related groups have been discussed previously [35].

## 2.1. Induced Representations of the Structure Group G.

The irreducible representations of a group $G$ may be induced from those of a subgroup $H$. These representations act on the sections [36] of a homogeneous vector bundle over the coset space $M = G/H$ with fiber the carrier space $U$ of the representations of $H$.

It is then convenient to consider the subgroups contained in SL(2,$\mathcal{Q}$), in particular the maximal compact subgroup. The higher dimensional simple subgroups are as follows:

1. The 10-dimensional group $P$, generated by $\kappa_\alpha$, $\kappa_{[\alpha}\kappa_{\beta]}$. This group is isomorphic to subgroups generated by $\kappa_{[\alpha}\kappa_{\beta]}\kappa_{\gamma]}$, $\kappa_{[\alpha}\kappa_{\beta]}$ and by $\kappa_{[a}\kappa_b\kappa_{c]}$, $\kappa_{[a}\kappa_{b]}$, $\kappa_5$;

2. The 6-dimensional group $L$, corresponding to the even generators of the algebra, $\kappa_{[\alpha}\kappa_{\beta]}$. This group is isomorphic to the subgroups generated by $\kappa_\alpha$, $\kappa_{[\alpha}\kappa_{b]}$ and by $\kappa_0\kappa_{[a}\kappa_{b]}$, $\kappa_{[a}\kappa_{b]}$;

3. There are two compact subgroups, generated by $\kappa_{[a}\kappa_{b]}$ and by $\kappa_0$, $\kappa_5$, $\kappa_1\kappa_2\kappa_3$.

The $P$ subgroup is, in fact, Sp(2,$\mathcal{Q}$), homomorphic to Sp(4,$\mathbb{R}$), as may be verified by explicitly showing that the generators satisfy the simpletic requirements [37]. This group is known to be homomorphic to SO(3,2), a De Sitter group. The $L$ subgroup is isomorphic to SL(2,$\mathbb{C}$). The compact subgroups are both isomorphic to SU(2) and therefore the maximal compact subgroup of the covering group $\bar{G}$ is SU(2)⊗SU(2) homomorphic to SO(4). The $L{\subset}P{\subset}G$ chain internal symmetry is SU(2)⊗U(1) and coincides with the symmetry of the weak interactions.

Irreducible representations of the covering group of SL(2,$\mathcal{Q}$) may be induced from its maximal compact subgroup. The representations are characterized by the quantum numbers associated to irreducible representations of both SU(2) subgroups.

These representations may be considered sections of an homogeneous vector bundle over the coset $\overline{\text{SL}(4,\mathbb{R})}$ /(SU(2)⊗SU(2)) with fiber the carrier space of representations of SU(2)⊗SU(2).

One of the SU(2), acting on spinor space, is associated to the rotation group, acting on vector space. Its irreducible representations are characterized by a quantum number $l$ of the associated Casimir operator $L^2$, representing total angular momentum squared. The other SU(2), as indicated in previous sections, may be associated to electromagnetism. It has a Casimir operator $C^2$, similar to $L^2$ but representing generalized total charge, with some quantum number $c$. The irreduciblerepresentations of SL(2,$\mathcal{Q}$) have a third label which may be associated to one of the Casimir operators of SL(2,$\mathcal{Q}$), for example, the quadratic operator $M^2$ on the symmetric space SL(4,R)/SO(4).

The states of these irreducible representations of SL(2,$\mathcal{Q}$) should be characterized by integers $m$, $q$ corresponding to physical particles with $z$ component of angular momentum $mh/2$ and charge $qe$. These quantum numbers are studied in detail in chapters 2 and 3. The price paid to obtain this charge quantization is only the association of electromagnetism to the second SU(2) in SL(2,$\mathcal{Q}$). In fact, this may not be a disadvantage in a unified theory of this type, where it may lead to new geometric representations of physical phenomena. It should be noted that Dirac's charge quantization scheme [38] requires the existence of magnetic monopoles. This requirement is not compatible with this theory.

Representations of SL(2,$\mathcal{Q}$) may also be induced from $L$, naturally including the spin representations of its SU(2) subgroup. Since $L$ is a subgroup of $P$, it is more interesting to look at the representations induced from $P$ which include, in particular, those of $L$. In some situations the holonomy group of the potential or connection may be $P=$ Sp(2,$\mathcal{Q}$) and we may expect that representations induced from it should play an interesting role. It should be clear that the coset $P/L$ is the De Sitter space $S$. [39]. The points of $S$ may be seen as translation operators on the space $S$ itself, in the same manner as the Minkowski space. The Laplace-Beltrami operators on $S$ have eigenvalues that should correspond to the concept of mass in this curved space. The isotropy subgroup at a translation (point) in $S$ is the rotation subgroup SO(3) for translations along lines inside the null 3-hypercone and the SO(2,1) subgroup for translations along the lines in the null 3-hypercone itself. The SL(2,$\mathcal{Q}$) representations induced from the De Sitter group $P$ are characterized by mass and spin or helicity and expressed as sections (functions) of a bundle over the mass shell 3-hyperboloid or the null 3-hypercone, respectively for massive or null mass representations as the case may be. These representations are solutions of Dirac's equation in this curved space and geometrically correspond to sections of bundles over subspaces of the De Sitter space $S$.

The group SO(3,2), homomorphic to Sp(2,$\mathcal{Q}$), is known [37] to contract to ISO(3,1) which is the Poincare group and we may have approximate representations of SL(2,$\mathcal{Q}$) related to the Poincare group. Then, the representations of the direct product of Wigner's little group [40] by the translations are introduced approximately in the theory. The SO(2,1) isotropy subgroup of $P$ contracts to the ISO(2) subgroup of ISO(3,1). Thus, we may work with representations characterized by mass and spin or helicity expressed as sections (functions) of a bundle over the mass shell 3-hyperboloid or the light cone, as the case may be. It should be clear from these considerations of groups that the *standard* Dirac equation, which is an equation for a representation of the Poincare group, plays an approximate role in the geometric theory.

We should point out that the isotropy subgroup of $P$ at a De Sitter translation (point) in a null subspace is the group of transformations that leaves invariant the null tangent vectors at the given translation. The isotropy subgroup is SO(2,1) acting on the curved De Sitter translations space or its contraction ISO(2) acting on the flat Minkowski translations space. In either case there is only one rotation generator, which corresponds to the common compact SO(2) subgroup in both isotropy subgroups. Therefore, the null mass representations induced from $P$ are characterized by the corresponding eigenvalue or SO(2) quantum number which is known as helicity. For the rest of the chapter we shall restrict ourselves to massive representations.

In order to study the irreducible representations of the group SL(2,$\mathcal{Q}$), we need to introduce the complex extension of this group, which is SL(4,$\mathbb{C}$). The relation of induced representations with those obtained in Cartan's approach [41], is discussed in [9]. The Cartan subspace of the complex algebra sl(4,$\mathbb{C}$), also known as the root space A$_3$, describes the commutation relations of the canonical Cartan generators of this complex algebra and all its real forms sl(4,$\mathbb{R}$), su(4), su(3,1), su(2,2) and su*(4) [9]. In particular, the real form sl(4,$\mathbb{R}$) is the least compact real form of sl(4,$\mathbb{C}$).

## 2.2. Relation Among Quantum Numbers.

The relation of the quantum numbers associated to the standard Cartan generators $G_i$ with the quantum numbers of the representations induced from the maximal compact subgroup may be seen by considering the corresponding Cartan subalgebras. There are other Cartan subalgebras in sl(4,$\mathbb{R}$). It is known that a Cartan subalgebra is not uniquely determined but depends on the choice of a regular element in the complete algebra. Nevertheless, it is also known that there is an automorphism of the *complex* algebra that maps any two Cartan subalgebras [41]. This implies there is a relation between the sets of quantum numbers associated to two different Cartan subalgebras within sl(4,$\mathbb{R}$). We may take the quantum numbers linked to spin and charge as the physically fundamental quantum numbers and consider the others that arise by use of different Cartan subalgebras as numbers which are functions of the fundamental ones.

For physical reasons, since electromagnetism and electric charge are associated to the SU(2) subgroup generated by $\kappa_0$, $\kappa_1\kappa_2\kappa_3$, $\kappa_0\kappa_1\kappa_2\kappa_3$ within our geometric theory, it is of interest to consider an algebra decomposition with respect to one of these generators giving a new set of generators $X$. Since all three of them commute with all the generators of the spin SU(2), $\kappa_1\kappa_2$, $\kappa_2\kappa_3$, and $\kappa_3\kappa_1$, we have different choices at our disposal. For convenience of interpretation we choose $\kappa_1\kappa_2$, and $\kappa_0\kappa_1\kappa_2\kappa_3$ as our starting point. The only other generator that commutes with them is $\kappa_0\kappa_3$. It has been shown [9] now that neither $\kappa_0\kappa_1\kappa_2\kappa_3$ nor $\kappa_1\kappa_2$ are regular elements of the total sl(4,$\mathbb{R}$) algebra in spite of being regular elements of the two su(2) subalgebras. Each of the two matrices, $\kappa_1\kappa_2$ and $\kappa_0\kappa_1\kappa_2\kappa_3$, generates a subspace $V_o$, corresponding to eigenvalues $\lambda=0$, which is seven-dimensional. Since the Cartan subalgebra of sl(4,$\mathbb{R}$) is tridimensional, it follows that neither of the two generators is a regular element.

Nevertheless the sum generator, $\mathrm{ad}\left(\kappa_1\kappa_2 + \kappa_0\kappa_1\kappa_2\kappa_3\right)$ has a $V_O$ subspace, for $\lambda=0$, which is tridimensional. Therefore, the sum generator is a regular element of the Lie algebra.

The $V_O$ space generated by this regular element is a Cartan subalgebra which is spanned by the generators

$$X_1 = \kappa_1 \kappa_2 \quad , \tag{2.2.1}$$

$$X_2 = \kappa_0 \kappa_1 \kappa_2 \kappa_3 \quad , \tag{2.2.2}$$

$$X_3 = \kappa_0 \kappa_3 \quad , \tag{2.2.3}$$

where the product is understood in the enveloping Clifford algebra.

It is clear that $X_1$, and $X_2$ are compact generators and therefore have imaginary eigenvalues. Because of the way they were constructed, they should be associated, respectively, to $z$-component of angular momentum and electric charge. Both of them may be diagonalized simultaneously in terms of their imaginary eigenvalues, leaving $X_3$ invariant. When dealing with compact elements of a real form, as spin, it is usual to introduce the standard notation in terms of the corresponding noncompact real matrices of the real base in the complex algebra.

The Clifford algebra matrices provide a geometric normalization of roots and weights in the Cartan subspace. The weight vectors in the base $X_i$ have the same structure of those in Cartan's canonical base except for a standard normalization factor of $(32)^{-1/2}$. Similarly, the roots in both bases have the same structure, differing by the standard normalization factor:

## 2.3. Spin, Charge and Flux.

It may be seen that, in the fundamental representation, one of the generators in the Cartan subalgebra may be expressed as a Clifford product (not a Lie product) of the other generators of the subalgebra,

$$X_1 = X_2 X_3 \quad . \tag{2.3.1}$$

This implies that, within the Clifford algebra, there is a multiplicative relation among the quantum numbers in the theory. In particular, the $z$-component angular momentum generator, spin, is the Clifford product of the electric charge generator and the $X_3$ generator. The quantum numbers associated to $X_3$ must have the physical meaning of angular momentum divided by electric charge, or equivalently, magnetic flux. Then the fundamental quantum of action should be the product of the fundamental quantum of charge times the fundamental quantum of flux,

$$h/2 = e\left(h/2e\right) \quad . \tag{2.3.2}$$

We may intuitively interpret the last equation as a quantum betatron effect, when a quantum change in magnetic flux is related to a quantum change in angular momentum.

We have taken the quantum of action in terms of $h$ rather than $\hbar$ because the natural unit of frequency is cycles per second. Then the quantum of flux $\phi_0$ is

$$\phi_0 = h/2e = \pi\hbar/e \quad . \tag{2.3.3}$$

The four members of the fundamental irreducible representation form a tetrahedron in the tridimensional $A_3$ Cartan space as shown in figure 1. They represent the combination of the two spin states and the two charge states of an associated particle, which we shall call a G-particle. One charge state represents a physical particle state and the other represents a charge conjugate state. The G-particle carries one quantum each of angular momentum, electric charge and magnetic flux and may be in one of the four state whose quantum numbers are:

| spin | charge | flux | | |
|------|--------|------|---|---|
| $+1$ | $-1$ | $-1$ | negative charge with spin up | $f_-^{\uparrow}$ |
| $-1$ | $-1$ | $+1$ | negative charge with spin down | $f_-^{\downarrow}$ |
| $+1$ | $+1$ | $+1$ | positive charge with spin up | $f_+^{\uparrow}$ |
| $-1$ | $+1$ | $-1$ | positive charge with spin down | $f_+^{\downarrow}$ |

The dual of the fundamental representation, defined by antisymmetric tensor products of 3 states of the fundamental representation or triads, corresponds to the inverted tetrahedron. A conjugate state or a dual state may be related to an antiparticle. This may be useful but is not a necessary interpretation. It is better to keep these mathematical concepts physically separate. We consider, on one hand, fundamental excitations and dual excitations and, on the other hand, excitations with particle and conjugate states.

The states of an irreducible representation of higher dimensions, built from the fundamental one, are also characterized by three integers: angular momentum $m$, electric charge $q$ and magnetic flux $f$. We may conjecture then that the magnetic moment is not as fundamental as the magnetic flux when describing particles.

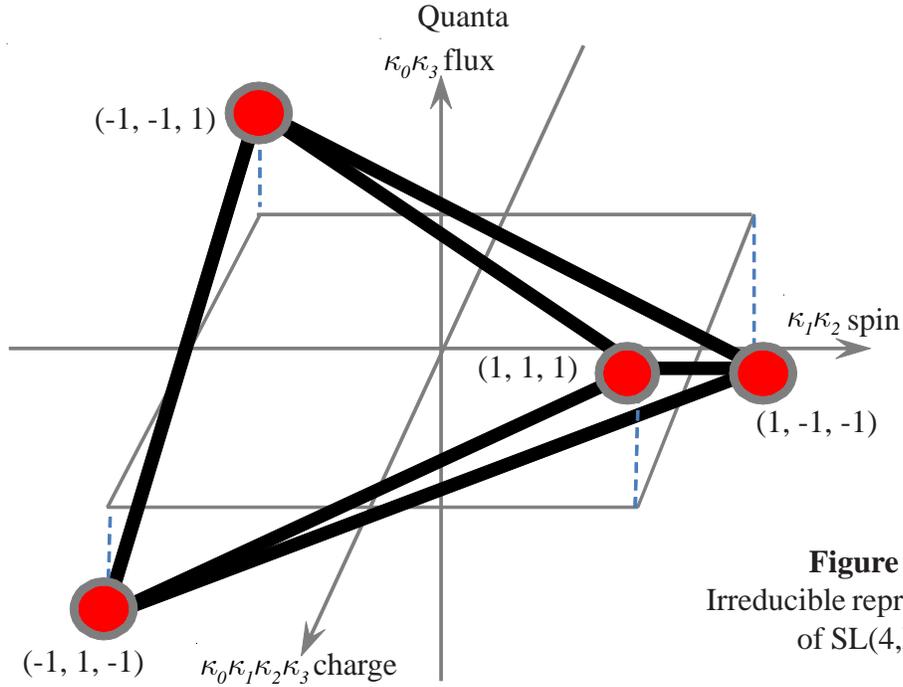

**Figure 1.**
Irreducible representation
of SL(4,R).

## 2.4. Representations of a Subgroup P.

It is known that the $SL(2,\mathbb{C})$ subgroup has a unidimensional Cartan subspace of type $A_1$ associated to the quantized values of angular momentum. The $Sp(4,\mathbb{R})$ subgroup has a bidimensional Cartan subspace of type $C_2$. In the same way we handled $SL(4,\mathbb{R})$, we may choose a regular element associated to the spin and flux generators in particular $\mathrm{ad}\left(\kappa_1\kappa_2 + \kappa_0\kappa_3\right)$, that annihilates the bidimensional subspace, with zero eigenvalues, spanned by the $\kappa_1\kappa_2$ and $\kappa_0\kappa_3$ generators. We may obtain the corresponding weight vectors and roots with the same previous normalization. Nevertheless we may also choose as regular element another element related to the charge in particular, $\mathrm{ad}\left(\kappa_1\kappa_2 + \kappa_0\right)$ which annihilates the bidimensional subspace, with zero eigenvalues, spanned by the generators $\kappa_1\kappa_2$ and $\kappa_0$.

In both cases the four members of the fundamental representation form a square in a bidimensional $C_2$ Cartan space, but in *different* Cartan subspaces of the $A_3$ root space. We may visualize the relation of these vectors with those of the full $SL(4,\mathbb{R})$ group recognizing that the tridimensional Cartan $A_3$ space collapses to a bidimensional $C_2$ subspace as indicated in figure 1. The tetrahedron which represents the states of the fundamental representation collapses to a square. The 4 tetrahedron vertices project to the 4 square vertices. The set of 4 weight vectors in $C_2$ may be obtained by projecting the 4 weight vectors in $A_3$, which correspond to the tetrahedron vertices, on the plane spanned by the vectors $\kappa_1\kappa_2$ y $\kappa_0\kappa_3$ or $\kappa_1\kappa_2$ and $\kappa_0$, forming squares in these Cartan subspaces. The 6 tetrahedron edges project to the 4 sides and 2 diagonals of each square. The opposite sides of a square are equivalent because they have the same directions. The set of 8 roots in $C_2$ may also be obtained by projecting the set of 12 roots in $A_3$, which correspond to the tetrahedron edges. In this case, 8 roots project on 4 degenerate root pairs (equivalent) corresponding to the square sides. The other 4 roots project to the 4 roots corresponding to the square diagonals. A collapse of the $C_2$ bidimensional spaces to the unidimensional $A_1$ Cartan space of $SL(2,\mathbb{C})$ produces a projection of the squares to the line segment that represents the 2 standard spin states and the 2 roots of the latter Cartan space, associated to an L-particle.

The fact that the Cartan bidimensional subspace is not uniquely determined has physical consequences within the given interpretation. The 4 states of the irreducible representation may be labeled by spin and flux or charge depending on the chosen regular element. But, as indicated before, both Cartan spaces are related by a complex algebra automorphism implying that one set of quantum numbers may be expressed as functions of the other set, in particular the flux quantum is a function of the spin and charge quanta. In this case, the relation is interpreted as the remnant of the multiplicative relation among spin, charge and flux in the fundamental representation of the parent group $SL(4,\mathbb{R})$. This indicates that a physical particle associated to this representation, which we shall call a P-particle, has the three quantum numbers. There are no continuous variables that represent the measurable values of spin, charge and flux of the P-particle. The difference between the G-particle and the P-particle is not displayed by these quantum numbers.

Three $sp(4,\mathbb{R})$ spaces may be injected as subalgebras into the three geometrically and algebraically independent sectors of the complete $sl(4,\mathbb{R})$ algebra of the group $G$. In this manner we may construct a section $p$ valued in $sl(4,\mathbb{R})$ by injecting three independent sections $(e_1, e_2, e_3)$ valued in the subalgebra $sp(4,\mathbb{R})$. Therefore, for any state $e$ of the fundamental representation of $P$, the dual $\tilde{e}$ formed by the triad $(e_1, e_2, e_3)$ of fundamental $P$-representation states determines a particle state $p$ of the fundamental representation of $G$. The corresponding states, equally charged, define a charge equivalence relation among these states,

$$\tilde{e} \cong (e_1, e_2, e_3) \cong p \quad . \tag{2.4.1}$$

We may choose any state in the fundamental representation of $G$ to physically represent the fundamental particle $p$ associated to this representation. The fundamental $(1, \pm 1, \pm 1)$ $SL(2, \mathcal{Q})$ states define a proton charge sign. Nevertheless, since the charges of corresponding $\tilde{e}$ and $p$ states are equivalent, we are free to define the charge of only one state in the $G$ y $P$ representations taken together. The charge corresponding to this chosen particle $p$ (the proton) may be defined to be positive. This **determines** an inequivalent negative charge for the corresponding state that physically represents the other fundamental particle $e$ (the electron) in the fundamental representation of $P$,

$$+1 \equiv Q(p) = Q(\tilde{e}) = -Q(e) \quad . \tag{2.4.2}$$

In other words the charge of the electron (dual) particle must have a sign opposite to the charge sign of the proton (dual) particle.

## 2.5. Magnetic Flux Quanta.

From the previous discussion it follows that any particle that is a fundamental representation of $SL(4, \mathbb{R})$ or $Sp(4, \mathbb{R})$ must be a charged particle with a quantum of flux. This quantum of flux is exactly the value determined experimentally by Deaver and Fairbank [42] and Doll and Nebauer [43], first predicted theoretically by London [44] with an extra factor of two. The experiment consisted in making a small cylinder of superconductor by electroplating a thin layer of tin on a copper wire. The wire was put in a magnetic field, and the temperature reduced until the tin became a superconducting ring. Afterwards the external field was eliminated, leaving a trapped minimum of flux through the ring.

Our result is consistent with the experimental result if we consider that the trapped flux through the cylindrical ring layer of superconducting tin is really associated to the discrete minimum intrinsic flux of a single electron inside the normal (not superconducting) copper wire which serves as nucleus for the superconducting tin. At present, it is generally believed that the quantization of flux is due to the topology of the superconducting material in the experiment and related to the charge of an electron pair inside the superconductor. The idea expressed here is that the quantum of flux is an intrinsic property of matter particles (electron, proton etc.) and only the possibility of trapping the flux depends on the topology of the superconductor.

The generalization of the Lorentz group (automorphisms of Minkowski space) to the group of automorphisms of the geometric algebra of Minkowski space and the corresponding physical interpretation indicate that any massive particle (or quasi particle) must carry not only quanta of angular momentum, but also quanta of charge $e$ and quanta of magnetic flux, $h/2e$, (*one or more*), both intrinsic and orbital,

$$\Phi = Nf\left(\frac{h}{2e}\right) \quad . \tag{2.5.2}$$

This equation provides a simple-minded interpretation for the Meissner effect [45] on superconductors. This expression should still be valid if the electrons are paired. If the pairing is such that the quanta of flux cancel each other, the value of $f$ should be zero for each pair and consequently the total magnetic flux inside the superconductor should also be zero.

Furthermore, if a particle crosses a line in a plane normal to its flux, there is a relation between the charge and flux crossing the line. If there are no resistive losses on a conductor along the line, the induced voltage leads to a transverse resistance which is fractionally quantized,

$$\frac{\Delta\Phi}{\Delta Q} = \frac{f\left(\frac{h}{2e}\right)}{qe} \quad . \tag{2.5.1}$$

This expression leads us to conjecture that the fractional quantized Hall effect (FQHE) [46, 47] gives evidence for the existence of these flux quanta instead of evidence for the existence of fractional charges. The FQHE experiment consists essentially of the measurement of the transverse Hall conductivity occurring at low temperature in bidimensional electron gas crystal interfaces in semiconductors.

## 2.6. Magnetic Energy Levels.

Usually the motion in a constant magnetic field is discussed in cartesian coordinates in terms of states with definite energy and a definite linear momentum component. The resultant Landau energy levels are degenerate in terms of the momentum component. Additionally, for the electron, the Landau energy levels are doubly degenerate except the lowest level. In order to use definite angular momentum states, the problem is expressed in cylindrical coordinates and the energy of the levels is [48, 49],

$$U = \left(n + m + \tfrac{1}{2} + s\right)\frac{eB\hbar}{Mc} \quad , \tag{2.6.1}$$

where $B$ is the magnetic field along the positive $z$ symmetry axis, $m$ is the absolute value of the orbital angular momentum quantum number, also along the positive $z$ axis, $n$ is a nonnegative quantum number associated to the radial wave function and $s$ is the spin. This expression has the peculiarity that even for zero $m$ there may be energy quanta associated to the radial direction.

The equations of motion of charged particles in a constant magnetic field, in quantum or classical mechanics, do not determine the momentum or the center of rotation of the particle. These variables are the result of a previous process that prepares the state of the particle. We can idealize this process as a collision between the particle and the field. Since the magnetic field does not do work on the particle, the energy of the particle is conserved if the system remains isolated after the initial collision. The angular momentum of the particle with respect to the eventual center of motion is also conserved. The values of angular momentum and energy inside the field equal their values outside. This implies there is no kinetic energy associated to the radial wave function, as the equation allows. Therefore, among the possible values of energy we must exclude all values of $n$ except the value zero. This means that the energy only depends on angular momentum, as in the classical theory.

The degeneracy of energy levels, apart from that due to the momentum component along the field, is the one due to the spin direction, as in the case of cartesian coordinates. Only moving electrons make a contribution to the Hall effect, hence we disregard zero orbital angular momentum states. Each degenerate energy level has two electrons, one with spin down and orbital momentum $m+1$ and the other with spin up and orbital angular momentum $m$. Each typical degenerate energy level may be considered to have a total charge $q$ equal to $2e$, total spin $0$ and a total integer orbital angular momentum $2m+1$. Using half integer units,

$$L_z = 2\left(2m+1\right)\left(\tfrac{\hbar}{2}\right) \quad . \tag{2.6.2}$$

In the classical treatment of motion in a magnetic field, it is known that the classical kinetic momentum of a particle has a circulation around the closed curve corresponding to a given orbit, that is twice the circulation of the canonical momentum. The latter, in turn, equals the negative of the external magnetic flux enclosed by the loop [50]. That is, for the given number of quanta of orbital angular momentum in a typical level, we should associate a number of orbital external magnetic flux quanta. In accordance with the previous equation, we assign a flux $\Phi_L$ to a degenerate level,

$$\Phi_L = -2\left(2m+1\right)\left(\tfrac{h}{2e}\right) \quad . \tag{2.6.3}$$

When we have an electron gas instead of a single electron, the electrons occupy a number of the available states in the energy levels depending on the Fermi level. The orbiting electrons are effective circular currents that induce a magnetic flux antiparallel to the external flux. The electron gas behaves as a diamagnetic body, where the applied field is reduced to a net field because of the induced field of the electronic motion. Associated to the macroscopic magnetization $M$, magnetic induction $B$ and magnetic field $H$ we introduce, respectively, a quantum magnetization flux $\Phi_M$, a quantum net $B$ flux $\Phi_B$ and a quantum bare $H$ flux $\Phi_H$ per level, which must be related by

$$\Phi_B = \Phi_H + \Phi_M \quad . \tag{2.6.4}$$

The first energy level of moving electrons is nondegenerate, corresponding to one electron with spin down and one quantum of orbital momentum. In order for the electron to orbit, there has to be a net (orbital) flux inside the orbit. If the flux is quantized, the minimum possible is one quantum of orbital flux for the electron. In addition, another quantum is required by the intrinsic flux asociated to the intrinsic spin of the electron. Therefore without further equations, the minimum number of net quanta in this level is 2 and its minimum net magnetic flux is

$$\Phi_B = \left(2\right)\frac{h}{2e} \quad . \tag{2.6.5}$$

It should be noted that the minimum flux attached to the electron, in this state, is twice the flux quantum. In other words, the attached flux is split into orbital and intrinsic parts.

The quantum magnetization flux is, considering its induced flux $\Phi_L$ opposed to $\Phi_H$, two negatve (-2) quanta for the orbital motion of the single electron and one (1) additional quantum for its intrinsic flux $\Phi_S$, which gives, for this nondegenerate level, a relation indicating an equivalent magnetic permeability of 2/3,

$$\left|\varPhi_B\right| = 2\left|\varPhi_M\right| \quad . \tag{2.6.7}$$

This reasoning is not applicable to the calculation of $\varPhi_B$ for the degenerate levels in the electron gas, rather, we somehow have to find a relation with $\varPhi_H$. To determine this exactly we would need to solve the (quantum) equation of motion for the electrons simultaneously with the (quantum) electromagnetic equation determining the field produced by the circulating electrons. Instead of detailed equations we recognize that, since flux is quantized, the orbital flux *must change by discrete quanta* as the higher energy levels become active. As the angular momenta of the levels increase, a proportional increase in any two of the variables in equation (2.6.4 ) would mean that the magnetic permeability does not vary from the value 2/3 determined by the nondegenerate level. Therefore, in general, the flux increases should obey the following quantum condition: the flux $\varPhi_B$ may deviate *discretely* from the proportionality indicated by the last relation, by an integral number of flux quanta according to,

$$\left|\varPhi_B\right| = 2\left|\varPhi_M\right| + \varDelta\varPhi \quad, \tag{2.6.8}$$

where $\varDelta\varPhi$ is an indeterminate quantum flux of the electron pair in the degenerate level. The magnetization flux $\varPhi_M$ has $\varPhi_L$ an upper bound and we may indicate the attached quanta by the inequality

$$\left|\varPhi_B\right| \leq 2\left[2\left(2m+1\right)h\!\!\!/_{2e} + \delta\left(h\!\!\!/_{e}\right)\right] \quad, \tag{2.6.9}$$

where $\delta$ is an integer indicating a jump in the flux associated to each electron in the pair.

Whatever flux $\varPhi_B$ results, it must be a function of only the flux quanta per electron pair. A degenerate energy level consists of the combination or coupling of two electrons with orbital momentum levels $m$ and $m+1$, in states of opposite spins, with $2\mu$ orbital flux quanta linked to the first electron and $2\mu+2$ linked to the other electron. The pair, has $2(2\mu+1)$ flux quanta, in terms of a nonnegative integer $\mu$. Hence the only independent variable determining the jump in the flux $\varPhi_B$ is $2(2\mu+1)$. In order for $\varPhi_B$ to be only a function of this variable, the integer $\delta$ must be even so that it adds to $m$, giving the possible values of the net flux of the level,

$$\varPhi_B = 4\left(2\mu+1\right)\left(h\!\!\!/_{2e}\right) \quad, \tag{2.6.10}$$

The physical interpretation of the last equation is the following: If the external magnetic field increases, both the effective net magnetic flux $\varPhi_B$ and the effective magnetization flux $\varPhi_M$ per electron increase discretely, by quantum jumps, and the equivalent permeability remains constant.

The number of possible quanta, indicated by $f$, linked to this level called a *superfluxed* level, is

$$f = 4\left(2\mu+1\right) \quad, \tag{2.6.11}$$

where the flux index $\mu$ has undetermined upper bound. The number of electric charge quanta per level, indicated by $q$, is

$$q = 2 \quad . \tag{2.6.12}$$

# 2.7. Fractional Quantum Hall Effects.

Each single electronic state is degenerate with a finite multiplicity. This gives the population of each typical energy half-level corresponding to a single electron, indicated by $N_0$, when it is exactly full. It is clear that if the degeneracy in the energy levels is lifted, by the mechanism described in the references, half levels may be filled separately and observed experimentally.

Since there is the same number of electrons $N_0$ in each sublevel we may associate one electron from each sublevel to a definite center of rotation, as a magnetic vortex. We have then, as a typical carrier model, a system of electrons rotating around a magnetic center, like a flat magnetic atom (quasi particle?). The validity of this vortex model relies in the possibility that the levels remain some how chained together, otherwise, the levels would move independently of each other. A physical chaining mechanism is clear: If we have two current loops linked by a common flux, and one loop moves reducing the flux through the other, Lenz's law would produce a reaction that opposes the motion of the first loop, trying to keep the loops chained together.

The total electronic population when the levels are exactly full is,

$$qN = N_0 + 2\nu N_0 = \left(2\nu+1\right)N_0 \quad, \tag{2.7.1}$$

where $\nu$ doubly degenerate Landau energy levels are full and where the first $N_0$ corresponds to the nondegenerate first Landau level. The charge index $\nu$ is an integer if the highest full level is a complete level, or a half integer if the highest full level is a half level or zero if the only full level is the nondegenerate first level.

The flux linked to this vortex, *whatever quantum theory wave functions* characterize the states of the electronic matter, should have quanta because the system is a representation that must carry definite quanta of charge, angular momentum and magnetic flux. In other words, the vortex carries quanta of magnetic flux. The flux linking a particular vortex is equal to the net flux $\Phi_B$ chained by the orbit of the electron pair corresponding to the *highest* level. The value of $f$, the number of net quanta linked to a vortex, is expressed by equation (2.6.12) depending on value of the flux index $\mu$ for the highest level in the system.

We now make the assumption that the condition for filling the Landau levels should be taken in the sense given previously [9, 48] by counting the *net* flux quanta linked to the orbital motion of all $N_0$ vortices. The filling condition is determined by conservation of flux (continuity of flux lines). The applied external flux must equal the total *net* flux linked *only* to the highest level. This determines a partial filling of the levels, in the canonical sense (only $h/e$ per electron). This filling condition implies, for a degenerate level with $N_0$ pairs,

$$\frac{\Phi}{N_0} = f\left(\frac{h}{2e}\right) \ . \tag{2.7.2}$$

The general expression for the Hall conductivity, in terms of the bidimensional density of carriers $N/A$ and the number of quanta of the carriers $q$, is

$$\sigma = \frac{qeN}{AB} \ . \tag{2.7.3}$$

Substituting, we obtain for the conductivity, for values of $q$ an $f$ given by equations (2.6.11), (2.6.12),

$$\sigma = \frac{e(2\nu+1)N_0}{\Phi} = \frac{2\nu+1}{2(2\mu+1)}\left(\frac{e^2}{h}\right) , \qquad \nu \geq \tfrac{1}{2} \ . \tag{2.7.4}$$

This expression is not valid if the highest level is the nondegenerate first level. For this special case, where $\nu$ is equal to zero, there are separate values for $q$ and $f$,

$$\frac{\Phi}{N_0} = f_0\left(\frac{h}{2e}\right) , \tag{2.7.5}$$

and, we get, using equation (2.6.5),

$$\sigma = \frac{eN_0}{\Phi} = \frac{1}{f_0}\left(\frac{2e^2}{h}\right) = \frac{e^2}{h} , \qquad \nu = 0 \ . \tag{2.7.6}$$

For half integer $\nu$ we may define another half integer $n$ which indicates the number of full levels of definite angular momentum number $m$ (*not energy levels*), by

$$n = \nu + \frac{1}{2} \ . \tag{2.7.7}$$

If we replace $\nu$ by $n$ we get for the conductivity, the equivalent expression,

$$\sigma = \frac{n}{(2\mu+1)}\left(\frac{e^2}{h}\right) , \qquad n \geq 1 \ . \tag{2.7.8}$$

The highest full level is characterized by the integer flux index $\mu$, indicating the number of flux quanta per electron in this highest level and the half integers charge index $\nu$, number of active degenerate energy levels, or $n$, number of active orbital angular momentum levels. As the magnetic intensity is increased, the levels, their population and flux quanta are rearranged, to obtain full levels. Details for this process depends on the microscopic laws. Nevertheless, the quantum nature of the magnetic flux requires a fractional *exact* value of conductivity whenever the filling condition is met *independent of details*. As the magnetic field increases, the resultant fractions for small integers $n$, $\mu$ are

$$\cdots \ 3, \ 7/3, \ 2, \ 5/3, \ 7/5, \ 4/3, \ 6/5, \ 1, \ 6/7, \ 4/5, \ 7/9, \ 5/7,$$
$$2/3, \ 7/11, \ 3/5, \ 4/7, \ 5/9, \ 6/11, \ 7/13, \ 6/13, \ 5/11, \ 4/9, \ 3/7, \ 2/5, \cdots \ . \tag{2.7.9}$$

These fractions match the results of the odd fractional quantum Hall effect [46, 51, 52]. The same expression, for integer $\nu$, indicates plateaus at fractions with an additional divisor of 2. For example, if $\mu$ is zero, 3/2 for $\nu$=1, 5/2 for $\nu$=2, (but not ½ since $\nu > 0$). This appears to be compatible with presently accepted values [53, 54, 55].

A partially filled level or superfluxed level occurs at a value of magnetic field which is fractionally larger than the corresponding value for a normal level with equal population because the carriers have fractionally extra flux.

The values of the conductivity are degenerate in the sense that one value corresponds to more than one set of indeces $\mu$, $\nu$. The magnetic field for any two sets with a given conductivity ratio is the same because the smaller number of carriers for one set is precisely compensated by the larger number of flux per carrier. Nevertheless, the two sets differ electromagnetically because of the extra flux for one of the carriers, and we may expect small energy differences between them, lifting this degeneracy. Of course, the proof of this difference would require detailed analysis using wave functions corresponding to a hamiltonian that includes appropriate terms. If all electrons are in Landau levels, the Fermi level would jump directly from one Landau level to the next and the conductivity curve would be a set of singular points. Localized states due to lattice imperfections allow a Fermi level between Landau levels, as discussed in the references. Since their small separation would make the two sets coalesce into a plateau. Since the value of the conductivity, at two sets $\mu$, $\nu$ with the same fraction, is exactly the same, the value of the conductivity should be insensitive to a small variation of the magnetic field (or energy) indicating a finite width plateau at the value given by equation (2.7.2) with very great precision. In particular for the ratio 1 there are many low numbered sets that coalesce producing a very wide plateau. This accounts for the plateaus seen at the so called fractional and integral fillings.

## 2.8. Results.

The geometry shows various mathematical triple structures which have physical interpretations. The geometric excitations are representations of the group. These representations may be displayed in a tridimensional space characterized by a triplet of quanta: charge, flux and spin. Therefore, in microscopic physics matter should be characterized by particles with these three fundamental quantum numbers. The sign of the P-particle charge should be opposite to the sign of the G-particle sign.

The existence of these quanta determine quanta of electrical resistance and provide a fundamental method to exactly calculate the numerical fractions that appear in the fractional quantum Hall effect.

# 3. Measurements and Motion.

It is possible to give a geometrical representation to the postulates of quantum mechanics. In most cases this is only a geometrical superstructure that may not contain new physical ideas. On the contrary, it may be possible to start from a geometrical physical theory, and obtain its quantum implications. In this manner new physical phenomena may arise.

Our theory considers excitacions of physical matter as representations of the structure group of the geometrical theory. From this idea it follows that there are certain discrete numbers associated with the states of microscopic matter. It was shown that these numbers may be interpreted as quanta of angular momentum, electric charge and magnetic flux, providing a plausible explanation to the fractional quantum Hall effect (FQHE) [14]. The theory also leads to a geometrical model for the process of field quantization [56], implying the existence of fermionic and bosonic operators and their rules of quantization. We also presented [57] a definition of mass in terms of the self energy of the nonlinear self interaction leading to the mass term in Dirac's equation. It should be clear that a process of physical measurement should display these discrete geometric numbers as experimental quanta, leading to a particle (atomic) description of matter. It is necessary, therefore, to discuss the process of measurements within the physical geometry [23].

It may be claimed that, *if* the uncertainty principle is taken as fundamental, the geometry at very short distances becomes a "fuzzy" geometry and the applications of differential geometry is questionable. Nevertheless, nothing prevents us, within our theory, to continue taking differential geometry as fundamental, providing the *germ* of quantum physics principles and represent particles by fluctuations or excitations around a geometric nonlinear substratum.

In particular, we consider the following questions: Can we define, within our theory, a geometric operation representing the process of physical quantum measurement? Are the results of this operation compatible with well known facts of experimental physics? Can we define and measure angular momentum and charge geometrically?

We are also faced with the concept of mass. This concept plays a fundamental role in relativity as shown by the relation between mass and energy and the principle of equivalence between inertial and passive gravitational mass. From a relativistic point of view, the rest mass of a system should be a unique concept defined in terms of the self energy of the system.

In quantum theory, a parameter in Dirac's equation is interpreted as the relativistic rest mass, by means of the correspondence principle, This parameter is considered an unmeasurable number, bare mass, and a process of renormalization is required to include self energy effects in a corrected physical mass. Apart from the infinities that appear in renormalization, there is no clear fundamentally derived relation between these masses and a relativistic definition based entirely in terms of energy. We consider that such a relation is only possible within a unified theory.

Our physical geometry leads to the equations of relativistic quantum mechanics and it was shown in chapter 1 that it provides a bare inertial mass for a generalized Dirac equation [14, 58]. Since the nonlinear field equation of the theory and its integrability condition determine simultaneously the evolution of the field and the motion of the sources, all self interaction effects are included in principle in any given solution. In other words, conceptually, it is not possible to have a solution for the field equation that does not satisfy the equations of motion of the sources. No additional self reaction forces are needed to describe the evolution of field and source.

As in general relativity, [17] the self interaction difficulties arise when separating the sources from the field while attempting an approximate solution. An approximation (linearization) that splits these equations in an infinite system of equations of different orders, requires the calculation of an infinite number of corrections due to self interactions. This renormalization is a consequence of the method of approximation and not due to the nonlinear theory. In particular, it is also possible to define and measure an inertial mass including self interaction without introducing *unmeasurable* bare masses.

# 3.1. Measurement of Geometric Currents.

We shall study the properties of fluctuations or excitations of the geometric elements of the unified theory. If as suggested before [14], a particle may be represented as an excitation of the geometry, its physical properties may be determined by its associated fluctuations. These fluctuations may be characterized mathematically by a variational problem. From a variational principle, if the equations of motion hold, it is possible to define the generator of the variation. There is a canonical geometric current associated with this generator which geometrically represents the evolution or motion of the excitation, and should be considered the subject of a physical measurement (observable).

The lagrangian density, in general, has units of energy per volume and the action has units of energy-time. In the natural units defined by the connection ($c=1$, $\hbar=1$, $e=1$), the action is dimensionless. In the standard relativistic units ($c=1$), the constant $\hbar$ arises as a factor in the action $W$.

It is well known that the variation of an action integral along a transformation of the variable $y$ with parameters $\lambda$ is

$$\delta W = \int_R \frac{\delta L}{\delta y} \delta y d^4 x + \int_{\partial R} \delta Q^\mu d\sigma_\mu \quad , \tag{3.1.1}$$

where the canonical current $\mathcal{I}$, and the conjugate momentum $\Pi$ are defined by

$$\delta Q^\mu = \boldsymbol{\mathcal{I}}^\mu \delta \lambda = \left( \Pi^\mu \frac{dy}{d\lambda} + L \frac{dx^\mu}{d\lambda} \right) \delta \lambda \quad , \tag{3.1.2}$$

$$\Pi^\mu = \frac{\partial L}{\partial y_{,\mu}} \quad . \tag{3.1.3}$$

The current $\mathcal{I}$ is a geometrical object field that represents an observable property of a physical excitation, for example, angular momentum density. In general, a measurement is not a point process but rather it is an interaction excitation with an apparatus over a local space-time region. In the geometric theory, the results of a physical measurement should be a number depending on a variational current $\mathcal{I}$ about a variation of a background section $e$, over a region $R(m')$ around a characteristic point $m$ on the base space $M$, with some instrumental averaging procedure over the region. Therefore, we shall make the hypothesis that a measurement on a geometrical excitation is represented mathematically by a functional $\mathcal{J}_m$ of the observable geometric current defined by the associated variation over a complementary hypersurface $\sigma$,

$$\boldsymbol{\mathcal{J}}_m \left( \boldsymbol{\mathcal{I}} \right) = \int_{\partial R} \boldsymbol{\mathcal{J}} \left( m', m \right) \boldsymbol{\mathcal{I}}^\mu \left( m' \right) d\sigma_\mu \quad . \tag{3.1.4}$$

In some cases, excitations may be approximated as point excitations with no extended structure. In order to show the relation of our geometric theory to other theories, without using any knowledge about the structure of excitations, we shall define a geometric measurement of a point excitation property by a process of shrinking the region $R(m')$ of the current to the point $m$. With this procedure, the local section representing the excitation shrinks to a singular section at $m$. We may express this mathematically by

$$\lim_{R(m') \to m} \boldsymbol{\mathcal{J}}_m \left( \boldsymbol{\mathcal{I}} \right) = \delta_m \left( \boldsymbol{\mathcal{I}} \right) \quad , \tag{3.1.5}$$

where the functional $\delta_m$ is the Dirac functional.

This process shrinks the current to a timelike world line. We may visualize the boundary $\partial R$ of region $R$ as an infinitesimal cylindrical pillbox pierced by the current at the bottom and top spacelike surfaces $\Sigma$. As the pillbox is shrunken to point $m$, the functionals of the current $\delta_m(\mathcal{I})$ at the top and bottom surfaces are equal, if the current is continuous. The functional $\delta_m(\mathcal{I})$ on either of the spacelike surfaces $\Sigma$ is the geometric measurement.

From the lagrangian given for the theory [8, 9] the geometric current of an arbitrary matter excitation has the general form

$$\mathcal{I}^{\mu} = \tilde{e}\iota^{\mu}Xe \ , \tag{3.1.6}$$

in terms of the frame $e$, an orthonormal set of the algebra $\iota$, the correlation on the spinor spaces $\sim$, and the group generator $X$ of the variation.

It should be noted that the frame $e$ is associated with a set of states forming a base of a representation of the structure group. It does not represent a single physical state, but a collection of physical states. For any operator $\Lambda$ in the algebra we may select, as the column vectors of the frame $e$, the eigenvectors $\phi$ corresponding to $\Lambda$. Then we may write

$$\Lambda e = \Lambda\left(\phi_1, \phi_2, \cdots \phi_i\right) = \left(\lambda^1 \phi_1, \lambda^2 \phi_2, \cdots \lambda^i \phi_i\right) = e\lambda \ , \tag{3.1.7}$$

where $\lambda$ is the diagonal matrix formed by the eigenvalues $\lambda_i$.

Accordingly, the results of the measurement given by eq. (3.1.6), in coordinates adapted to the 4-velocity $u$, is

$$\delta_m\left(\mathcal{I}\right) = \mathcal{I}^0 = \tilde{e}\iota^0 Xe = \tilde{e}\Lambda e \ , \tag{3.1.8}$$

which defines an associated operator $\Lambda$.

In this expression, we should note $\tilde{e}$ is the group inverse of $e$, and the correlated product in the spinor space is a scalar. The product $\tilde{e}\,e$ gives a unit matrix of scalars, and the measurement values of the current $\mathcal{I}$ coincide with the diagonal matrix formed with its eigenvalues.

This result agrees with one of the postulates of quantum mechanics. The mathematical content of the last equation is truly independent of a physical interpretation of the frame $e$. In particular, it does not require, but allows that $e$ is a probability amplitude.

The result of the measurement essentially equals the value of the current at the representative point $m$. Equivalently it is the average over a characteristic 3-volume $V$ of $\Sigma$,

$$\langle \mathcal{I} \rangle = \frac{1}{V}\int_V \frac{\delta Q^{\mu}}{\delta\lambda}\, d\sigma_{\mu} = \frac{1}{V}\int_V \mathcal{I}^{\mu} d\sigma_{\mu} \ . \tag{3.1.9}$$

This averaging, indicated by $\langle\ \rangle$, is similar to the operation of taking the expectation value of an operator in wave mechanics.

The different generators of the group produce excitations whose properties may be investigated by measuring the associated geometric currents. In particular, we are interested here in currents associated with generators of the compact subgroups, which were used to characterize the induced representations.

## 3.2. Geometric Spin.

The concept of spin is related to rotations. In the geometric theory the compact even generators form an su(2) subalgebra which is related to the rotation algebra. The group homomorphism between this SU(2) subgroup and rotations is

$$R_b^a = \frac{1}{2}\text{tr}\left(\sigma^a g^{\dagger}\overline{\sigma}_b\, g\right) \ , \tag{3.2.1}$$

where $g \in$ SU(2) and $R \in$ SO(3). The isomorphism between this SU(2) and the compact even subgroup of SL(2,$\mathcal{Q}$) is the well known isomorphism between the complex numbers and a subalgebra of the real 2 × 2 matrices.

The isomorphism given by these expressions is not accidental, but is part of the conceptual definition of the geometric Clifford algebras as a generalization of the complex numbers and the quaternions. These algebras, and the spinor spaces on which they act, have well defined complex structures.

If we consider that the generators $\kappa^i \kappa^j$ that belong to su(2) are the rotation generators, the associated geometric current is the angular momentum along the third axis. For example, the result of the measurement of this current in a preferred direction $\mathcal{I}^3$, using as illustration eq. (3.1.9) for the expectation value is

$$\langle \mathcal{I}^3 \rangle = \frac{1}{V}\int d\sigma_{\mu}\left(\Pi^{\mu}\frac{de}{d\lambda} + L\frac{dx^{\mu}}{d\lambda}\right) \ , \tag{3.2.1}$$

where the variation $\delta e$ is generated by $\kappa^1 \kappa^2$. Then, for a flat metric,

$$\langle \mathcal{I}^3 \rangle = \frac{1}{V} \int d\sigma_\mu \, \tilde{e} \iota^{\hat\alpha} u_{\hat\alpha}^\mu \frac{de}{d\lambda} = \frac{1}{V} \int d^3 x \, \tilde{e} \iota^0 \frac{de}{d\lambda} \quad . \tag{3.2.2}$$

Since $\iota^0$ commutes with $\kappa^1 \kappa^2$ and $\iota^0 \iota^0$ is $-1$, we identify the matrices and express the variation generated by $\kappa^1 \kappa^2$ as the differential of eq. (3.2.1). This rotation induces a change in the functions on tridimensional space, giving a total variation for $e$ of

$$\delta e = \left( \frac{\kappa^1 \kappa^2}{2} e + \left( x \partial_y - y \partial_x \right) e \right) \delta\lambda = \left( \frac{i}{2} \sigma^3 + \left( x \partial_y - y \partial_x \right) \right) e \, \delta\lambda \quad , \tag{3.2.3}$$

which leads to

$$\langle \mathcal{I}^3 \rangle = \frac{1}{V} \int d^3 x \, \tilde{e} \left( \tfrac{1}{2} \sigma^3 - i \left( x \partial_y - y \partial_x \right) \right) e \quad . \tag{3.2.4}$$

The factor ½ indicates, of course, that the SU(2) parameter $\lambda$ is half the angular rotation, due to the 2-1 homomorphism between the two groups.

This calculation was done, for simplicity, with only one component. It is clear that if we use the three spatial components we get the expression for the angular momentum vector operator $\Lambda$ in quantum mechanics. The frame sections play the role of wave functions and the group generators play the role of quantum operators. These similarities between our geometric theory and quantum mechanics provide essentially equivalent results. There are differences; in particular, wave functions have a complex structure and our frame sections have a Clifford structure. Rather than a contradiction, this difference is a generalization, since there are complex structures in different subspaces of the geometric Clifford algebra. It is possible to introduce spaces of sections, but they certainly may have a structure more general than a Hilbert space. The geometrical and group elements in the theory actually determine many of its physical features.

## 3.3. Geometric Charge.

The geometric source current $J$ is a generalization of electric current. The three compact generators $\kappa_0$, $\kappa_5$, and $\kappa_1 \kappa_2 \kappa_3$ are equivalent as electromagnetic generators within the theory because there are automorphisms that transform any one of them into any other. It follows that the set $\iota^\mu$, that enters into the current, is defined up to an algebra automorphism. This allows us to take the $\iota^0$ element as any of three electromagnetic generators without changing the physical content of the theory.

The generalized source current $J$ is the canonical current $\mathcal{I}$ corresponding to a variation generated by an electromagnetic generator. In order to see this, we choose the set $\kappa^\mu$ for $\iota^\mu$, and look for a variation generator that results in an automorphism of the current. In other words, we look for a generator that gives an equivalent set to the set $\kappa^\mu$ by right multiplication. A generator that accomplishes this is $\kappa^5$,

$$\kappa^\mu \kappa^5 = \exp\left( -\frac{\pi}{4} \kappa^5 \right) \kappa^\mu \exp\left( \frac{\pi}{4} \kappa^5 \right) \quad . \tag{3.3.1}$$

It is possible to find another one that results in a different automorphism but it would lie in the electromagnetic sector. Thus it is clear that the current $J$ corresponds to variations generated by the electromagnetic sector.

When we make a measurement of this canonical current $J$, we are measuring the charge associated with the fluctuation of $e$ related to a fundamental irreducible representation of the group. We repeat the same calculation done in the previous section for angular momentum current. If we neglect the gravitational part, the metric is flat, the expression for the measurement on the charged current is

$$\langle \mathcal{I} \rangle = \frac{1}{V} \int \mathcal{I}^\mu d\sigma_\mu = \frac{1}{V} \int \tilde{e} \kappa^{\hat\alpha} u_{\hat\alpha}^\mu \dot{e} d\sigma_\mu \tag{3.3.2}$$

where it is understood that we are working in the bundle $SM$ which is the Whitney sum of the associated spinor vector bundle $VM$ and its conjugate. Explicitly, in terms of elements of $VM$ the last equation is written as

$$\langle j \rangle = \frac{1}{V} \int d^3 x \begin{bmatrix} e^{-1} & 0 \\ 0 & \overline{e} \end{bmatrix} \begin{bmatrix} \kappa^0 \kappa^5 & 0 \\ 0 & -\kappa^0 \kappa^5 \end{bmatrix} \begin{bmatrix} e & 0 \\ 0 & \overline{e}^{-1} \end{bmatrix} \quad . \tag{3.3.3}$$

It should be emphasized that the matrices in the last equation are 8×8 real matrices, the double-dimensional representation used in the $SM$ bundle [9, 14]. The even generators may be written as 4×4 complex matrices using the isomorphisms mentioned after eq.(3.2.1). We may substitute any equivalent generator for $\kappa^0 \kappa^5$.

It is possible to choose a frame section $e$ corresponding to eigenvectors of the fundamental representation of SU(2). In fact, since $\kappa^5$ commutes with $\kappa^1\kappa^2$, it is also possible to choose a frame corresponding to common eigenvectors of these two antihermitian generators belonging to the two su(2) subalgebras in sl(4,$\mathbb{R}$). The eigenvalues correspond to the quanta of spin and charge. Of course it is usual to work with the associated hermitian operators obtained by multiplication by $i$, with real eigenvalues $\pm 1$. Nevertheless the use of the antihermitian expression is natural since they are generators of the two compact su(2) subalgebras

If $e$ is an eigenframe of the generator, we get

$$\langle j \rangle = \pm i \quad . \tag{3.3.4}$$

In other words, the result of this measurement for the fundamental representation is a quantum number equal to $\pm 1$. This conserved number may be interpreted as the quantum of charge.

It is known that the electron charge plays two roles, one as the quantum of charge and the other as the square root of the fine structure coupling constant $\alpha$. In order to be able to reduce the theory to electromagnetism we must account for Coulomb's coupling constant $k/4\pi$ and understand the relationship of these constants. This $k$ may be absorbed into the definition of the current in eq.(1.2.2), but at the end it must be identified. It is better to show it explicitly and to keep the frame $e$ separate, as a section in the principal bundle, so that the conjugate $\tilde{e}$ is the dual inverse of $e$, and the product $\tilde{e}\, e$ is the unit matrix.

The dimensionless fine structure constant $\alpha$ is expressed in physical units as $ke^2/4\pi\hbar c$. The units of the arbitrary constant $k$, which allows us to define the electromagnetic units from the mechanical ones, are $ml^3t^{-2}q^{-2}$. If we set $k=4\pi$, the units correspond to the gaussian system, where Coulomb's constant is $1$. If we set $k=1$, we obtain the Heaviside-Lorentz system where Coulomb's constant is $1/4\pi$. If we set $k=4\pi c^2 \times 10^{-7}$, we obtain the rationalized MKSA system of units, where Coulomb's constant is $c^2 \times 10^{-7}$. In these systems, the minimal coupling determines that the connection $\Gamma$ corresponds to $eA$, in terms of the potential $A$ and the electron charge $e$. On the other hand, rather than setting the value of $k$, it seems better to consider that the geometric theory introduces a *natural unit* of charge by defining the electromagnetic potential, in energy units, *equal* to the connection. In this form the new geometric unit of charge (the "*electron*") equals $e$ Coulombs, $k$ is determined to be $4\pi\alpha$ and Coulomb's constant becomes the fine structure constant $\alpha$. With our definition of current and coupling constant ($4\pi\alpha$) the calculated charge of the electron is $\pm 1$, in these geometric units. In arbitrary units, the calculated quantum of charge $e$ is $\pm(4\pi\alpha c/k)^{1/2}$.

In other words, this value is the minimum quantum of measurable charge changes. This geometric prediction explains the two roles played by the electron charge, as coupling constant and as quantum. In the previous section the calculation lead to the well known values of angular momentum. In this section we obtain a new theoretical result.

When there is only a U(1) electromagnetic field in flat space, our field equations reduce to[9]

$$d\,{}^*dA = 4\pi\alpha\,{}^*j \quad . \tag{3.3.5}$$

A particular solution for a static spherically symmetric potential $A$ is

$$A_0 = \frac{\alpha q}{r} \quad , \tag{3.3.6}$$

where $q$ is the charge in "electrons". If we now change our units to the gaussian system (1 electron = $e$ coulombs), where $\alpha = e^2$,

$$\frac{A_0}{e} = \varphi = \frac{\alpha q}{er} = \frac{qe}{r} \qquad \alpha = e^2 \ \text{ (Gauss) } \quad , \tag{3.3.7}$$

which is Coulomb's law in terms of the charge $qe$ in Coulombs.

# 3.4. The Concept of Mass.

In relativity, the inertial rest mass of a particle is the norm of its four momentum. In quantum mechanics, momentum is related to derivatives in space-time. To show the consistency of the concept of mass within our geometric theory, we shall consider the variation generated by a translation in the base space along the integral curve of the vectors of the space-time tetrad $u_\alpha$. We obtain, in this way, four canonical geometric currents, as defined previously, whose average values over the volume $V$ are

$$\langle \theta_{\hat{\alpha}} \rangle = \frac{1}{V} \int \Pi^\mu \mathcal{L}_{\hat{\alpha}} e\, d\sigma_\mu \quad , \tag{3.4.1}$$

where $\mathcal{L}_a$ indicates the four Lie derivatives with respect to the vectors $u_a$ and $\Pi$ is the canonical momentum. In adapted coordinates we get

$$\langle \theta_{\tilde{\alpha}} \rangle = \frac{1}{V} \int e^{-1} \kappa^\mu \partial_{\tilde{\alpha}} e \, d\sigma_\mu = \frac{1}{V} \int e^{-1} \kappa^0 \partial_{\tilde{\alpha}} e \, d^3x \quad . \tag{3.4.2}$$

In particular, consider the trace of the time current and using the equation of motion assuming for the moment that $e$ is constant, express the integral in terms of the connection $\Gamma_\mu$ or potential,

$$\operatorname{tr} \langle \theta_{\hat{0}} \rangle = \frac{1}{V} \int \operatorname{tr} e^{-1} \kappa^\mu e \Gamma_\mu d^3x = \frac{1}{V} \int \operatorname{tr} J^\mu \Gamma_\mu d^3x \quad . \tag{3.4.3}$$

It is clear from the analogy with $j^m A_m$ that $J^\mu \Gamma_\mu$ should have the meaning of energy and that $<\theta_0>$ is the corresponding measured value. Furthermore we recognize that the integrand in eq. (3.4.3) is in accordance with the fundamental mass-energy eq. (1.2.5). Later we shall come back to indicate the meaning of the condition assumed.

This average energy leads to the concept of Dirac mass parameter for a particle. It was pointed out above, that mass may be defined as a parameter related to the potential. For geometrical reasons the unit of connection is inverse length, the same as the operator $\partial_\mu$. This provides a geometric (natural) unit of mass in terms of inverse length. Initially [14] we defined the mass parameter by

$$m = \frac{1}{4} \operatorname{tr} \left( \kappa^\mu \Gamma_\mu \right) \quad . \tag{3.4.4}$$

It was noted that since we use time (length) as units of interval, the metric and its related $\kappa^\mu$ matrix are dimensionless. Then the mass $m$ has units of inverse length. We recognize now that a more general definition in accordance with section 1.3 is

$$m = \frac{1}{4} \operatorname{tr} \left( e^{-1} \kappa^\mu e \Gamma_\mu \right) = \frac{1}{4} \operatorname{tr} \left( J^\mu \Gamma_\mu \right) = \mathcal{M} \tag{3.4.5}$$

which reduces to the previous one, eq. (3.4.4) under the simplifying assumptions given previously.

The inertial mass of a particle is defined as a parameter related to a pair of connection and current that solve the nonlinear self interacting system. If the sign of the current in the source is changed it is conjectured that the sign of the connection also changes and the mass remains positive. For an excitation around the geometric inert substratum solution this nonlinear solution itself provides a parameter $m_0$ for the linear excitation equation that may be considered the bare inertial mass parameter of the particle associated to the excitation.

Since the only element in the algebra with nonzero trace is the unit, we find that

$$e^{-1} \kappa^\mu e \Gamma_\mu = m I + \cdots \tag{3.4.6}$$

and we may write the equation of motion (1.2.7)

$$\kappa^\mu \nabla_\mu e = \kappa^\mu \left( \partial_\mu e - e \Gamma_\mu \right) = 0 \quad , \tag{3.4.7}$$

as a Dirac equation,

$$\kappa^\mu \partial_\mu e = m e + \cdots \quad . \tag{3.4.8}$$

If we used the standard unit of mass, instead of the geometric unit, a constant appears in front of the differential operator, which is Planck's constant $\hbar$. The geometric nature of Planck's constant is determined by the connection in the same manner that the geometric nature of the velocity of light $c$ is determined by the metric.

## 3.5. Invariant Mass.

There is one difficulty with the given definition. The connection $\Gamma$ is not a tensor and under arbitrary change of reference frame the mass is not invariant [9]. In words we may say that, in order to have an invariant mass, we must restrict the change of the reference frame in such a way that the generator of the transformation is orthogonal to the current $J$. For example, this would mean for standard electrostatics that the scalar potential should not change in the transformation. If the current $J$ is odd, the mass given by eq. (3.4.5) is invariant under $SL_1(2, \mathbb{C})$ and is, therefore, a Lorentz invariant.

Nevertheless, we realize that the theory applies to matter in the whole universe. If a particle is associated to an excitation on a matter local frame in a dynamic cosmic geometric background, we expect that the corresponding mass should be related to the part of the connection responsible for the nonlinear local interaction with the local matter current.

It is convenient to separate the current into a part $J_l$ corresponding to a local system representing a particle and another part $J_b$ corresponding to the cosmic background that borders the system. The field equation in terms of the curvature takes the form,

$$D^*\Omega_t = 4\pi\alpha\left(\,^*\!J_l + \,^*\!J_b\right)\,,\tag{3.5.1}$$

where $\Omega_t$ is the generalized curvature of the total connection $\Gamma_r$. If there is no local matter, we have a background equation,

$$D^*\Omega_b = 4\pi\alpha\,^*\!J_b\,,\tag{3.5.2}$$

where $\Omega_b$ is the curvature of a cosmic background connection $\Gamma_b$.

Far away from the region of the local material frame we may consider that its effect is a small perturbation with respect to the dynamic cosmic background, but this is not the case very close to a local material frame. In fact, close to it, the dynamic cosmic background may be considered a perturbation, with respect to the nonlinear self interaction in the area of the tetrad section.

With this in mind, we may define the local material potential tensorial form $\Lambda_l$, the difference between the total and background connections,

$$\Lambda_l = \Gamma_t - \Gamma_b\,,\tag{3.5.3}$$

as the element responsible for the self energy of the interaction in the system.

It is clear that a difference of connections is a tensor and that the last equation is valid even in the case where the effect of the background is zero. In this case $\Gamma_b$ would be a flat inertial connection $\Gamma_l$, which would be zero in some reference frames but may have nonzero values in arbitrary reference frames. In general this is not the case due to the presence of far away macroscopic matter.

The invariant expression for the mass is obtained from the total dynamic potential tensorial form $\Lambda$ defined by eq. (1.3.2). Thus, we define

$$m \equiv \frac{{}^{C}\!g\left(J^{\mu}\Lambda_{\mu}\right)}{{}^{C}\!g\left(-\kappa^{0}\kappa_{0}\right)} = \frac{\mathrm{tr}\left(J^{\mu}\Lambda_{\mu}\right)}{\mathrm{tr}\left(I\right)} = \tfrac{1}{4}\mathrm{tr}\left(J^{\mu}\left(\Lambda_{l\mu} + \Lambda_{b\mu}\right)\right)\,.\tag{3.5.4}$$

This mass parameter is determined by the scalar product, using the Cartan-Killing metric, of the potential and current that solve the nonlinear self interacting system. Since the Cartan-Killing metric depends on the representation [59] of the algebra $A$ used in the Dirac equation, the parameter $m$ depends on the chosen representation. In some representations the products involved are convolutions rather than matrix multiplication.

If the background is negligible, $\Gamma_b$ may be taken as a flat connection and this equation reduces to the original one in those reference frames where $\Gamma_b$ is zero. The extra mass like terms that appear in Dirac's equation when an improper reference frame is used are similar to the inertial effects that appear in accelerated reference systems. These effects appear in eq.(3.4.8), apart from the mass, as part of a "fictitious" inertial interaction. In fact, the flat $\Gamma_l$ related to the background connection is called the inertial connection since it is the part responsible for all these inertial effects.

It should be noted here that a particle is associated to an excitation of the matter local frame in a very small region compared with the universe. The local material frame is the dominant element and the dynamic cosmic background should be treated as a perturbation which we consider negligible. We shall say that the local matter frame is *the inert substratum* for the excitation. In this case the substratum is the dominant element, and the excitation should be treated as a linear perturbation on this nonlinear substratum solution. The value of the mass parameter for the excitation should be related to the constant $\mathfrak{M}$ associated to the inert substratum solution.

# 3.6. Equation of Motion.

After justifying the identification of the parameter in eq. (3.4.4) as the inertial mass we take a closer look at the geometric motion associated to the current $J$. Using the potential relative to the inert substratum defined by equation (1.3.3), in induced representations, the new equation of motion (1.3.4) explicitly displays the term depending on the invariant substratum mass, required by the Dirac equation,

$$\kappa^{\mu}\nabla_{\mu}e = \kappa^{\mu}\left(\partial_{\mu}e - e\,^{0}\Gamma_{\mu}\right) = \kappa^{\mu}\left(\partial_{\mu}e - e\hat{\Gamma}_{\mu} - e\left(\frac{-m}{4}J_{\mu}\right)\right) = \kappa^{\mu}\hat{\nabla}_{\mu}e - me = 0\,.\tag{3.6.1}$$

We now express all geometric objects by their even and odd parts. The geometric equation breaks in two complex equations for the even and odd parts of the G-system. We then designate different vector potentials in the connection with the following notation: $^{+}\!Ai$ is the even part of the $SU(2)_Q$ sector $\bar{\Gamma}(Q)$ of the connection and corresponds to the electromagnetic potential; $\bar{A}$ is the complementary odd part in the same sector; $\Gamma$ is, in this section, the SL(2,$\mathbb{C}$) sector of the connection which determines an $L$ covariant derivative; $\Upsilon$ is the complementary part. Both $\bar{A}$ and $\Upsilon$ are complex matrices, respectively proportional to the identity and the Pauli matrices, along the odd direction $\kappa^0$.

The odd and even parts of the frame, $\xi$ and $\eta$, are elements of sl(2,$\mathbb{C}$) and sl(2,$\mathbb{C}$)$\oplus$u(1) respectively. The u(1) component contributes an overall phase and should be neglected. Using conjugation the first equation may be written as

$$\sigma^\mu \left( i\nabla_\mu + {}^+\!A_\mu \right)(i\xi) = \sigma^\mu \, \bar{A}_\mu \, \eta^\dagger + \sigma^\mu \eta^\dagger \Upsilon_\mu + m\eta \ . \tag{3.6.2}$$

Similarly we get for the second equation

$$-\overline{\sigma}^\mu \left( i\partial_\mu \eta + {}^+\!A_\mu \eta - \eta \left( i\Gamma_\mu \right) \right) = \overline{\sigma}^\mu \, \bar{A}_\mu \left( i\overline{\xi}^\dagger \right) + \overline{\sigma}^\mu \left( i\overline{\xi}^\dagger \right) \Upsilon_\mu - m\left( i\xi \right) \ . \tag{3.6.3}$$

If we define

$$\varphi \equiv \tfrac{1}{\sqrt{2}} \left( \eta + i\xi \right) \ , \tag{3.6.4}$$

$$\chi \equiv \tfrac{1}{\sqrt{2}} \left( \eta - i\xi \right) \ , \tag{3.6.5}$$

we can add and subtract the even and odd equations and write the generalized Dirac's equations in the following form

$$\left( i\nabla_0 + {}^+\!A_0 \right)\varphi - \sigma^m \left( i\nabla_m + {}^+\!A_m \right)\chi = \bar{A}_0\,\chi^\dagger + \sigma^m \, \bar{A}_m\,\varphi^\dagger + \chi^\dagger \Upsilon_0 + \sigma^m \varphi^\dagger \Upsilon_m + m\varphi \ , \tag{3.6.6}$$

$$\left( i\nabla_0 + {}^+\!A_0 \right)\chi - \sigma^m \left( i\nabla_m + {}^+\!A_m \right)\varphi = -\bar{A}_0\,\varphi^\dagger - \sigma^m \, \bar{A}_m\,\chi^\dagger - \varphi^\dagger \Upsilon_0 - \sigma^m \chi^\dagger \Upsilon_m - m\chi \ . \tag{3.6.7}$$

## 3.6.1. Agreement with Standard Quantum Mechanics.

If we fully simplify eqs. (3.6.2) and (3.6.3) to an even abelian $iA$ potential along $\kappa_5$ and let

$$\xi \to i\xi \ , \tag{3.6.8}$$

we obtain

$$\sigma^\mu \left( i\partial_\mu + A_\mu \right)\xi = m\eta \ , \tag{3.6.9}$$

$$\overline{\sigma}^\mu \left( i\partial_\mu + A_\mu \right)\eta = m\xi \ , \tag{3.6.10}$$

which are the standard Dirac equations with electromagnetic coupling if we identify the connection with the minimal coupling $eA$.

The 4×4 real matrices are mapped into 2×2 complex matrices. The resultant columns may be taken as a pair of 2 component spinors and we see that the original frame decomposes into complex spinors. These equations are pairs of Dirac equations, in the standard 2 component form. The natural interpretation is to say that the Dirac fields may be represented geometrically by a spinor frame in the associated vector bundle. The corresponding Dirac "field equations" are the equations of motion or covariant transplantation equations for the spinor frame.

We shall consider first the quantum mechanics of free particles. By this we mean that there is no explicit coupling to an external or interacting field, except possibly that self interaction giving rise to the mass term in Dirac's equations. This implies, in our theory,  that the connection is zero except for the mass parameter.

The equations of the theory are essentially relations between matrices that represent generalized spinor frames. We have introduced 2-spinors matrices $\eta$, $\xi$ equal to the even and odd parts of the frame, respectively.

We have the following equations for the $\eta$, $\xi$ parts:

$$i\sigma^\mu \partial_\mu \xi = m\eta \ , \tag{3.6.11}$$

$$i\overline{\sigma}^\mu \partial_\mu \eta = m\xi \ , \tag{3.6.12}$$

implying that a frame for a massive particle must have odd and even parts. In our case, if we set the odd part $\xi$ equal to zero we obtain also that  $m$ is zero,

$$\bar{\sigma}^\mu \partial_\mu \eta = 0 \quad . \tag{3.6.13}$$

It should be clear that a wave moving along the positive $z$ axis according to this equation only admits eigenfunctions with negative eigenvalues of $\sigma^3$. This means that the zero mass field associated to an even frame has negative helicity. In other words, this equation is the one normally associated with a neutrino field. The nonexistence of a positive helicity neutrino is due, within the theory, to the impossibility of having a pure odd frame, within the theory. The geometric reason is that a principal bundle can not be defined with only the odd components because they do not form a subgroup.

If a field excitation corresponds to a representation of a subgroup with specific quantum numbers, it may be associated to only one of the spinor columns of the frame, the one with the corresponding quantum numbers. Accordingly, we may restrict the excitations or fluctuations of frames to matrices that have only one column in each of the two parts of the frame, the even $\eta$ and the odd $\xi$.

$$\eta = \begin{bmatrix} \eta_1^{\hat{1}} & 0 \\ \eta_1^{\hat{2}} & 0 \end{bmatrix} \quad , \tag{3.6.14}$$

$$\xi = \begin{bmatrix} \xi_1^{\hat{1}} & 0 \\ \xi_1^{\hat{2}} & 0 \end{bmatrix} \quad . \tag{3.6.15}$$

We now restrict to the even simple subgroup SL(2,C), homomorphic the Lorentz group. The $\eta$, $\xi$ parts have inequivalent transformations under this group. We may form a four dimensional (Dirac) spinor by adjoining the two spinors, where the components $\eta$, $\xi$ are two complex 2-spinors. We may combine the 2 columns into a single column Dirac 4-spinor,

$$\psi = \begin{bmatrix} \xi_1^{\hat{1}} \\ \xi_1^{\hat{2}} \\ \eta_1^{\hat{1}} \\ \eta_1^{\hat{2}} \end{bmatrix} \quad . \tag{3.6.16}$$

We now show that the even and odd parts of a frame are related to the left and right handed components of the field. We calculate the left handed and right handed components, and obtain, omitting the indices,

$$\psi_L = \tfrac{1}{2}\left(1 + \gamma^5\right)\psi = \tfrac{1}{2}\left(1 + \gamma^5\right)\begin{pmatrix} \xi \\ \eta \end{pmatrix} = \begin{pmatrix} 0 \\ \eta \end{pmatrix} \quad , \tag{3.6.17}$$

$$\psi_R = \tfrac{1}{2}\left(1 - \gamma^5\right)\psi = \tfrac{1}{2}\left(1 - \gamma^5\right)\begin{pmatrix} \xi \\ \eta \end{pmatrix} = \begin{pmatrix} \xi \\ 0 \end{pmatrix} \quad . \tag{3.6.18}$$

We have that the left handed component is equivalent to the $\eta$ field which in turn is defined in terms of the even part of the frame field. Similarly, we see that the right handed component is equivalent to the $\xi$ field and consequently to the odd part of the field. Therefore, an even frame corresponds to a left handed particle, as should be for a neutrino. This Lorentz frame excitation has neutrino properties.

It is easy to calculate the trace of the corresponding matrices. It is clear that the equations combine, with the usual definitions, to produce the standard form of the Dirac equation. Since the nonrelativistic limit of this equation is Schroedinger's equation which represents the free motion of a particle of mass $m$, it is clear that Ehrenfest's theorem [9] is valid. As a result, the measurable values obey Newton's second law of motion. In a sense, our theory says that there is a correspondence with mechanics, and we must interpret $<-i\hbar\partial_m>$ as the classical momentum and $<-i\hbar\partial_0>$ as the classical energy. It is clear that the operators representing momentum and position satisfy the Heisenberg commutation relations,

$$\left[p,x\right] = -i\hbar \quad , \tag{3.6.19}$$

$$[E,t] = i\hbar \ . \tag{3.6.20}$$

Now we are in a position to indicate that the hypothesis of constant $e$ after eq. (3.4.2) means, in this theory, that the particle associated to the specified fluctuation has zero lineal momentum and its energy equals its rest mass, as should be the case in relativistic quantum mechanics.

## 3.7. Results.

Measuremens on geometric excitations determine integral expressions similar to the expectation values in quantum mechanics. These expressions display the triple structure of charge, flux and spin quanta. The associated operators satisfy the Heisenberg commutation relations. In fact, it appears that this geometry is the germ of quantum physics including its probabilistic aspects. The mass may also be similarly defined in an invariant manner in terms of energy, depending on the potential and matter current. The equation of motion is a geometric generalization of Dirac's equation. The geometric nature of Planck's constant $h$ and of light speed $c$ is determined by their respective relations to the connection and the metric.

# 4. Masses.

The fundamental dynamic process in the theory is the action of the connection or potential on the frame. Since the connection is valued in the Lie algebra of the group $G$ and the frame is an element of $G$, the dynamic is realized by the action of the group on itself. The principal bundle structure [36] of the group, $(G, K, L)$, provides a natural geometric interpretation of its action on itself. A particular subgroup $L$ defines a symmetric space $K$, the left coset $G/L$, the base space of the bundle. The subgroup $L$, the fiber of the bundle, acts on itself on the right, and also is the isotropy subgroup of the coset $K$. The complementary coset elements act as translations on the symmetric coset.

This geometric interpretation may be transformed into a physical interpretation if we choose $L$ to be homomorphic to the spinor group, SL(2,$\mathbb{C}$), related to the Lorentz group. The action of $L$ is then interpreted as a Lorentz transformation (pseudorotation) in *external* space, the tangent space $TM$ of the physical space-time manifold $M$, defining its metric. The action of the complementary coset $K$ is interpreted as a translation in the *internal* space, the symmetric coset $K$ itself. There is then a nontrivial geometric relation between the internal and external spaces determined by the Clifford algebra structure of the manifold. The space $K$ is the exponentiation of the odd sector of the Clifford algebra and is related to local copies of the tangent space $TM$ and its dual cotangent space $T^*M$. It may be interpreted as a generalized momentum space. The states of momentum $k$ would correspond to the points of $K$.

It follows that the frame excitations are also acted by the potential and evolve as representations of $G$. The constant parameter $m$ characterizes an eigenvalue of a differential operator $\mathcal{P}$, defined by the equation of motion, acting on excitations $\delta e$ of a substratum solution over a locally symmetric space-time, seen by some definite observer and therefore $m$ also characterizes the eigenvalue of the quadratic differential operator $\mathcal{P}^2$ on space-time.

Different observers would measure different relative momenta $k$ for a given excitation. A measurement for each $k$ corresponds to a function in momentum space. An abstract excitation is an equivalence class of these functions, under the relativity group. Since the group space itself carries its own representations, the realization of excitations as representations defined on the group space have a fundamental geometric character. The geometric action of the $K$ sector is a translation on itself and the functions on $K$ are the internal linear representations that may be observed (observables). The action of the $L$ sector is a Lorentz transformation and sl(2,$\mathbb{C}$) spinors are the observable external representations. For these reasons, we *must represent physically observable excitations by classes of spinor valued functions on the symmetric $K$ space*. In particular, we realize them on a vector bundle, associated to the principal bundle $(G, K, L)$, taking as fiber the sl(2,$\mathbb{C}$) representation valued functions on the symmetric $K$ space. Essentially, this is, in fact, done in particle physics when considering representations of the Poincare group.Our representations are representations of SL(4,$\mathbb{R}$) induced from SL(2,$\mathbb{C}$).

It was indicated in chapters 1 and 2 that the base manifold, the space-time $M$, is a manifold modeled on $K$. Given an observer, the excitations on $M$ may be locally idealized as functions on the symmetric space $K$ locally tangent to $M$ at a point $u$ in an neighborhood $U \subset M$. For the moment we restrict $G$ to its 10 dimensional $P$ subgroup. The tangent symmetric modelspace is isomorphic to the hyperbolic space $P/L$ which has a LaPlace-Beltrami differential operator corresponding to the quadratic Casimir operator $\mathcal{C}^2$. Because of the local isomorphism between the tangent spaces of $M$ and $P/L$, we may take $\mathcal{P}^2$ as the corresponding space-time image of $\mathcal{C}^2$.

The constant $m$ would then characterize the eigenvalues of $\mathcal{C}^2$. It is known that the representations of $P$ induced on $P/L$ are characterized by the eigenvalues of this Casimir operator [60, 36] and therefore they would also be characterized by $m$. In general we would then have that the parameter $m$ associated to an excitation (wave) function on $M$, as seen by some definite observer, characterizes an induced representation of $G$ on $G/L$. Because of this pure geometric relation between the $\mathcal{C}^2$ and $\mathcal{P}^2$ operators the induced representations should play a significant physical role. It should be clear the geometrical origin of this well known physical hypothesis in elementary particle physics.

Choosing a particular induced representation of $G$ on $G/L$ and definite local observers in a neighborhood around a given point in $M$ determines the excitation (wave) function. Since we deal with a class of relativistic equivalent observers, we actually have a class of (wave) functions on $M$, one for each transformation linking valid observers. The section of the fiber bundle $E$ defining a particular induced representation has for support the mass hyperboloid in $K$ characterized by the rest mass value $m$. Once we have chosen observers at points $u$ in a neighborhood $U$ in $M$, corresponding to points $k$ on the mass hyperboloid, we have (wave) functions on $M$ with definite values of $k(u)$ in the neighborhood $U$. The isotropy group of the mass hyperboloid, the one that leaves a point (momentum) invariant ("little" group), transforms the (wave) functions on the mass hyperboloid. If we have the flat trivial bundles used in standard quantum theories, for example, if $M$ is flat Minkowski space, $P$ is the Poincare group and $L$ is the Lorentz group then the quotient $P/L{\equiv}K$ is a flat Minkowski (momentum) space distinct but isomorphic to (space-time) $M$ and the little group on the mass hyperboloid in $K$ is SO(3) or ISO(2). The (wave) function becomes the standard particle wavefunction depending on a single definite value of $k$ throughout $M$.

Due to the preceding arguments, when we try to calculate the physical values $m$, corresponding to excitations on a substratum, we *must* use the induced representations characterized by $m$, which are the ones that represent geometric excitations of definite momentum $k$ by (wave) functions on space-time, carrying geometrical and physical significance.

From our geometric point of view, it has been claimed [61] that the proton, the electron and the neutrino are representations of SL(2,$\mathcal{Q}$) and its subgroups, induced from the subgroups SL$_1$(2,$\mathbb{C}$) and SL(2,$\mathbb{C}$). This is a generalization of particles as representations of the Poincare group induced from its Lorentz subgroup [62]. Using our definition of mass [57, 63], it is possible to find an expression for the mass of these geometric excitations and compare them with the proton electron mass ratio (a not fully explained geometrical expression is known [64, 65, 66]). It should be noted that there is no contradiction in this calculation with present physical theories, which may be considered as effective theories derived under certain conditions and limits from other theories. The structure group $G$ of the theory, SL(4,$\mathbb{R}$), has been used in an attempt to describe particle properties [67, 68, 69] in another approach.

These results ratify that quantum mechanics effects may be contained in the unified theory. A curved substratum may also be a mechanism to give masses to the potential excitations. This is only possible for nonlinear theories like the one under discussion and is interesting, since we have found a simple inert substratum solution to the nonlinear field equation. Here we find particular equations for the case of a potential field excitation around this substratum solution [63]. In addition we will discuss under what conditions the excitations are massless.

# 4.1. Bare Inertial Masses for Frame Excitations or Fermions.

The definition of the mass parameter $m$, in terms of a connection or potential on the principal fiber bundle $(E, M, G)$, has been given in the fundamental defining representation of SL(4,$\mathbb{R}$) in terms of $4{\times}4$ matrices, but in general, may be written for other representations using the Cartan-Killing metric $^cg$ [70], defined by the trace. The definition of this metric may be extended to the Clifford algebra $A$, which is an enveloping algebra of both sl(4,$\mathbb{R}$) and sp(4,$\mathbb{R}$). The Clifford algebra $A$ is a representation and a subalgebra of the universal enveloping algebra $U$ of these Lie algebras. We have normalized the mass, within a fixed group representation, by the dimension of the vector space carrying the representation, given by the trace of the representative of the identity in $A$. We may write the definition of the mass parameter from eqs. (1.2.5) or (3.5.4) in any representation of the algebras sl(4,$\mathbb{R}$) and sp(4,$\mathbb{R}$) and the corresponding representation of the common enveloping Clifford algebra $\mathcal{D}(A)$ as

$$m = \frac{1}{4}\operatorname{tr} J^\mu \Gamma_\mu = \frac{\operatorname{tr} J^\mu \Gamma_\mu}{\operatorname{tr} I_A} \equiv \frac{^cg\left(J^\mu \Gamma_\mu\right)}{^cg\left(I_{\mathcal{D}(A)}\right)} = \frac{\operatorname{tr}\left(\mathcal{D}(J){\bullet}\mathcal{D}(\Gamma)\right)}{\operatorname{tr}\left(-\mathcal{D}\left(\kappa^0\right)\mathcal{D}\left(\kappa_0\right)\right)} \quad . \tag{4.1.1}$$

It is known that the Cartan metric depends on the representations, as indicated in section 3.5, but we shall only apply this expression to find ratios within a particular fixed induced representation of the enveloping algebra.

If we consider geometric excitations on a substratum, this mass may be expanded as a perturbation around the substratum in terms of the only constant of the theory, the small fine structure parameter $\alpha$ characterizing the excitation,

$$J = J_0 + \alpha J_1 + \alpha^2 J_2 + \cdots \ , \tag{4.1.2}$$

$$\Gamma = \Gamma_0 + \alpha \Gamma_1 + \alpha^2 \Gamma_2 + \cdots \ , \tag{4.1.3}$$

$$m = \frac{1}{4}\operatorname{tr}\left(J_0^\mu \Gamma_{0\mu} + \alpha J_1^\mu \Gamma_{0\mu} + \alpha J_0^\mu \Gamma_{1\mu} + \mathcal{O}\left(\alpha^2\right)\right) \ , \tag{4.1.4}$$

indicating that the zeroth order term, that we shall call the bare mass, is given entirely by the substratum current $J_0$ and connection $\Gamma_0$, with corrections depending on the excitation self interaction.

The variation of $J$ is orthogonal to $J$. In the same manner as the 4-velocity has unit norm in relativity and the 4-acceleration is orthogonal to the 4-velocity, we have in the defining representation of tetradimensional matrices,

$$J_G \bullet J_G = e^{-1} \kappa_\mu e e^{-1} \kappa^\mu e = -4I = J_H \bullet J_H = J'_G \bullet J'_G = J'_H J'_H \quad . \tag{4.1.5}$$

This implies that $J_I$ is orthogonal to $J_0$,

The first order mass correction is

$$\Delta m = \frac{\alpha}{4} \operatorname{tr} \left( J_I \bullet \Gamma_0 + J_0 \bullet \Gamma_I \right) = \frac{\alpha}{4} \operatorname{tr} \left( m_g J_I \bullet J_0 + J_0 \bullet \Gamma_I \right) = \frac{\alpha}{4} \operatorname{tr} J_0 \bullet \Gamma_I \quad . \tag{4.1.6}$$

The first order equation for the potential is linear, expressed using a linear differential matricial operator. If we take the projection along $J_0$, its source becomes zero and since $J_0$ is constant there are no $\Delta m$ solutions that depend physically on the source $J$. It follows that the first order correction must be zero and the mass correction should be of order $\alpha^2$, or $10^{-5}$. These corrections correspond to a geometric quantum field approximation (QFT) [56]. In this section we limit ourselves to the zeroth order term which we consider the bare mass of QFT.

The structure group $G$ is $SL(2,\mathcal{Q})$ and the even subgroup $G_+$ is $SL_1(2,\mathbb{C})$. The subgroup $L$ is the subgroup of $G_+$ with real determinant in other words, $SL(2,\mathbb{C})$. There is another subgroup $H$ in the group chain $G \supset H \supset L$ which is $Sp(2,\mathcal{Q})$. The corresponding symmetric spaces and their isomorphisms are discussed in a previous publication [63]. We are dealing with two quotients which we shall designate as $C$ and $K$,

$$K \equiv \frac{G}{G_+} \cong \frac{SL(4,\mathbb{R})}{SL(2,\mathbb{C}) \otimes SO(2)} \cong \frac{SO(3,3)}{SO(3,1) \otimes SO(2)} \quad , \tag{4.1.7}$$

$$C \equiv \frac{H}{L} \cong \frac{Sp(4,\mathbb{R})}{SL(2,\mathbb{C})} \cong \frac{SO(3,2)}{SO(3,1)} \quad . \tag{4.1.8}$$

These groups have a principal bundle structure over the cosets and themselves carry representations. The geometric action of the $K$ generators are translations on the coset $K$. The functions on $K$ are the natural internal representations.

We shall consider, then, the representations of $SL(2,\mathcal{Q})$ and $Sp(2,\mathcal{Q})$ induced from the subgroups $SL_1(2,\mathbb{C})$ and $SL(2,\mathbb{C})$ over the symmetric spaces $SL(2,\mathcal{Q})/SL_1(2,\mathbb{C})$ and $Sp(2,\mathcal{Q})/SL(2,\mathbb{C})$, respectively. These geometric induced representations may be realized as sections of a homogeneous vector bundle $(D, K, \mathcal{D}[SL_1(2,\mathbb{C})], G_+)$ with $SL_1(2,\mathbb{C})$ representations $\mathcal{D}$ as fiber $F$ over the coset $K$ [71], as shown in figure 2. To the induced representation of $SL(2,\mathcal{Q})$ on $D$, there corresponds an induced representation of the enveloping Clifford algebra $A$ on $D$ [72]. Furthermore, to the latter also corresponds a representation of the subgroup $Sp(2,\mathcal{Q})$ on $D$. In other words, the vector bundle $D$ carries corresponding representations of $A$, $SL(2,\mathcal{Q})$ and $Sp(2,\mathcal{Q})$. These three representations are functions on $K$ valued on representations of $SL_1(2,\mathbb{C})$. At each point of the base space $M$ we consider the function space $\Sigma$ of all sections of the homogeneous vector bundle $D$. Define a vector bundle $S \equiv (S, M, \Sigma, G)$, associated to the principal bundle $E$, with fiber the function space $\Sigma$ of sections of $D$. The fiber of $S$ is formed by induced representations of $G$.

There is an induced potential acting, as the adjoint representation of $G$, on the bundle $S$. The potential or connection $\omega$ is represented by Lorentz rotations on $L$ and translations on $K$. The induced potential $\omega$ may be decomposed in terms of a set of state functions characterized by a parameter $k$, the generalized spherical functions $Y_k$ on the symmetric space [73]. If $K$ were compact, the basis of this function space would be discrete, of infinite dimensions $d$. The components, relative to this basis would be labeled by an infinite number of discrete indices $k$. The Cartan-Killing metric expresses the equipartition of energy in the induced representations and we have the mass parameter, in terms of the $\Gamma$, $J$ components,

$$m = \frac{1}{4d} \operatorname{tr} \sum_{k,k'} J^{k\mu}_{k'} \Gamma^{k'}_{k\mu} \quad . \tag{4.1.9}$$

All states equally contribute to the mass. Since the spaces under discussion are noncompact, the discrete indices $k$ that label the states, become continuous labels and the summation in matrix multiplication becomes integration over the continuous parameter $k$ or convolution of functions $J(k)$, $\Gamma(k)$. In addition, if we work with tetradimensional matrices and continuous functions on the coset $K$, the Cartan-Killing metric in $A$ is expressed by trace and integration, introducing a $4V(K_R)$ dimension factor for the common representation space $D$, giving

$$m = \frac{1}{4V(A_R)} \operatorname{tr} \int dk \int J(k, k_2) \Gamma(k_2, k) \, dk_2 \quad . \tag{4.1.10}$$

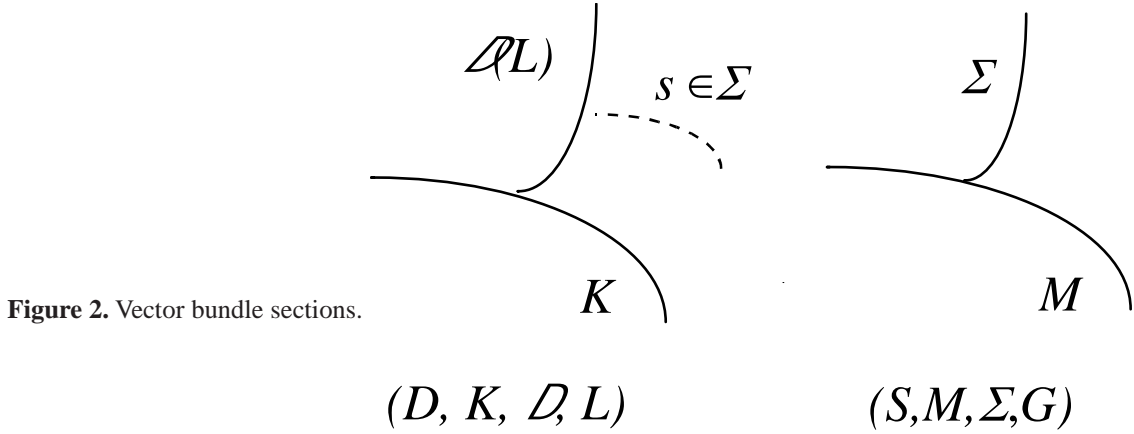

**Figure 2.** Vector bundle sections.

where $V(A_R)$ is a characteristic volume determining the dimension of the continuous representation of $A$. We interpret the value of a function at $k$ as the component with respect to the basis functions $Y(kx)$ of the symmetric space $K$, parametrized by $k$, as usually done in flat space in terms of a Fourier expansion. We may say that there are as many "translations" as points in $K$. It should be noted that these "translations" do not form the well known abelian translation group.

The $G$-connection on $E$ induces an SO(3,1)-connection on $TM$. The combined action of the connections, under the even subgroup $G_+$, leaves the orthonormal set $\kappa_\mu$ invariant [9], defining a geometric relativistic equivalence relation $R$ in the odd subspace $K$. Each element of the coset is a group element $k$ that corresponds to a space-time moving frame. *Physically* a class of equivalent moving frames $k$ is represented, up to a boost transformation, by a single rest frame, a point $k_0$ which corresponds to the rest mass $m_0$. The decomposition of $\Gamma$ and $J$ over the coset $K$ is into equivalence classes of state functions $Y(kx)$. The number of classes of state functions $Y(kx)$ (independent bases), is the volume of a subspace $K_R \subset K$ of classes (nonequivalent points under a relativistic boost transformation integral $I(\beta)$). Physically the integral represents summation of the potential × current product, over all inequivalent observers, represented by a rest observer.

There is a constant inert substratum solution [74], discussed in section 1.3, for the nonlinear differential equations that provides an SL(2,$\mathcal{Q}$) valued 1-form inert potential $\Lambda$ to the principal fiber bundle $(E,M,G)$. The only contribution of eq. (4.1.10) to the inert rest mass $m_0$ is through $\Lambda$ which represents the class of equivalent local potential forms $\Gamma$. At some particular frame $s$, that we may take as origin of the coset, the local expression for $\Lambda$ is constant.

All points of $K$ or $C$ may be reached by the action of a translation by $k$, restricted to the corresponding subgroup, from the origin of the coset. As the reference frame changes from $s$ at the origin to $sk$ at point $k$ of the coset, the local potential form changes as eq. (1.3.2), corresponding to the equivalence class of constant solutions $\Lambda$. All changed potentials correspond to the same constant solution class $\Lambda$, seen in the different reference frames of the coset.

In the principal bundle the constant potential $\Lambda$ combines with the constant $J$ to produce a constant product over the coset $K$ as shown in section 3.5. As indicated before, the mass variation produced by the last term in the equation, due to an arbitrary choice of frame, corresponds to inertial effects. The dominant noninertial effects are due to the first term in the right hand side of the equation because the current $J$ is a tensorial form that corresponds to the substratum potential tensor $\Lambda$. It is clear that the subtraction of $\Lambda$ from the potential, the last term in the equation that only has a frame $k$ dependence, transforms as a connection, and corresponds to the inertial connection. The physical contribution to the bare mass parameter may be calculated in terms of $\Lambda$, in the special frame $s$, defining an invariant expression in the group $G$ and its subgroups valid for a given representation.

We are interested in the representations of SL(2,$\mathcal{Q}$) and Sp(2,$\mathcal{Q}$) induced by the *same* representation of SL$_1$(2,$\mathbb{C}$) and corresponding to the *same* inert substratum solution. In the defining representation of tetradimensional matrices, the product $JJ$ contained in the $\Lambda J$ term is

$$J_G \bullet J_G = e^{-1}\kappa_\mu e e^{-1}\kappa^\mu e = -4I = J_H \bullet J_H = J'_G J'_G = J'_H J'_H \ , \tag{4.1.11}$$

invariant under a SL(2,$\mathcal{Q}$) transformation on coset $K$ and equal to the unit in the algebra $A$ for both the $G$ group and any $H$ subgroup. There is a representation of $A$ on the bundle $S$ corresponding to the induced representation. The invariance (equality) of the product $JJ$ must be valid in any representation of $A$, although the value of the product may differ from one representation to the other. For the induced function representations valued in the sl(2,$\mathbb{C}$) algebra (Pauli matrices), the invariant product becomes the $k$ integration,

$$\int J_\mu^{\hat{\alpha}}\left(k_1,k_2\right)\kappa_0\kappa_{\hat{\alpha}}J_{\hat{\beta}}^\mu\left(k_2,k_3\right)\kappa^0\kappa^{\hat{\beta}}dk_2 \equiv F\left(k_1,k_3\right)=F_0=\int J_\mu^{\hat{\alpha}}\left(k_1',k_2'\right)\kappa_0\kappa_{\hat{\alpha}}J_{\hat{\beta}}^{\mu'}\left(k_2',k_3'\right)\kappa^0\kappa^{\hat{\beta}}dk_2' \ , \tag{4.1.12}$$

which, must be a constant $4\times4$ matrix $F_0$ on $K$, independent of $k_1$ and $k_3$.

The trace expression for the mass becomes

$$m = \frac{\mathcal{M}}{V(A_R)} \operatorname{tr} \int_{K_R} F(k,k) dk \quad . \tag{4.1.13}$$

The integrand is the same $F_0$ constant for both groups, but the range of integration $K_R$ differs. Integration is on a subspace $K_R \subset K$ of relativistic boost inequivalent points of $K$ for the group $G$ and on a subspace $C_R \subset C \subset K$ for the group $H$. The expressions for the masses corresponding to $G$-excitations and $H$-excitations become,

$$m_G = \frac{\mathcal{M}}{V(A_R)} \operatorname{tr} \int_{K_R} F(k,k) dk = \frac{V(K_R)}{V(A_R)} \mathcal{M} \operatorname{tr}(F_0) \quad , \tag{4.1.14}$$

$$m_H = \frac{\mathcal{M}}{V(A_R)} \operatorname{tr} \int_{C_R} F(k,k) dk = \frac{\mathcal{M}}{V(A_R)} \operatorname{tr}(F_0) V(C_R) = \frac{m_G V(C_R)}{V(K_R)} \quad . \tag{4.1.15}$$

The bare mass parameters $m$ are related to integration on coset subspaces, depending proportionally on their volume. In other words, the ratio of the bare mass parameters for representations of $SL(2,\mathcal{Q})$ and $Sp(2,\mathcal{Q})$, induced from the same $SL(2,\mathbb{C})$ representation as functions on cosets $K$ and $C \subset K$ should be equal to the ratio of the volumes of the respective subspaces. The values of mass are related by the equation

$$m = \tfrac{1}{4} \operatorname{tr}\left(J^\mu \Gamma_\mu\right) = \tfrac{1}{4}\operatorname{tr}\left(J^\mu \Lambda_\mu\right) = \tfrac{1}{4}\operatorname{tr}\left(\left(e^{-1}\kappa^\mu e\right)\left(\frac{-\mathcal{M}}{4}e^{-1}\kappa_\mu e\right)\right) = \tfrac{1}{4}\operatorname{tr}\left(\mathcal{M}I\right) = \mathcal{M} \quad . \tag{4.1.16}$$

# 4.2. Symmetric Cosets.

## 4.2.1. Volume of $C$ Space.

First consider, the volume of the tetradimensional symmetric space $Sp(2,\mathcal{Q})/SL(2,\mathbb{C})$ which coincides with the quotient $SO(3,2)/SO(3,1)$. In the regular representation, it has the matricial structure determined by the groups [75, 76]. In particular the volume of $C$ is obtained by an integration over this curved minkowskian momentum space using coordinate relations given by the group. We split the integration into the angular integration on the compact sphere $S^2$, the radial boost $\beta$ and the energy parameter $\chi$,

$$V(C) = \int_0^\pi d\chi \sin^3\chi \int_0^\beta d\beta \sinh^2\beta \int_0^{4\pi} d^2\Omega = \frac{16\pi}{3} \int_0^\beta d\beta \sinh^2\beta = \frac{16\pi}{3} I_C(\beta) \quad , \tag{4.2.1}$$

obtaining the result in terms of a boost integral $I(\beta)$.

## 4.2.2. Volume of $K$ space.

For the volume of $K$, the integration is over an 8 dimensional symmetric space. This space $G/G_+$ has a complex structure and is a nonhermitian space [77]. Using its complex coordinates we define an equivalence relation $S$ in the points on $K$ by defining equivalent points as points with equal coordinate moduli,

$$S = S^1 \times S^1 \times S^1 \times S^1 . \tag{4.2.2}$$

The volume of $K$ will then be the product of the volumes of $S$ and the volume of the coset $K/S$. We split the integration into the integration on the 4 compact phase spaces $S$ and integration on the tetradimensional coset $K/S$, parametrized by the coordinate moduli. The last integration is further split into integration on the compact 3 spheres $S^3$ and integration on the complementary noncompact direction, which corresponds to boosts $\beta$, obtaining

$$V(K) = \int_K \sqrt{-g}\, d^4V d\psi_0 d\psi_1 d\psi_2 d\psi_3 = V(K/S) \times V(S) = \int_0^\beta d\beta \sinh^3\beta \int_0^\pi d\zeta \sin^2\zeta \int_0^{4\pi} d^3\Omega \times \left(\int_0^{2\pi} d\psi\right)^4 , \qquad \textbf{(4.2.3)}$$

$$V(K) = (2\pi^2)(2\pi)^4 \int_0^\beta \sinh^3\beta\, d\beta = 2^5\pi^6 I_K(\beta) , \qquad \textbf{(4.2.4)}$$

in terms of another boost integral $I(\beta)$. This result shows, as should be expected, that the relativistic equivalent subspace of $K$ is of higher dimension than the relativistic equivalent subspace of $C$, as indicated by the respective integrals of the hyperbolic sine of $\beta$.

### 4.2.3. Ratio of Geometric Volumes

We expect that the ratio of the volumes $V$ of the inequivalent subspaces of $K$ and $C$, corresponding to fundamental representation of spin ½, should be related to the ratio of the corresponding geometrical masses. We have to eliminate these equivalent points by dividing by the equivalence relation $R$ under the boosts of SO(3,1). Equivalent points are related by a Lorentz boost transformation of magnitude $\beta$. There are as many equivalent points as the volume of the orbit developed by the parameter $\beta$.

The respective inequivalent volumes are, for $C$,

$$V(C_R) = \frac{V(C)}{V(R(\beta))} = \frac{\frac{16\pi}{3} I_C(\beta)}{I_C(\beta)} = \frac{16\pi}{3} , \qquad \textbf{(4.2.5)}$$

and for $K$,

$$V(K_R) = \frac{V(K)}{V(R(\beta))} = \frac{2^5\pi^6 I_K(\beta)}{I_K(\beta)} = 2^5\pi^6 . \qquad \textbf{(4.2.6)}$$

Although the volumes of the noncompact spaces diverge the ratio of the compact subspaces, taking in consideration the equivalence relation $R$, has a well defined limit, obtaining

$$\frac{V(K_R)}{V(C_R)} = 6\pi^5 . \qquad \textbf{(4.2.7)}$$

We actually have shown a theorem that says: *The ratio of the volumes of K and C, up to the equivalence relation R under the relativity subgroup, is finite and has the value $6\pi^5$.*

## 4.3. The $p$, $e$ and $\nu$ Mass Ratios.

For a constant solution, the masses are proportional to the respective volumes $V(K_R)$. The constants of proportionality only depend on the specified inducing SL(2,ℂ) representation. In particular for any two representations induced from the spin 1/2 representation of $L$, the respective constants are equal.

Armed with the theorem of the previous section, the ratio of bare masses of the fundamental representations of the groups $G$ and its subgroup $H$ induced from the spin 1/2 representation of $L$, for the constant solution, equals the ratio of the volumes of the inequivalent subspaces of the respective cosets $G/L_1$ and $H/L$, and has the finite exact value

$$\frac{m_G}{m_H} = \frac{V(K_R)}{V(C_R)} = 6\pi^5 = 1836.1181 \approx \frac{m_p}{m_e} = 1836.153 , \qquad \textbf{(4.3.1)}$$

which is a very good approximation for the experimental physical proton electron mass ratio, in confirmation of the relation of the $G$ group to the proton and the $H$ group to the electron. The only other physical subgroup $L$=SL(2,ℂ) of $G$ should lead to a similar mass ratio. Previously we have related $L$ to the neutrino. In this case the quotient space is the identity and we get

$$\frac{m_L}{m_H} = \frac{V_R\left(L/L\right)}{V_R\left(C\right)} = \frac{V_R\left(I\right)}{V_R\left(C\right)} = 0 = \frac{m_\nu}{m_e} \quad , \tag{4.3.2}$$

which is in accordance with the zero bare mass of the neutrino. The physical mass of the neutrino may include a small correction term due to the excitation.

# 4.4. The Equation for the Potential Excitations or Bosons.

We consider now the equations for the geometric potential excitations around the fixed substratum geometry. In a particular case we shall show that the excitation equations take the form of a Yukawa equation. The corresponding solutions behave as short range fields, with a range given by a constant associated to the substratum potential solution. This constant, which in general depends on the particular representation, may also be interpreted as the mass of the particle associated to the potential excitation.

The field equation of the theory,

$$D \, {}^*\Omega = 4\pi\alpha \, {}^*J \quad , \tag{4.4.1}$$

is expanded [78] by writing the exterior product in terms of differential forms and group generators. The expression for the field equation becomes

$$\frac{-2}{\sqrt{-g}} \partial_\rho\left(\sqrt{-g}g^{\rho\mu}g^{\alpha\nu}\partial_{[\mu}\omega^a_{\nu]}\right)E_a + \frac{1}{\sqrt{-g}} \partial_\rho\left(\sqrt{-g}\omega^{a\rho}\omega^{b\alpha}\right)[E_a,E_b] +$$
$$2g^{\rho\mu}g^{\alpha\nu}\omega^a_\rho\partial_{[\mu}\omega^b_{\nu]}[E_a,E_b] + \omega^c_\rho\omega^{a\rho}\omega^{b\alpha}\left[E_c,[E_a,E_b]\right] = 4\pi\alpha \, J^{a\alpha}E_a \quad . \tag{4.4.2}$$

The commutator in the expressions introduce the structure constants and the trace of products of the Lie algebra base introduces the Cartan-Killing metric.

We shall proceed formally in the defining $4\times4$ representation. The equation may be written in terms of components referred to these bases. We note that the terms quadratic in the structure constant also include terms depending on the Cartan-Killing metric. This may be seen from the trace expression.

To obtain field excitations we perform perturbations of the geometric objects in the equation. Then the linear differential equation for the perturbation of the potential takes the general form

$$\frac{-2}{\sqrt{-g}} \partial_\rho\left(\sqrt{-g}g^{\rho\mu}g^{\alpha\nu}\partial_{[\mu}\delta\omega_{\nu]d}\right) + 4\omega^c_c\omega^\rho_c\delta\omega^\alpha_d + {}^1L^{\alpha c}_{d\rho}\delta\omega^\rho_c + {}^2L^{\gamma\mu\nu}_d\delta g_{\mu\nu} = 4\pi\alpha \, \delta J^\alpha_d \quad , \tag{4.4.5}$$

where ${}^1L$ and ${}^2L$ are linear first and second order operators, respectively, with variable coefficients which are functions of $\omega$. The second term, that arises from the cubic self interaction term, may provide a mechanism to give effective masses to the potential excitations in the curved substratum, in terms of a parameter that represents the self energy determined using the Cartan-Killing metric in the defining representation,

$$\omega^2 = \omega^c_\rho\omega^\rho_c \quad . \tag{4.4.6}$$

# 4.5. Massive Particular Solutions.

We have constructed a pair of conjugate constant substratum solutions [78], in terms of the geometrical fundamental unit of energy $\mathcal{M}$, by extending the real functions in eq. 1.3.2 for the substratum to complex functions,

$$\omega^\mp = -dx^{\hat\alpha}\left(\kappa_{\hat\alpha} \pm \kappa_5\kappa_{\hat\alpha}\right)\mathcal{M}/4 = -dx^{\hat\alpha}\kappa_{\hat\alpha}\left(1 \pm i\right)\mathcal{M}/4 \quad . \tag{4.5.1}$$

In particular if we assume that we can obtain a potential solution with algebraic components only in the complex Minkowski plane $K_L$ generated by the Clifford algebra orthonormal set $\kappa$, we have that the complete base $E$ may be replaced by the orthonormal set $\kappa$. Since the potential has no even components, the metric in the base space-time $M$ is flat. Using the trace relations we obtain the excitation equation using the complex metric $g$ of the complex Minkowski plane $K_L$, which involves lowering the indices and taking the complex conjugate,

$${}^*d \, {}^*d\delta\omega^\nu_\delta + 4\left(\delta\omega^{\hat\chi}_\rho\omega^{\hat\alpha\rho}\omega^{\hat\beta\nu} + \omega^{\hat\chi}_\rho\delta\omega^{\hat\alpha\rho}\omega^{\hat\beta\nu} + \omega^{\hat\chi}_\rho\omega^{\hat\alpha\rho}\delta\omega^{\hat\beta\nu}\right)\times\left(g_{\hat\delta\hat\beta}g_{\hat\chi\hat\alpha} - g_{\hat\delta\hat\alpha}g_{\hat\chi\hat\beta}\right) = 4\pi\alpha \, \delta J^\nu_\delta \quad , \tag{4.5.2}$$

$$^*d\,^*d\delta\omega_{\delta}^{\alpha}+\left|\omega^{\pm}\right|^{2}\delta\omega_{\delta}^{\alpha}+2\left|\omega^{\pm}\right|^{2}\delta\omega_{\rho}^{\hat{\rho}}\delta_{\delta}^{\alpha}=4\pi\alpha\,\delta J_{\delta}^{\alpha}\quad. \tag{4.5.3}$$

Given a representation, a solution to these coupled linear equations may always be found in terms of the Green's function of the differential operator and the current excitation $\delta J$, which may be an extended source. In order to decouple the equations, it is necessary to assume that $\delta\omega^{\prime}{}_{\rho}$ vanishes. If this is the case then all equations are essentially the same and the solution is simplified.

We shall restrict ourselves to consider the equation for a unit point excitation of the quantum of physical charge and its solution which is the Green's function. If we assume a time independent excitation with spherical symmetry, the only relevant equation would be the radial equation. Since $\mathcal{M}$ is constant, it represents an essential singularity of the differential equation, an irregular singular point at infinity. The corresponding solutions have the exponential Yukawa [79] behavior. Let us interpret that the curved substratum gives an effective range $\omega^{-1}$ to the linear excitations. Equation (4.5.3) when time independent for a point source becomes, designating the fluctuation as a weak field $W$,

$$-\nabla^{2}W\left(x\right)+\omega^{2}W\left(x\right)=4\pi\alpha^{2}g^{2}\delta\left(x-x'\right)\quad, \tag{4.5.4}$$

where we explicitly recognize that the current is of order $\alpha$ or equivalently of order charge squared. Furthermore, we realize that the assumed excitation is the odd part of an su(2)$_Q$ representation and should explicitly depend on the formal odd part of the charge. The charge that enters in this first order perturbation equation is the su(2)$_Q$ charge, eq. (3.3.4), defined by the original nonlinear unperturbed equation. This is the quantum of the only physical charge: the electric charge quantum $e=1$. We should define a factor $g$ to expresses the geometric charge quantum in units of an undefined formal odd charge (or weak charge). The $\alpha$ and $g$ factors are not part of the $\delta$ function which represents a unit odd charge. This equation may be divided by $(\alpha g)^{2}$, obtaining the equation for the odd excitations $W$ produced by the unit odd point charge or equation for the Green's function

$$\frac{1}{\left(\alpha g\right)^{2}}\nabla^{2}W\left(x\right)-\left(\frac{\omega}{\alpha g}\right)^{2}W\left(x\right)=-4\pi\delta\left(x-x'\right)\quad. \tag{4.5.5}$$

We may define a new radial coordinate $r=\alpha g x'$, rationalized to a new range $^{-}\mu^{-1}$ for odd excitations, which incorporates the $\alpha g$ constant. Taking in consideration the covariant transformations of $W$ and $\delta$, the radial equation is then

$$\frac{1}{r}\frac{\partial^{2}}{\partial r^{2}}\left(rW\right)-^{-}\mu^{2}W=-4\pi\delta\left(r-r'\right)\quad. \tag{4.5.6}$$

Since $^{-}\mu$ is constant, the Green's function for this differential operator is

$$\mathcal{G}=\frac{1}{4\pi}\frac{e^{-^{-}\mu|x-x'|}}{\left|x-x'\right|}=\frac{1}{4\pi}\frac{e^{-^{-}\mu r'}}{r'}\quad. \tag{4.5.7}$$

whose total space integration introduces, in general, a range factor,

$$\frac{1}{4\pi}\int_{0}^{\infty}dr'r'e^{-^{-}\mu r'}\int_{0}^{4\pi}d^{2}\Omega'=\frac{1}{^{-}\mu^{2}}\quad, \tag{4.5.8}$$

The range $^{-}\mu$ may be evaluated using eqs. (4.5.1) and (4.5.3),

$$\omega^{2}=g_{CC}\left(\omega_{\mu}\omega^{\mu}\right)=\tfrac{1}{4}\operatorname{tr}\omega_{\mu}^{*}\omega^{\mu}=\mathcal{M}^{2}/2\quad, \tag{4.5.9}$$

obtaining

$$^{-}\mu=\frac{\omega}{\alpha g}=\frac{\mathcal{M}}{\sqrt{2}\alpha g}\quad. \tag{4.5.10}$$

In order to obtain a complete excitation, instead of only its odd part, set $g$ equal to 1.

## 4.6. Massive SU(2) Bosons.

According to physical geometry we may consider excitations of the SU(2)$_Q$ subgroup of $G$ around the substratum. The complex substratum solution indicates that this excitation acquires an effective mass. Let us consider the three components of an $SU(2)_Q$ potential $A$ as three classically equivalent electromagnetic potentials.

The electromagnetic subgroup is $SU(2)_Q$, similar to the spin subgroup $SU(2)_S$. The group itself, as a fiber bundle $(SU(2), S^2, U(1))$ carries its own representations [78, 80]. The base space is the coset $SU(2)/U(1)$ which is the bidimensional sphere $S^2$. The fiber is an arbitrary even subgroup $U(1)$. The action of this electromagnetic $SU(2)_Q$ is a multiplication on the fiber by an element of the $U(1)$ subgroup and a translation on the base space $S^2$ by the action of the $SU(2)_Q$ group Casimir operator, representing a squared total $SU(2)_Q$ rotation.

This action is not as simple as translations in flat spaces, but rather has complications similar to those associated with angular momentum due to the $SU(2)$ group geometry. The orientation of directions in $SU(2)$ is quantized. In particular only one component of the electromagnetic rotation generator $E$ commutes with the group Casimir operator $E^2$. This operator acts on the symmetric coset $S^2$ becoming the Laplace-Beltrami operator on the coset. Its eigenvalues are geometrically related to translations on $S^2$ in the same manner as the eigenvalues of the usual Laplace operator are related to translations on the plane. There are definite eigenvalues, simultaneous with the Casimir operator, only along any arbitrary *single* su(2) direction, which we have taken as the direction of the even generator $^+E$. The generator $E$ may be decomposed in terms of the even part and the complementary odd component $^-E$. We can not decompose $E$ into components with definite expectation values. The splitting into even and odd parts represents, respectively, the splitting of the group action into its vertical even action on the fiber and a complementary odd translation on the base $S^2$.

Consider that the exponential functions $e^{k \cdot x}$ form a representation of the translation group on the plane. The magnitude of the translation $k$ is determined by the eigenvalue of the Laplace operator $\triangle$,

$$\Delta e^{k \cdot x} = \lambda e^{k \cdot x} = k^2 e^{k \cdot x} \quad , \tag{4.6.1}$$

where the absolute value of $k$ is

$$|k| = \left( \delta^{mn} k_m k_n \right)^{1/2} \quad . \tag{4.6.2}$$

The odd subspace of $su(2)_Q$, spanned by the two compact odd electromagnetic generators in $^-E$, is isomorphic to the odd subspace of the quaternion algebra, spanned by its orthonormal subset $q^a$. Associated to this orthonormal subset we have the Dirac operator $q \cdot \nabla$ on a curved bidimensional space. This Dirac operator represents the rotation operator $L^2$ on the vector functions on the sphere which also corresponds to the laplacian bidimensional component. We obtain for this action, if we separate the wave function $\varphi$ into the $SU(2)_S$ eigenvector $\phi$ and the $SU(2)_Q$ eigenvector $\psi$, and use the fact that the Levi-Civita connection is symmetric,

$$\left( q^a \nabla_a \right)^2 \psi = q^a q^b \nabla_a \nabla_b \psi = q^{(a} q^{b)} \nabla_a \nabla_b \psi + q^{[a} q^{b]} \nabla_a \nabla_b \psi = -g^{ab} \nabla_a \nabla_b \psi = -\Delta \psi = L^2 \psi \quad . \tag{4.6.3}$$

This equation shows that the squared curved Dirac operator is the Laplace-Beltrami or Casimir operator. Therefore, we may define the odd electromagnetic generator $^-E$ as the quaternion differential operator

$$^-E \equiv q^a \nabla_a = \sqrt{-\Delta} = \sqrt{C^2} \quad . \tag{4.6.4}$$

The $^-E$ direction in the odd tangent plane is indeterminable because there are no odd eigenvectors common with $^+E$ and $E^2$. Nevertheless the absolute value of this quaternion must be the square root of the absolute value of the Casimir quaternion. The absolute values of $^+E$ and $^-E$ define a polar angle $\Theta$ in the su(2) algebra as indicated in figure 3.

The angle $\Theta$ is a property of the algebra representations, independent of the normalization as may be verified by substituting $E$ by $NE$. The generators have twice the magnitude of the standard spin generators. This normalization introduces factor of 2 in the respective commutation relations structure constants and determines that the the generator eigenvalues, characterized by the charge quantum numbers $c$, $n$, are twice the standard eigenvalues characterized by the spin quantum numbers $j$, $m$. Nevertheless, it appears convenient to use the two different normalizations for the $SU(2)_Q$ and $SU(2)_S$ isomorphic subgroups in accordance with the physical interpretation of the integer charge and half-integer spin quanta.

The electromagnetic generator has an indefinite azimuthal direction but a quantized polar direction determined by the possible translation values. Therefore we obtain, since the absolute values of the quaternions are the respective eigenvalues,

$$\frac{\left| ^-E \right|}{\left| ^+E \right|} = \frac{\left| E^2 \right|^{1/2}}{\left| ^+E \right|} = \frac{\sqrt{j(j+1)}}{m} \equiv \tan \Theta_m^j = \frac{\sqrt{c(c+2)}}{n} \equiv \tan \Theta_n^c \quad . \tag{4.6.5}$$

The internal directions of the potential $A$ and the current $J$ in $su(2)_Q$ must be in the possible directions of the electromagnetic generator $E$. In the near zone defined by $^-\mu r \ll 1$ the $A$ components must be proportional to the possible even and odd translations. The total $A$ vector must lie in a cone, which we call the electrocone, defined by a quantized polar angle $\Theta_n^c$ relative to an axis along the even direction and an arbitrary azimuthal angle.

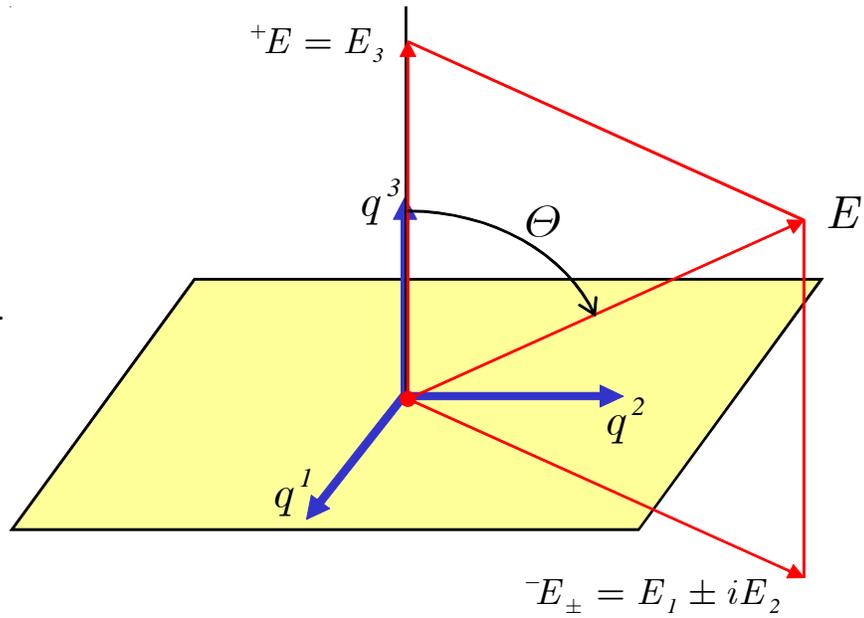

**Figure 3.**
Electromagnetic
polar angle in su(2).

The fundamental state $|c,n\rangle$ that represents a charge quantum is $|1,1\rangle$ and corresponds to the $|\frac{1}{2},\frac{1}{2}\rangle$ electromagnetic rotation state. The corresponding $\Theta_{\frac{1}{2}}^{\frac{1}{2}}$ angle is

$$\frac{\left|{}^-E\right|}{\left|{}^+E\right|} = \frac{\sqrt{\frac{1}{2}\left(\frac{1}{2}+1\right)}}{\frac{1}{2}} = \sqrt{3} = \tan\left(\frac{\pi}{3}\right) \equiv \tan\Theta_{\frac{1}{2}}^{\frac{1}{2}} \quad . \tag{4.6.6}$$

The su(2) potential excitations are functions over the bidimensional sphere SU(2)/U(1) generated by the $^-E$ odd generators. We may define the complex charged raise and lower generators $^-E^{\pm}$, which are different from the odd generator $^-E$, in terms de the real generators. Therefore, these charged excitations, defined as representations of the $SU(2)_Q$ group, require a substratum with a preferred direction along the even quaternion $q^3$ in the $su(2)_Q$ sector. Only the complex substratum solution, [78, 1] is adequate.

The odd complex substratum solutions in the su(2) sector reduce to

$$^-\omega = {}^-\omega^{0\pm}\left(\kappa_0 \pm i\kappa_1\kappa_2\kappa_3\right) = {}^-\omega^{0\pm} \, {}^-q^{\pm} \tag{4.6.7}$$

and should correspond to the two $^-E$ odd sphere generators. This odd su(2) subspace generated by the complex solution is physically interpreted as an SU(2) electromagnetic odd substratum, a vacuum, with potential $^-\omega$ which determines ranges which may be interpreted as masses for the excitations of the electromagnetic potential in its surroundings. The range $^-\mu^{-i}$ is determined by the $^-\omega$ fundamental bidimensional component in the substratum equatorial plane.

The even and odd components of the su(2) excitation in the $E$ direction have to obey the relations,

$$\left|{}^+A\right|^2 + \left|{}^-A\right|^2 = |A|^2 \quad , \tag{4.6.8}$$

Due to the quantization of the su(2) potential, in the near zone there only are two possible definite absolute values, associated to a fundamental representation, which are related by

$$\frac{\left|{}^-A\right|}{|A|} = \frac{\left|{}^-E\right|}{|E|} = \sin\Theta_{1/2} \quad . \tag{4.6.9}$$

A fundamental $su(2)_Q$ potential excitation should be the spin 1 boson ($SU(2)_S$ representation which also is a fundamental $SU(2)_Q$ representation where the three component generators keep the quantized relations corresponding to electromagnetic rotation eigenvalues ½, ½, similar to the frame excitation representation (proton). Of course, they differ regarding the spin $SU(2)_S$ because the former is a spin 1 representation and the latter is a spin ½ representation. We shall call this excitation by the name complete fundamental excitation. When the angle $\Theta$ is written without indices it should be understood that it refers to this representation. A fluctuation of the odd substratum does not provide a complete $su(2)_Q$ excitation.

The $su(2)_Q$ currents are similarly quantized and define the formal odd charge. The $g$ factor that expresses the geometric charge quantum in units of the formal odd charge (or weak charge) is

$$g = \csc\Theta_{1/2} \quad . \tag{4.6.10}$$

The only possible ranges associated with an su(2)$_Q$ excitation should be proportional to the only possible values of the quantized potential

$$\frac{\left|{}^-A\right|}{|A|} = \frac{1}{g} = \sin\Theta = \frac{{}^-\mu}{\mu} \quad . \tag{4.6.11}$$

The possible values of the range are

$${}^-\mu = \frac{\omega}{\alpha g} = \frac{2\sqrt{2}\, m_g \sin\Theta_{1/2}}{\alpha} \tag{4.6.12}$$

and

$$\mu = \frac{\mathcal{M}}{\sqrt{2}\alpha} \quad . \tag{4.6.13}$$

On the other hand, the excitation ranges should be provided by the excitation field equation (4.5.3) as the absolute value of some substratum potential. The odd complex substratum admits a related complete substratum with an additional even component ${}^+\omega$ potential that should provide the value of the energy difference. This even part ${}^+\omega$ increases the modulus of the total potential. The possible value of ${}^+\omega$ should correspond to the allowed value of the total substratum potential absolute value $\mu$. We shall call this substratum the complete complex substratum.

The value $\mu^2$ represents the square of the bound energy of a complete (with its three components) su(2)$_Q$ fundamental excitation around the complete complex substratum with equation,

$$\nabla^2 A(x) - \mu^2 A = -4\pi\delta(x - x') \quad . \tag{4.6.14}$$

Since the orientations of the potential $A$ and the substratum connection $\omega$ are quantized, their vectorial decompositions into their even and odd parts are fixed by the representation of the connection $\omega$. Therefore this $\mu^2$ energy term may be split using the characteristic electrocone angle $\Theta$,

$$\mu^2 = \mu^2\left(\cos^2\Theta + \sin^2\Theta\right) = \left(\mu\cos\Theta\right)^2 + \left(\mu\sin\Theta\right)^2 \equiv {}^+\mu^2 + {}^-\mu^2 \quad . \tag{4.6.15}$$

The energy parameter ${}^+\mu$ corresponds to the even component ${}^+\omega$ associated with the U(1) group generated by the $\kappa_5$ even electromagnetic generator.

The energy $\mu$ produced by the excitation components may not be decomposed without dissociating the su(2) excitation due to the orientation quantization of its components. If the complete excitation is disintegrated in its partial components, the corresponding even equation is separated from the odd sector in $K$ as indicated in eq. (4.5.3) and has an abelian potential. Therefore, no mass term appears in the even equation, which is physically consistent with the zero mass of the photon. The energy ${}^+\mu$ associated to ${}^+\omega$ in the $\kappa_5$ direction is available as free energy. For a short duration the disintegration takes energy from the substratum. On the other hand, the energy ${}^-\mu$ corresponds to the odd complex substratum and when the complete excitation is disintegrated, the energy appears as the mass term in the coupled equations (4.5.3) associated to the pair of odd generators ${}^-E^\pm$

$$\nabla^2\, {}^-A(x) - \left(\mu^2 - {}^+\mu^2\right)\, {}^-A + L_1\left(\delta\omega\right) = -4\pi\delta(x - x') \quad . \tag{4.6.16}$$

We may neglect the coupling term $L_1$, as was done previously in eq. (4.5.3)), so that the equation may approximately be written

$$\nabla^2\, {}^-A(x) - {}^-\mu^2\, {}^-A = -4\pi\delta(x - x') \quad . \tag{4.6.17}$$

## 4.6.1. Mass Values in Free Space.

These SU(2)$_Q$ potential excitations are also spin 1 representations of $G$ or $P$ induced from the $L$ subgroup. The considerations in section 4.1 apply to them, in particular the expression for the physical mass, eq. (4.1.10). This amounts to a calibration of the geometric mass $\mathcal{M}$ which appears in equations (4.5.9) for the potential excitation range and (4.1.16) for the stable fermions [78, 9].

The only way of calibrating the geometrical $\mathcal{M}$ mass for $\omega$ is through the only two physical masses proportional to $\mathcal{M}$ by the integration required for the induced representations as described in section 4.1. The calibration with the $\mathrm{Sp}(2,\mathcal{Q})$ representation mass $m_e$ is inadequate for a free $^-A$, su(2) potential particle excitation, because the latter requires a full $\mathrm{SL}(2,\mathcal{Q})$ matter source $\delta j$ current representation. If we calibrate $\mathcal{M}$ with the mass $m_p$ we obtain

$$\omega = \left| \frac{1}{4}\operatorname{tr}\omega_\mu^*\omega^\mu \right| = \frac{\mathcal{M}}{\sqrt{2}} = \frac{m_p}{\sqrt{2}} = \alpha g \mu \ . \tag{4.6.18}$$

Considering these relations, the $\Theta$ electrocone angle also determines the ratio of the two energies or masses associated to the fundamental excitation,

$$\frac{m_{^-A}}{m_A} = \sin\Theta \ . \tag{4.6.19}$$

These relations determine that the SU(2) potential excitation energies or masses are proportional to the proton mass $m_p$,

$$m_A = \mu = \frac{\mathcal{M}}{\sqrt{2}\alpha} = \frac{m_p}{\sqrt{2}\alpha} = 90.9177 \ \text{ Gev.} \approx m_Z = 91.188 \ \text{ Gev.} \tag{4.6.20}$$

$$m_{^-A} = \ ^-\mu = \frac{m_p \sin\Theta}{\sqrt{2}\alpha} = 78.7370 \ \text{ Gev.} \approx m_W = 80.42 \ \text{ Gev.} \tag{4.6.21}$$

These values indicate a relation with the weak intermediate bosons [81]. They also indicate a relation of this electroconic angle with Weinberg's angle: Weinberg's angle would be the complement of the angle $\Theta$. They admit corrections of order $\alpha$, in particular the charged boson masses need electromagnetic corrections. For example if we use a corrected value for $\Theta$, obtained from its relation to the experimental value of Weinberg's angle,

$$m_{^-A} = \frac{m_p \sin\Theta}{\sqrt{2}\alpha} = 79.719 \ \text{ Gev} \approx m_W = 80.42 \ \text{ Gev} \ . \tag{4.6.22}$$

Let us define a monoexcitation as an excitation that is not a complete or odd $\mathrm{SU}(2)_Q$ representation and is associated to a single generator. A collision may excite a resonance at the $A$ fundamental potential excitation energy $m_A$ determined by the complete complex substratum. The fundamental potential excitation $A$ which is an SU(2) representation may also be dissociated or disintegrated in its three components ($^+A$, $^-A^\pm$), when the energy is sufficient, as two $^-A^\pm$ free monoexcitations, each one with mass (energy) $m_{^-A}$ determined by the odd complex substratum, and a $^+A$ third free U(1) monoexcitation, with null mass determined by the even complex substratum component $^+\omega$ which is abelian.

This simply determines the existence of four spín-1 excitations or boson particles associated to the SU(2) potential and their theoretical mass values:

1. The resonance at energy $\mu$ of the fundamental potential $A$, which has to decay neutrally in its components ($^+A$, $^-A^\pm$) and may identified with the $Z$ particle.
2. The pair of free monoexcitations $^-A^\pm$, charged $\pm 1$ with equal masses $^-\mu$, which may identified with the $W^\pm$ particles.
3. The free excitation $^+A$, neutral with null mass which may identified with the photon.

In this manner we may give a geometric interpretation to the $W$, $Z$ intermediate bosons and to Weinberg's angle which is then equivalent to the complement of the $\Theta$ electrocone angle.

## 4.6.2. Potential Excitations in a Lattice.

The equation for the odd $\mathrm{SU}(2)_Q$ excitation, eq. (4.5.4) displays a geometric current automatically produced by the same field excitation. This current component, with charged potential excitations, may dominate the total effective current. In these cases the original wave equation (4.5.3), decoupled and time independent, determines a Helmholtz equation [82] for a coherent self sustained collective magnetic su(2) potential wave where the perturbation current $\delta j$ has a charged potential excitation component in addition to the material component,

$$\nabla^2 A(x) = \omega^2 A(x) - 4\pi\alpha^2 \ ^-\delta j\big(A(x)\big) \ . \tag{4.6.23}$$

As an aplication, we may consider excitations around a lattice with periodic potentials in some media. Any wave in a periodic medium should be described as a Bloch wave [83]. The properties of this waves are essentially based on the Floquet theory [84] of differential equations. If we have a many excitation system composed by different particle species in the medium, we should consider the Bloch wave equations for Helmholtz's potential and Pauli's particle equation for the different species. The wave functions behave as Bloch waves consisting of a plane wave envelope function and a periodic Bloch function $u_{nk}(r)$ with the same periodicity of the external potential,

$$\psi_{n\vec{k}}(\vec{r}) = e^{i\vec{k}\cdot\vec{r}} u_{n\vec{k}}(\vec{r}) \ . \tag{4.6.24}$$

The index $n$ characterizes the energy bands of the solutions. A Bloch wave vector $k$ represents the conserved momentum, modulo addition of reciprocal lattice vectors. The wave group velocity is also conserved. Wave associated particles can propagate without scattering through the medium almost like free particles. There are possible transfers of energy and momentum among the different waves.

For the collective $A$ field the domain of the Bloch functions $u_{nk}$ are lattice cells where disturbances arise. These disturbances should be produced by first order $\alpha$ terms in the $\delta j$ current, eq. (4.6.23) and would determine the range of the excitations.

Equation (4.6.12) for the range $\mu$, using the mass $m_p$ gives $m_W$ as shown in the previous section. For a collective state of excitations, the self sustained $^-A$ su(2) total potential excitation from all particle species does not require a particular SL(4,R) matter source $\delta j$ current representation and may additionally couple to the $\kappa^0$ generator of an Sp((4,R) representation with the other fundamental mass $m_e$. The same expression (4.6.12), using this mass $m_e$, gives a lower energy mass

$$m_{^-A} = \ ^-\mu = \frac{m_e \sin\Theta}{\sqrt{2}\alpha} = 83.9171 m_e = 42.8814 \text{ MeV} \ . \tag{4.6.25}$$

Therefore the massive collective $SU(2)_Q$ connection excitations can be in these two states of diferent energy. These excitations obey Bose-Einstein statistics and under certain conditions may condense into a collective coherent state corresponding to the lowest energy. The mass given by eq. (4.6.25) should be related to energy terms in the $A$ potential Bloch functions $u_{nk}$. Using this collective mass instead of the free space mass $m_W$ we obtain the quantum relation required by the coherent global complete su(2) algebra between its odd and even components,

$$\left|\ ^-A\right|^2 = \tan^2\Theta \ e^{-2 m_A r} \left|\ ^+A\right|^2 \ . \tag{4.6.26}$$

# 4.7. Results.

We are able to identify three classes of fundamental excitations, corresponding to the three stable physical particles: neutrino, electron and proton. The quotient of bare inertial masses of these stable particles are calculated using integration on the corresponding symmetric spaces and leads to a surprising geometric expression for the proton electron mass ratio, previously known but physically unexplained.

There are massive potential excitations of the Yukawa equation for the triple electromagnetic potentials. The electromagnetic subgroup is $SU(2)_Q$, similar to the spin subgroup. The action of this electromagnetic group is a multiplication by an element of the U(1) subgroup and a translation on a sphere. This action is not as simple as translations in flat spaces, but rather has complications similar to those associated with angular momentum. The orientation of directions in SU(2) is quantized. The calculated masses correspond to the weak boson masses and allow a geometric interpretation of Weinberg's angle. It is possible that collective excitation states of the potential may have a smaller Yukawa mass.

# 5. Nuclear Energy and Interaction.

The role that Clifford algebras play in the geometrical structure of the theory provides a link to the nonclassical interactions theories. It should also be possible to represent nuclear interactions within these geometric ideas, in the same manner as gravitation and electromagnetism are represented by the unified theory. The potential corresponding to the $SU(2)_Q$ generators represents a generalized nuclear electromagnetism and is subject to quantization in the same manner that angular momentum. In particular, this geometric coupling provides a short range attractive magnetic nuclear potential with a $1/r^4$ strong radial dependence which is an important nonlinear nuclear interaction. The interest in electromagnetic effects on the nuclear structure is old [85]. Nevertheless, here we present new effects due to this geometric nuclear interaction. For certain applications it is not necessary to use the complete theory and it is sufficient to take the triple electromagnetic coupling of the $SU(2)_Q$ subgroup.

It is also be possible to represent electron neutrino weak nuclear interactions and relate it to the Fermi theory. An even frame obeys the neutrino equation. The holonomy groups of the potential or connection may be used geometrically to classify the interactions contained in the theory. The subgroup chain $SL(2,\mathcal{Q}) \supset Sp(2,\mathcal{Q}) \supset SL(2,\mathbb{C})$ characterizes a chain of subinteractions with reducing sectors of nonclassical interactions and has a symmetry $SU(2)\otimes U(1)$ equal to that of weak interactions.

# 5.1. Nonrelativistic Motion of an Excitation.

The general equations of motion were discussed in section 3.6. We now take the equations in that section as linearized equations around the substratum solution with a potential fluctuation representing self interactions. For a nonrelativistic approximation we may neglect the low velocity terms of order $v/c$, which correspond to the boost sector of the algebra, in other words the boost $\Upsilon$ and the hermitian parts of $\eta$ and $\xi$,

$$\eta^\dagger = -\eta \quad , \tag{5.1.1}$$

$$\xi^\dagger = -\xi \quad . \tag{5.1.2}$$

As usually done in relativistic quantum mechanics [86] in a nonrelativistic approximation, we let $\varphi$ and $\chi$ in eqs. (3.6.6) and (3.6.7) be slowly varying functions of time. Neglecting $\Upsilon$, the equations become after recognizing the space component $^-A_m$ as an odd magnetic vector potential,

$$\left(i\nabla_0 + {}^+\!A_0 + {}^-\!A_0\right)\varphi - \sigma^m\left(i\nabla_m + {}^+\!A_m - {}^-\!A_m\right)\chi = 0 \quad , \tag{5.1.3}$$

$$\left(i\nabla_0 + {}^+\!A_0\right)\chi - \sigma^m\left(i\nabla_m + {}^+\!A_m + {}^-\!A_m\right)\varphi = -2m\chi + {}^-\!A_0\,\chi \quad , \tag{5.1.4}$$

where $\chi$ is a small component, of order $v/c$ relative to the large component $\varphi$. We neglect the small terms which are, as usual, the $\chi$ terms in equation (5.1.4) unless multiplied by $m$, and substitute the resultant expression for $\chi$ in equation (5.1.3). The result is

$$\left(i\nabla_0 + {}^+\!A_0 + {}^-\!A_0\right)\varphi - \frac{\sigma^m\left(i\nabla_m + {}^+\!A_m - {}^-\!A_m\right)\sigma^n\left(i\nabla_n + {}^+\!A_m + {}^-\!A_m\right)\varphi}{2m} = 0 \quad . \tag{5.1.5}$$

Since we have the well known relation

$$\sigma.a\sigma.b = a.b + i\sigma.\left(a\times b\right) \quad , \tag{5.1.6}$$

substitution in equation (5.1.11 ) gives, denoting the 3-vector magnetic potential by $A$,

$$i\nabla_0\varphi = \Big[\frac{\left(i\nabla + {}^+\!A - {}^-\!A\right)\bullet\left(i\nabla + {}^+\!A + {}^-\!A\right)}{2m} - \left({}^+\!A_0 + {}^-\!A_0\right) +$$

$$-\frac{\sigma\bullet\left(\nabla\times\left({}^+\!A + {}^-\!A\right)\right)}{2m} + \frac{\sigma\bullet\left({}^-\!A\times\nabla\right)}{m} - \frac{i\sigma\bullet\left({}^-\!A\times{}^+\!A\right)}{m}\Big]\varphi \tag{5.1.7}$$

which is a generalized Pauli's equation [87] depending on the even and odd magnetic vector potentials $^+A$, $^-A$. The vector $^-A$ decays exponentially in the characteristic Yukawa form as seen in section 4.5. Distances $m_W r \ll 1$ define a subnuclear zone where the exponential approximates to 1. Distances $m_W r \gg 1$ define a far zone where the exponencial is negligible. In the far zone we obtain Pauli's equation,

$$i\nabla_0\psi = \left[\frac{\left(i\nabla + A\right)^2}{2m} - V - \frac{\sigma\cdot\nabla\times A}{2m}\right]\psi \quad . \tag{5.1.8}$$

# 5.2. Magnetic Moments of $p$, $e$ and $n$.

According to physical geometry [7, 8] a $G$-solution should include the three $SU(2)_Q$ generators. The equation of motion (5.1.7) displays the remarkable geometrical structure of triplets. We may associate the effect of a combination of three $SU(2)_Q$ potential components, one for each possible $P$ in $G$, as three classically equivalent electromagnetic potentials $A$'s. It has been shown [8, 9], in this geometry, that all long range fields correspond to fields associated to the fiber bundle obtained by contracting the structure group $SL(4,\mathbb{R})$ to its even subgroup $SL_1(2,\mathbb{C})$ which in turn correspond to classical fields. Therefore the long range component of the $sl(4,\mathbb{R})$-potential coincides with the long range component of an $sl_1(2,\mathbb{C})$-potential corresponding to an even subalgebra $sl(2,\mathbb{C})\oplus u(1)$ of $sl(4,\mathbb{R})$ related to gravitational and electromagnetic fields. In fact any direction in the tridimensional geometric electromagnetic $su(2)_Q$ subalgebra may be identified as a valid direction corresponding to this remaining long range classical electromagnetic

u(1). There is no preferred direction in $su(2)_Q$. If we observe a long range electromagnetic field, we may always align the classical field $A$ with any of the 3 geometric electromagnetic $\kappa$ generators in $su(2)_Q$, or a linear combination, by performing an $SU(2)_Q$ transformation. Nevertheless the two extra $A$'s should make additional contributions to the magnetic energy of the short range $G$-system, as shown in equation (5.1.7), and therefore to the corresponding magnetic moment [80]. In order to study this energy we shall restrict ourselves to the subnuclear zone. See section 4.6 for more details about the electromagnetic $SU(2)_Q$ subgroup which is similar to the spin subgroup $SU(2)_S$.

The internal direction of the potential $A$ is along the possible directions of the electromagnetic generator $E$ in $su(2)_Q$. The $A$ components are proportional to the possible even and odd translations. In the subnuclear zone the total $A$ vector lies in the electrocone defined by the quantized polar angle $\Theta$ [78] relative to the even direction axis and an arbitrary azimuthal angle, as shown in fig. 3.

The action of the 4-potentials $^-A_\mu$ or $^-A_\mu$ over a common proper vector in the algebra may be simply expressed in terms of the action of $^+A_\mu$,

$$A_\mu \psi = \left( {}^+A_\mu + {}^-A_\mu \right) \psi = \left( n + e^{-{}^-\mu r} \lambda(i) \right) \psi_\mu$$

$$= \left( I + e^{-{}^-\mu r} \lambda(i)/n \right) n \psi_\mu \equiv \left( I + \tan \Theta_n^{t_i}(\mu, r) \right) {}^+A_\mu \psi \equiv \left( I + \Xi \right) {}^+A_\mu \psi \,. \tag{5.2.1}$$

The 4-potential $A_\mu$ has an orientation in the Clifford algebra with the angle $\Theta$' respect to the standard electromagnetic potential, which reduces in the subnuclear zone $m_W r \ll 1$ to the electroconic angle $\Theta_n^{t_i}$ required by the quantization of the $SU(2)_Q$ generator in the representation $(n, i)$, as indicated in section 4.6. The angle $\Theta_{1/2}^{1/2}$ of the fundamental representation is approximately equal to the complement of Weinberg's angle and it may suggest its geometric interpretation. When the angle $\Theta$ is written without indices it should be understood that it refers to the fundamental representation.

The magnetic energy coupled to the spin in eq. (5.1.7) in the subnuclear zone is

$$U = -\frac{\sigma}{2m} \bullet \left( \nabla \times \left( {}^+A + {}^-A \right) \right) \varphi = -\frac{\sigma \bullet B}{2m} (n + \lambda) \varphi = -\frac{\sigma \bullet B}{2m} \left( I + \frac{\lambda}{n} \right) n \varphi$$

$$= -\left( I + \frac{\left( i(i+I) \right)^{1/2}}{n} \right) \frac{\sigma}{2m} \bullet \left( \nabla \times {}^+A \right) \varphi = -\left( I + \tan \Theta_n^i \right) \frac{\sigma}{2m} \bullet \left( \nabla \times {}^+A \right) \varphi \,. \tag{5.2.2}$$

We have introduced the electroconic angle $\Theta$, defined by eq. (4.6.5), representing the direction of the total generator $A$ for a $G$-system, with respect to the even direction in the $su(2)$ algebra.

The fundamental state representing a proton is the $SU(2)_Q$ state with charge $+1$, corresponding to electromagnetic rotation eigenvalues $\frac{1}{2}$, $\frac{1}{2}$. The value of $\Theta$ is $\pi/3$. In terms of the even magnetic field the energy becomes

$$U = -\left( I + \frac{\left( \frac{1}{2}\left( \frac{1}{2} + I \right) \right)^{1/2}}{\frac{1}{2}} \right) \frac{\sigma \bullet {}^+B}{2m} \varphi = -\frac{\left( I + \sqrt{3} \right)}{2m} \sigma \bullet {}^+B \varphi \,. \tag{5.2.3}$$

The first term in the parenthesis is related to the $P$-system which has only a U(1) electromagnetic subgroup and the complications due to SU(2) are not present. The complementary odd subspace corresponding to $S^2$ does not exist. The orientation of the complete electromagnetic potential $A$ can always be taken along the definite even direction defined by the physical u(1) algebra. The angle $\Theta$ may be taken equal to zero. It corresponds to a $P$-system, associated to the electron, with only a $\kappa^0$ electromagnetic component. In this case the energy reduces to the first term.

If it were possible to make a transformation that aligns the internal direction $\Theta$ along the even direction everywhere, we really would be dealing with a $P$-system because we actually would have restricted the potential to a $P$ subgroup. A $P$-system provides a preferred direction in the subgroup $SU(2)_Q$ of $G$, the only electromagnetic generator $\kappa^0$ of the associated $P$ group. If we use a $P$-test-particle (an electron) to interact with an external magnetic field $B$, we actually align the long range component $A$ with this preferred direction. In this manner we may explain the long-range physical experiments using an abelian electromagnetic field equation with a remote source and a Dirac equation. If we use a $G$-test-particle (a proton) to interact with an external field, the most we can do is to align the classical long-range component $A$ with a random direction in su(2). Part of the total internal field $^-B$ is only observable in a short range region. To each $B$ direction related by an $SU(2)_Q$ transformation, there corresponds a "*partner*" spin direction, defining associated scalar products $\sigma.B$. The additional scalar products, relative to a $P$-system, arise from the two additional geometric fields, and/or equivalently, from the two additional copies of Pauli matrices $\kappa^n \kappa$ in the universal Clifford algebra, determined by the additional spin-magnetic directions $\kappa$, which originally were along $\kappa^1 \kappa^2 \kappa^3$ and $\kappa^5$. The additional spin operators, introduced by the short-range "*non-classical*" electromagnetic potentials, represent extra internal current loops that generate an "anomalous" increase of the intrinsic magnetic energy.

We have calculated the magnetic energy of a free $G$-system, with zero external field $B$, in terms of the observable even internal field $^+B$ whose internal direction coincides with the eventual internal direction of an external field $B$. In this gedenken experiment, the magnetic moment *is defined* as the partial derivative of the magnetic energy, produced by the total internal generalized magnetic field $^iB$, with respect to the even component $^+B$, in whose internal direction would align the internal direction of the external $B$ field,

$$\mu^a \equiv -\frac{\partial\,U}{\partial\,^+B_a} \quad . \tag{5.2.4}$$

The standard physical methods of magnetic moment measurement are nuclear paramagnetic resonance [88, 89], molecular beams and optical spectroscopy [90]. In a real physical experiment, when the sample is placed in the external field, there is a change in the magnetic energy of either a $G$-system or a $P$-system. For both systems, our test particle will respond to an external field, sensing a variation of the electromagnetic field linked to the test particle, in the internal direction of the even $^+B$ field which is the only observable component at long range. The change in the magnetic energy, after restoring in the equation the fundamental physical constants $e$, $\hbar$, $c$ which are all equal to 1 in our geometric units, is

$$\Delta U = \frac{\partial\,U}{\partial\,^+B}\,\Delta B = -2\left(1 + \tan\Theta_i\right)\left(\frac{e\hbar}{2m_i c}\right)\frac{S\bullet B}{\hbar} = -g_i\mu_i\,\frac{S\bullet B}{\hbar} \tag{5.2.5}$$

where the variation seen by the test-particle is equal to the external field $B$, the angle $\Theta_i$ is zero for the electron or $\pi/3$ for the proton and $m_i$ is the respective mass. This expression defines the magneton (atomic or nuclear) $\mu_i$ and the anomalous Landé gyromagnetic factor $g_i$.

The magnetic moments of the proton and electron result,

$$\mu_p^a = 2\left(1 + \sqrt{3}\right)\left(\frac{e\hbar}{2m_p c}\right)\frac{\sigma^a}{2} = g_p\mu_N\,\frac{S^a}{\hbar} \quad , \tag{5.2.6}$$

$$\mu_e^a = -2\left(\frac{e\hbar}{2m_e c}\right)\frac{\sigma^a}{2} = g_e\mu_B\,\frac{S^a}{\hbar} \quad , \tag{5.2.7}$$

determining the anomalous gyromagnetic factors of the proton, $2(2.732)$ and electron, $-2$.

In QED the calculated values of the zero order magnetic moment for the electron and the proton, given by its external Coulomb field scattering diagrams, are singular. In each case, the interaction hamiltonian for the external Coulomb field differs by the respective additional anomalous coefficient $g$, determined by equation (5.2.7). The effect of this difference is to adjoin the respective coefficient $g$ to the external vertex in the diagram. Thus after renormalization, the zero order magnetic moment values for the proton and the electron are proportional by their respective zero order factors $g_0$. The radiative corrections for the electron have been calculated by Schwinger [91, 92, 93]. For both, electron and proton, the first order corrections, determined exclusively by the vertex part of the Coulomb scattering diagram, are proportional to the corresponding zero order terms. For the electron the vertex correction diagram, formed by an internal photon line between the two internal fermion lines, gives Schwinger's $\alpha/2\pi$ correction factor. For the proton the triple electromagnetic U(1) structure present in the SU(2) sector of the geometric interaction operator $J\bullet\Gamma$, determines three standard $j.A$ coupling terms. The triple structure is also present in the noncompact sector of the algebra. This indicates that a full proton excitation description requires three boost momenta $k_i$. Consequently a full description of the external Coulomb field scattering diagram requires a pair of internal fermion boost momentum triplets, instead of simply the pair of electron momenta. There are additional U(1) radiative processes between the *six* internal fermion boost lines corresponding to the two triplets. Therefore, there are multiple additional internal vertex diagrams, obtained by permutation of the internal photon line among the six necessary internal fermion lines, that contribute to the first order correction. The multiplicity of the vertex corrections of the *6* lines taken *2* at a time is 15. Each correction is equal to Schwinger's value, $\alpha/2\pi$, due the SU(2)$_Q$ group equivalence. Up to first order, the proton magnetic moment is

$$\frac{g_1}{2} = \frac{g_0}{2}\left(1 + \frac{15\alpha}{2\pi}\right) = 2.7321(1 + 0.01742) = 2.7796 \quad . \tag{5.2.8}$$

Let us calculate the magnetic moment of a combination of $G \supset P \supset L$ chain excitations. The total electromagnetic potential is the sum of the SU(2)$_Q$ $A$ potential and a ***different*** U(1) $A_U$ potential. Let $\varphi$ be a ($\frac{1}{2}$) eigenfunction of $A$ and $A_U$ [63]. Including the radiative correction, with the 28 vertex multiplicity corresponding to 4 $P$-excitations in the resultant su(2)+u(1) algebra, the magnetic moment of the combination is

$$\mu^a = -2\left(1+\sqrt{3}/2\right)\left(1+\frac{28\alpha}{2\pi}\right)\left(\frac{e\hbar}{2m_p c}\right)\frac{\sigma^a}{2} = g_n \mu_N \frac{S^a}{\hbar}$$

$$= -2\left(1.967\right)\mu_N \frac{S^a}{\hbar}\ . \tag{5.2.9}$$

These results have a *0.5%* discrepancy with the experimental values of the proton magnetic moment (*2.7928*) and the neutron magnetic moment (*1.9130*). The neutron may be considered a combination of the proton, the electron and the antineutrino.

# 5.3. The Modified Pauli Equation.

If the generalized Pauli equation (5.1.7) in the subnuclear zone is useful in calculating the magnetic moments of the nucleons we may expect that this equation is involved in the nuclear interaction. In particular we are interested in applying it to the study the properties of the light stable nuclides [94]. A nonlinear substratum solution implies the self interaction between its excitation field and motion. It is normal to define the self potential of a charge as positive. With the adequate definition of the classic potential $A$, in accordance with section 3.6, and using eqs. (5.2.1) and (5.1.7) for the quantum states of the potential $A$ we obtain,

$$A_\mu \equiv i\Gamma_{+\mu}(Q) = -\ ^+A_\mu\ , \tag{5.3.1}$$

$$[-\nabla^2 - iA\cdot\left(-\hat{r}m_W e^{-m_W r}\tan\Theta + 2\nabla\right) - \left(e^{-2m_W r}\tan^2\Theta - 1\right)A^2 + 2m\left(1+e^{-m_W r}\tan\Theta\right)V$$
$$-2e^{-m_W r}\tan\Theta\left(\sigma\cdot A\times\nabla\right) + \sigma\cdot\left(\left(1+e^{-m_W r}\tan\Theta\right)\nabla\times A - m_W e^{-m_W r}\hat{r}\times A\tan\Theta\right)]\psi = 2mE\psi\ . \tag{5.3.2}$$

The relation of this fundamental excitation with its substratum determines its electromagnetic potentials which depend on the bosonic mass $m_W$. The excitation decomposes linearly into local fundamental excitations, with relative signs of charge. These signs correspond to the action of the operator $A$, which now is considered an external field, on the wavefunction $\psi$. Both are eigenvectors of the $SU(2)_Q$ physical charge operator with signs corresponding to their representations.

The $SU(2)_Q$ potential is a short range field [94]. At atomic distances in the far zone $m_W r \gg 1$, $\Theta$ is zero. For the atom, the wavefunction $\psi$ corresponds to the negatively charged electron excitation and the scalar potential $A_0$ corresponds to the positively charged nucleus excitation. We get the standard Pauli equation (5.1.8) with the classic U(1) electromagnetic potential $A$ when we introduce the $-e$ charge in the usual units.

At small distances in the subnuclear zone $m_W r \ll 1$ the tangent of the fundamental quantum electroconic angle $\Theta$, which is related to Weinberg's angle, is $3^{\frac{1}{2}}$. The presence of the odd $A$ sector changes the sign of the potential term $A^2$, in accordance with equation (5.1.7), so that this strong potential term is attractive. In terms of the magnetic moment $\mu$ and the positive charge $q$ of a point source, we obtain an equation which shows that the dominant effects are due to the magnetic terms which include an attractive nuclear $r^{-4}$ potential,

$$[-\vec{\nabla}^2 - 2i\frac{\vec{\mu}\times\vec{r}}{|r|^3}\cdot\vec{\nabla} - 2\frac{\left(\vec{\mu}\times\vec{r}\right)^2}{|r|^6} + \left(1+\sqrt{3}\right)\vec{\sigma}\cdot\vec{\nabla}\times\frac{\vec{\mu}\times\vec{r}}{|r|^3} -$$
$$\sqrt{3}m_W\frac{\vec{\sigma}\cdot\hat{r}\times\vec{\mu}\times\vec{r}}{|r|^3} - 2\sqrt{3}\frac{\vec{\sigma}\cdot\vec{\mu}\times\vec{r}\times\vec{\nabla}}{|r|^3} + \frac{2m\left(1+\sqrt{3}\right)q}{r}]\psi = 2mE\psi\ . \tag{5.3.3}$$

It should be indicated that the gyromagnetic factor $g$ is an internal geometric property of a particle or excitation which is determined by the geometric (quantum) electromagnetic operators $A$ present in the internal magnetic energy term of its internal motion equation. Therefore, it is convenient to recognize a geometric factor $\gamma$ in the magnetic moment $\mu$ which includes these factors but excludes the mass which represents a secondary inertial reaction in the charge motion. Depending on the particular system considered this mass may correspond to the system reduced mass or to a total mass.

# 5.4. The Proton-Electron-Proton Model for the Deuteron.

Consider that equation (5.3.3) represents a system of $G$ and $P$ excitations, in particular a system of one electron and two protons moving about the system center of mass. The fields are dominated by the electron magnetic field because its higher magnetic moment $\mu$. The resultant SU(2) field is characterized by the system reduced mass which is the electron mass $m$. There are flux quanta associated to the protons $p$ and the electron $e$, one for each particle, that should link among themselves. This makes convenient to use the flux lines or magnetic strings to characterize the links between particle magnetic moments. The dominant potential is a strong attraction in the equatorial radial direction which may also be considered as the attraction among magnetic flux lines linked to flux quanta. The three indicated quantum flux strings cannot be linked so that they fully attract themselves as a whole without breaking the Pauli exclusion principle of the two $p$. To fully attract themselves the only possibility is the necessary presence of a fourth string. This string must be supplied by the only neutral fundamental excitation in the theory, an $L$ excitation or neutrino with a linked flux quantum. The resultant system $\left(p, e, \bar{\nu}, p\right)$ is a model for the deuteron. The only possible way to attractively link the flux lines determines that the subsystem $\left(e, \bar{\nu}\right) \equiv e'$ has the $e$ and $\bar{\nu}$ spins in the same direction. Hence $e'$ has spin 1, the charge, mass and the magnetic moment of the electron. The operators $^{+}A_{e}$ of the electron $e$ and $A_{p}$ of the protons $p$ in the stable system $\left(p, e', p\right)$ constitute a single coherent total field operator $A$, which depends on the reduced mass, and determines its fundamental SU(2) excitation. In other words, the resultant 4-potencial $A$ has the Clifford algebra orientation required by the fundamental SU(2)$_{Q}$ quantum representation of the generator. This determines the geometric electroconic angle $\Theta$ of $A$ with respect to the even electromagnetic potential $^{+}A$. The resultant field affects the system components and in particular the protons. The dominant internal magnetostatic energy is the one that holds the excitation system $\left(p, e', p\right)$ together and determines its energy proper value $E$. We may say that this is a coherent symmetric quasi-static disposition of the two protons around the excited electron in a proton-electron-proton formation that facilitates a stable solution.

Because of the symmetry we may use cylindrical coordinates with the $z$ axis along the electron magnetic moment $\mu$. It is convenient to approximate the potentials to very small $\rho$ by only keeping the dominant $\rho^{-4}$ potential term. The energy eigenvalue is expressed as a mass fraction $E = m\varepsilon^{2}$. In this manner we obtain [94] the equation

$$\left[-\nabla^2 - 2m^2\varepsilon^2 - \frac{2\mu^2}{\rho^4\left(1 + \dfrac{z^2}{\rho^2}\right)^3}\right]\psi = 0 \ . \tag{5.4.1}$$

This equation is not separable [95] due to the spatial polar angle $\theta$ dependence contained in the cross products.

Nevertheless, since orbital and spin angular momenta tend to align with the magnetic $B$ field we should expect that the wave function concentrates around the equatorial plane and most of the energy is in the equatorial region. A calculation of the negative gradient of the dominant potential term confirms this consideration. The negative gradient has direction toward the equatorial plane $z = 0$ for all values of $z$ and toward the origin in an equatorial region defined by $z^2 < 2\rho^2$. This equatorial region is limited by two cones near the equator, characterized by an equatorial angle or latitude $\theta_c$. In the limit of very small $\rho$ the problem reduces to a bidimensional problem on the equatorial plane. Therefore, near this limit we should simplify by setting $\sin\theta \approx 1$, $z \approx 0$., obtaining a separable approximate equation,

$$\left[\nabla^2 + 2m^2\varepsilon^2 + \frac{2\mu^2}{\rho^4}\right]\psi = 0 \ , \tag{5.4.2}$$

$$\left[\frac{\partial^2}{\partial\rho^2} + \frac{1}{\rho}\frac{\partial}{\partial\rho} + \frac{1}{\rho^2}\frac{\partial^2}{\partial\varphi^2} + \frac{\partial^2}{\partial z^2} + 2m^2\varepsilon^2 + \frac{2\mu^2}{\rho^4}\right]\psi = 0 \ . \tag{5.4.3}$$

Since the $z$ dependence is originally through the spatial polar angle $\theta$, it is convenient to introduce in its place, in the equatorial zone, an angular latitude coordinate $\zeta$,. In terms of this coordinate in the equatorial region the equation becomes

$$\left[\rho^2\frac{\partial^2}{\partial\rho^2}+\rho\frac{\partial}{\partial\rho}+\frac{\partial^2}{\partial\varphi^2}+\frac{\partial^2}{\partial\zeta^2}+2m^2\varepsilon^2\rho^2+\frac{2\mu^2}{\rho^2}\right]\psi=0\ .$$  (5.4.4)

We assume a separable solution in the equatorial region, of the form

$$\psi=R(\rho)\Lambda^\pm(\varphi)Z(\zeta)\ .$$  (5.4.5)

The separated equation for $Z$, which does not produce eigenvalues, is

$$\frac{\partial^2 Z}{\partial\zeta^2}=\alpha^2 Z$$  (5.4.6)

where $\alpha$ is only an approximate separation constant.

The cylindrical symmetry implies the conservation of the angular momentum azimuthal component. The Pauli spinor $\psi$, or large component of the Dirac spinor, is a common eigenspinor of the generators of rotations $L_z$, spin $S_z$ and total rotation $J_z$. The eigenfunctions may be constructed using the exponential functions,

$$\psi_+(\nu)=\left|\nu,+\tfrac{1}{2}\right\rangle=\begin{pmatrix}e^{i\nu\varphi}\\0\end{pmatrix}\ ,$$  (5.4.7)

$$\psi_-(\nu)=\left|\nu,-\tfrac{1}{2}\right\rangle=\begin{pmatrix}0\\e^{i\nu\varphi}\end{pmatrix}\ .$$  (5.4.8)

We require that the spatial part of the eigenfunction be single valued, which determines that $\nu$ is an integer. With this understanding, the separated equation for the angular function $\Lambda^\pm$ is

$$\frac{\partial^2}{\partial\varphi^2}\left|\nu,\pm\tfrac{1}{2}\right\rangle=-\nu^2\left|\nu,\pm\tfrac{1}{2}\right\rangle\ .$$  (5.4.9)

Values of $\nu$ mean that the system has orbital angular momentum around its center of mass. From a mechanical point of view, this implies a possible rotation of the lighter electron $e$ around the center of mass of the two protons $p$. A nonzero $\nu$ may be inconsistent with the quasi-static mechanism, indicating an instability, and we should rule out all values $\nu\neq0$. The stable model might only be possible for a static system.

In accordance with the indication at the end of section 5.2, the magnetic moment is determined by a geometric factor $\gamma$ that depends on the fields and charges in motion. The magnetic moment is inversely proportional to the mass, which tends to oppose the motion. It is convenient to make a change of variables to a radial dimensionless complex variable $z$ rationalized to the mass,

$$\rho^2=\frac{\mu}{m\varepsilon}z^2=\frac{\gamma}{m^2\varepsilon}z^2\ ,$$  (5.4.10)

where the electron $\gamma$ is ½. We obtain the separated radial equation which has irregular singular points at zero and infinity [95],

$$z^2R''+zR'+\left[2\gamma\varepsilon\left(z^2+\frac{1}{z^2}\right)-\left(\nu^2-\alpha^2\right)\right]R=0\ .$$  (5.4.11)

If we let $z=e^u$ we obtain

$$R''+\left[2\gamma\varepsilon\left(e^{2u}+e^{-2u}\right)-\left(\nu^2-\alpha^2\right)\right]R=0\ ,$$  (5.4.12)

which is the modified Mathieu equation [95, 96, 97] with the parameters

$$q=2\gamma\varepsilon\ ,$$  (5.4.13)

$$a=\nu^2-\alpha^2\ .$$  (5.4.14)

In the last equation the parameters $a$, and $\alpha$ should be considered effective parameters because, independently of $\nu^2$ and $\alpha^2$, $a$ may include contributions from equation (5.4.1). Nevertheless, the eigenvalues of the angular and radial equations are the ones that directly determine the possible values of the pair $\nu$, $a$ and consequently the effective value $\alpha^2$.

This approximate result regarding the application of the Mathieu equation for the system under study might also be obtained using spherical coordinates.

# 5.5. Binding Energy of the Deuteron.

According to Floquet's theorem [84], a solution of eq. (5.4.12) includes an overall factor, $e^{su}$ where $s$ is a complex constant. If $s$ is an integer we get the Mathieu functions of order $s$. A Mathieu function [95] is one of the few special functions which are not a special case of the hypergeometric function and therefore the determination of its proper values differs. The Mathieu functions can be expressed as a Fourier series. The series expansion coefficients are proportional to the characteristic root values $a_c$ of a set of continued fraction equations obtained from the three-term recursion relations among the expansion coefficients. These characteristic roots are obtained as a power series in $q$.

A solution of the Mathieu equation can also be expressed in terms of a series of Bessel functions with the same expansion coefficients of the Fourier series. This Bessel series is convenient to find the asymptotic behavior of the solution. By replacing the Bessel functions with the Neumann functions we obtain a solution of the second kind. Solutions of the third and fourth kind are obtained by combining the Bessel and Neumann functions to form Hankel functions. We want a regular solution at zero that vanishes at infinity. We require that $s = 0$ to avoid a singular or oscillatory asymptotic behaviour. The appropriate negative exponential behavior at infinity is provided by the radial Mathieu function of zero order, of the third kind, or Mathieu-Hankel function, indicated by $He_0(q, u)$.

It is known that the characteristic roots and the corresponding series expansion coefficients for the Mathieu functions have branch cut singularities on the imaginary $q$ axis[98, 99] and the complex plane $a$ of the corresponding Riemann surfaces. Therefore, the expansion coefficients are multiple-valued. In the standard radial equation eigenvalue problem, the energy is determined by eliminating the singularities of the hypergeometric series by requiring its coefficients to vanish after a certain order. In our case, the energy should be determined by eliminating the branch cut singularity of the Mathieu series by choosing appropriate coefficients. We should require that the solution coefficients of the $R$ function be single valued as we required for the $\Lambda$ function. To eliminate the possibility of multiple-valued coefficients and obtain a regular solution we must disregard all points $q$ on all branches of the Riemann surface except the points $q_0$ common to all branches. We require that $q$ be the two common points on the Riemann surface branches corresponding to the first two characteristic roots $a_0$ and $a_2$ which determine the rest of the expansion coefficients of the even Mathieu functions of period $\pi$.

The values of the constants $q_0$ and $a$ have been calculated [98, 99, 100] with many significant digits. The parameter $q$ determines a single degenerate negative eigenvalue corresponding to the states of the solution, as follows,

$$q = 2\gamma\varepsilon = \varepsilon = \pm iq_0 \ , \tag{5.5.1}$$

and we have a stable solution with the value of binding energy per state

$$E_0 = \varepsilon^2 m_e = -q_0^2 m_e = -\left(1.4687686\right)^2 m_e \approx -1.10237 \ \ \text{Mev.} \tag{5.5.2}$$

According to the model, both $\psi = \left|0, \pm\frac{1}{2}\right\rangle$ states are occupied and the total energy is

$$E = -2.20474 \ \ \text{Mev.} \ \approx -2.2246 \ \ \text{Mev.} \ = -U_d \tag{5.5.3}$$

where $U_d$ is the deuteron binding energy. It is the necessary energy to destroy the symmetric quasi-static arrangement. It is required to desintegrate the system into a proton and an electron-proton excitation or neutron, as indicated in the next section. This negative binding energy is due to the $SU(2)_Q$ strong electromagnetic interaction. Therefore, the physical reaction of a proton with a neutron, which produces a deuteron, must simultaneously release this cuantity $U_d$ of electromagnetic energy in radiation form. This determines the known $p$ $n$ reaction.

Since the component protons are in opposite angular momentum states, the deuteron spin is equal to the $e'$ system spin. If we find the correct equation for the spin 1 $e'$ system it would display a magnetic energy term similar to the ones found previously where the coefficient of the $B$ field is equal to the Bohr magneton times a spin 1 matrix $S$. This expression rules out the spin 0 state for $e'$ and for the deuteron. The deuteron can only be in the $\pm1$ spin states. It also shows that $e'$ does not have an anomalous gyromagnetic factor.

In general the nuclear magnetic potential energy barrier for this deuteronic "$pep$ bond" is approximately given by the $A$ boson in the equatorial radial equation,

$$[-\partial_\rho^2 - \frac{\left(3e^{-2m_A\rho} - 1\right)}{4m^2\rho^4} - \frac{\varepsilon}{m\rho^3} - \frac{2m}{\rho}]\psi = 2mE\psi \quad,$$ (5.5.4)

where $\varepsilon = \pm 1$. The SU(2) barrier height critically depends on the value $2\rho m_A = 1.3984$ which determines the location of the $\rho^{-4}$ term maximum. If the system is in a periodic collective medium, equation (4.6.25) for the SU(2) potential range $1/m_A$ determines that the $pep$ equatorial barrier decreases by a factor of $8.8 \times 10^{-14}$, relative to its free space values, facilitating a proton penetration of the barrier in a low energy nuclear reaction.

# 5.6. The Electron-Proton Model for the Neutron.

If only one state $p$ of the deuteron system is occupied we could expect an unstable model for the neutron. Comparing the possible 1-proton and the bound deuteron 2-proton system equations, we consider that the average of the dominant electron magnetic fields $B_c$ in the corresponding equations should remain approximately equal at the respective proton positions. If the stable 2-proton system suddenly looses one proton, by a collision for example, we may assume that the potential terms, produced by the electron, remain constant for the duration of a short transition, while the residual system becomes unstable. The symmetric quasi-static disposition of the two protons around the electron in the proton-electron-proton formation is destroyed, leading to a strongly-dynamic asymmetric disposition in an electron-proton system. When the system looses a proton, the motion instability breaks $^+A_c$ away from $^+A_p$ in such a way that their combination does not constitute the fundamental SU(2) representation. In its place there form separate electron U(1) and proton SU(2) representations respectively. In other words, this combination now implies that the sum of the respective proper values determine the neutron anomalous gyromagnetic factor and its energy proper value. During this transition, the parameters associated to the magnetic properties of the wavefunction must adjust themselves to the new situation in the unstable 1-proton system. One of these parameters is the anomalous gyromagnetic factor $g$ that we recognize present in one of the potential terms as a coefficient of the $B$ field

$$2\left(1 + \sqrt{3}\right)\frac{\mu}{\rho^3} = g_p\frac{\mu}{\rho^3} \quad.$$ (5.6.1)

We now understand that the gyromagnetic factor $g$, as indicated previously, is determined by the SU(2) interaction. The other parameter is the energy eigenvalue $E$. The only potential terms in equation (5.3.3) that should undergo changes are those depending on these magnetic parameters. Thus, we concentrate on the expression

$$W = 2m^2\varepsilon^2 + 2\left(1 + \sqrt{3}\right)\frac{\mu}{\rho^3} = 2mE + g\frac{\mu}{\rho^3}$$ (5.6.2)

which is the dominant expression containing these parameters, while keeping $\rho$ and the other potential terms constant in equation (5.3.3). Therefore the total value of $W$ should also remain constant while only $E$ and $g$ change. Then the last equation relates the change of the energy eigenvalue to the change of the gyromagnetic factor while the su(2) magnetic fields are rearranged to form a neutron. This means that $E$ and $g$ should proportionally evolve toward the unstable neutron state and $W$ is zero after the collision and during the evolution toward the unstable state. Therefore, the energy eigenvalue is proportional to the gyromagnetic factor and to the intrinsic magnetic moment potential term which is a fraction of total effective potential

$$\frac{E}{g} = \frac{E_n}{g_n} = \frac{E_0}{g_p} \quad.$$ (5.6.3)

The ratio of the two nucleon energy eigenvalues is then approximately equal to the ratio of the gyromagnetic factors. We know from section 5.2 that both anomalous proton and neutron magnetic moments are determined and theoretically calculated from two different su(2) magnetic field arrangements in nucleon excitation systems with mass $m_p$. The proton-electron system undergoes su(2) magnetic field changes toward an unstable neutron system that eventually decays. The energy of the latter system is then theoretically determined by these gyromagnetic factor theoretical values

$$E_n = \frac{-1.967}{2.780}\left(-1.10237\right) = 0.780 \ \text{Mev.} \approx m_n - m_p - m_e = 0.782 \ \text{Mev.}$$ (5.6.4)

This system has a mass-excess equal to $E_n$ which is given-up as energy when it decays into its constituents proton and electron.

The genralized Pauli equation (5.3.3) for the internal motion relative to the center of mass, with reduced mass $m$, under the internal magnetic field, serves to calculate gyromagnetic effects and binding energies. The Pauli equation also serves to study the motion of the neutron system center of mass under an external magnetic field, now with the resultant total mass.

The system $(p,e,\bar{\nu})$ is a spin ½ system because the proton spin is opposite to the $(e,\bar{\nu}) \equiv e'$ spin. Thus, the neutron and proton forming the deuteron have their spins in the same direction. The fact that the deuteron magnetic moment is the approximate sum of the proton and neutron moments may be explained because the presence of an external magnetic field forces one proton to react by itself while the rest of the system, $p$, $e'$ reacts adjusting its fields as a neutron.

# 5.7. The Many Deuteron Model.

We follow here the idea of nuclear pairings by Bohr, Mottelson and Rainwater [101]. The presence of orbital angular momentum, $\lambda \neq 0$, breaks the stable quasi-static mechanism of the model. The equation determines only one stable bond excitation corresponding to the spin ±½ states of frame field excitations bound by the strong magnetic interaction. This deuteron "$pep$ bond" is a $(p,e',p)$ fundamental excitation and supplies another pairing mechanism which also allows the combination of more than 2 protons. The alpha particle, or excitation $\alpha$, may be considered as a 2-deuteron excitation [102].

The magnetic interaction range defines the subnuclear zone $r \sim 1/m_W \ll r_p$, very small in relation to the proton radius $r_p$. Mathieu's equation determines a single $pep$ bond energy in a proton pair. In order to feel the strong attraction the proton centers must be inside the subnuclear zone. Therefore the protons are essentially superposed. In the $\alpha$ model there are 4 protons and 2 electrons which are also superposed and must share the magnetic field. The stationary state, of minimum energy, is the symmetrized quantum superposition, as in the hydrogen molecular ion. The electrons and protons are shared by all possible $pep$ bonds in the system. The possible $pep$ bonds obey the Pauli exclusion principle and are only those sharing a single electron. All $p$ participate in as many $pep$ bonds as proton pairings.

One difference in the $\alpha$ model is that the reduced mass $m$ of the system is approximately half the electron mass $m_e$ due to the presence of two electrons in the system. The required coordinate transformation to obtain Mathieu's equation is then different from eq. (5.4.10),

$$\rho^2 = \frac{\mu}{m\varepsilon}z^2 = \frac{\gamma}{m_e m\varepsilon}z^2 = \frac{\gamma}{2m^2\varepsilon}z^2 \ , \tag{5.7.1}$$

which changes $q$ proportionally and determines twice the binding energy $U_d'$ per pair,

$$q = \varepsilon/2 \ , \tag{5.7.2}$$

$$E_d' = 2E_0' = 2\varepsilon^2 m = 2\left(2q\right)^2 m = -4q_0^2 m_e = -2U_d \ . \tag{5.7.3}$$

The energy $E_d' = 2E_d$ is the unique Mathieu equation eigenvalue even if $e'$ is shared and is the same for each $pep$ magnetic bond. This fact together with the symmetry simplifies the energy calculation of the $n$-body problem. There are symmetric cluster states whose energies are determined by their $pep$ bonds. The symmetric $\alpha$ cluster associated to 4 protons has 6 $pep$ bonds and the energy of these bonds minus the electronic mass difference,

$$\sum_{p,e'}^{4p,2e'} m_i = \sum_{p,e'}^{4p,6e'} m_i - 4m_{e'} \ , \tag{5.7.4}$$

$$E_{(4p,2e')} = E_{4p,6e'} - 4m_e = 6E_d' - 4m_e = -28.501 \text{ Mev.} \equiv E_\alpha \approx -U_\alpha \ . \tag{5.7.5}$$

Similarly we have the same considerations for 3 protons. The symmetric cluster associated to 3 protons has 3 $pep$ bonds with a unique Mathieu energy $E_d$,

$$E_{(3p,e')} = E_{3p,3e'} - 2m_e = 3E_d - 2m_e = -7.636 \text{ Mev.} \equiv E_{^3He} \approx -U_{^3He} \ . \tag{5.7.6}$$

The exclusion principle prohibits $2e$ in a $pep$ bond. An extra $e$ may link to a $p$ which is not in $pep$ bond at that moment, which is impossible in the $4p$ $\alpha$ system, with the energy of eq. (5.6.4),

$$\Delta E = m_e - \left(m_n - m_p\right) \ , \tag{5.7.7}$$

$$E_{(3p,2e)} = E_{^3He} + \Delta E = -8.416 \text{ Mev.} \equiv E_{^3H} \approx -U_{^3H} \ . \tag{5.7.8}$$

These equations determine a mass relation identity for the A=3 isobars. Since our notation uses the correspondence

$(3p, e') = (2p, n)$ we may write

$$m_{{}^3H^+} = m_p + 2m_n + E_{{}^3He} + \Delta E = 2m_p + m_n + m_e + E_{{}^3He}$$

$$m_{{}^3H^+} \equiv m_{{}^3He^{2+}} + m_e$$

(5.7.9)

which indicates a small $\beta$ decay possibility of ${}^3H^+$ into ${}^3He^+$ from total energy considerations in spite of the unfavorable binding energy difference. It also implies that the corresponding atoms ${}^3He$ and ${}^3H$ have essentially equal masses, up to the negligible Coulomb electron interaction.

## 5.7.1. Nuclear Structure, Fusion and Fission.

We have obtained the binding energies for ${}^4He$ and the isobars ${}^3H$, ${}^3He$ within 1%. We are lead by the model to consider that the $pep$ bonds or deuterons act as the essential components of more complex clusters in electromagnetic interaction with additional nucleons. This is consistent with the protonic and neutronic numbers of the nuclides. Any proton outside a $pep$ bond when associated with $e'$ may be considered a neutron. Therefore the energy bond $U_d$ determines a "nuclear" force of electromagnetic origen among nucleons, which is independent of a neutron interpretation. These simple facts about deuterons and alpha particles would determine the general nuclear structure. The inclusion of neglected interaction terms may be sufficient, without any other force, to explain the properties of nuclides and their fusion and fission reactions.

We have built these models using the geometric ideas, in particular the electromagnetic $SU(2)_Q$ subgroup and its attractive subnuclear $r^{-4}$ potential term. It should be clear that the numbers obtained also depend on other particular model features and they do not constitute by themselves a proof of the theory. Nevertheless the overall picture lends them credibility. From a practical point of view these results should not be overlooked because the implications they might have in the understanding of the process of nuclear fusion of light nuclei. The electromagnetic field would be an active element of a fusion generator. In particular it may provide possible quasi static fusion processes for the development of clean fusion energy without the present problems of high radiation and radioactivity.

The binding energy of the deuteron and other light nuclides would be essentially magnetostatic and the nuclear fission and fusion reactions in a nuclide system would be magnetic multipolar transitions. The electromagnetic field may interchange electromagnetic gamma radiation of type M1 through the corresponding quantum or gamma photon. If the system interacts with another (external) system, in addition to the radiative channel there should be other possible channels in the $S$ dispersion matrix. The total potential may have energy transfers in the non radiation near zone. Quasistatic channels may open between the nuclide system and the other system through the field in this electromagnetic zone. In particular, if the potential is periodic and *coherent* in a collective system the cuantum magnetic barrier for quasistatic exchanges may be reduced, as indicated in section 5.5. In this manner magnetic energy would be tranferred with low radiation, as it usually happens between near electromagnetic and matter systems.

# 5.8. Geometric Weak Interations.

In addition to the strong effects shown in the previous sections it is also possible to find weak effects. The field equation implies a conservation law for the geometric current, which determines a generalized Dirac equation (3.6.1) in terms of the local frames. This equation, for the even and odd parts of a frame $f$ under the presence of a pure odd potential reduces [7] as seen in chapter 3 to

$$\kappa^\mu \partial_\mu f_+ = \kappa^\mu A_{\mu-} f_- = m f_- \ ,$$

(5.8.1)

$$\kappa^\mu \partial_\mu f_- = \kappa^\mu A_{\mu-} f_+ = m f_+ \ ,$$

(5.8.2)

implying that a frame for a massive corpuscle must have odd and even parts. For an even frame,

$$f_- = 0 \ \Rightarrow \ m = 0 \ .$$

(5.8.3)

Therefore, for an even frame we have, multiplying by $\kappa$, we obtain the equation normally associated with a neutrino field. A fluctuation of $f_+$ on the fixed substratum obeys also the last equation.

From previous discussion, [103] we consider that an electroweak interaction may be related to the action of the $Sp(2, \mathcal{Q})$ holonomy group. The total interaction field should correspond to a potential $A$, which is a representation of $Sp(2, \mathcal{Q})$. The total matter current should be associated to an $Sp(2, \mathcal{Q})$ frame $f$ representing both the electron field $e$ and the neutrino field $\nu$. We should then be able to describe neutrino electron interactions.

At a point, the total frame excitation $f$ of the interacting $e, \nu$ is related to an element of the group $Sp(2, \mathcal{Q})$, a subspace of the geometric algebra $R_{3,1} = R(4)$. The frame $f$ may be decomposed into fields associated to the particles $e, \nu$ by means of the addition operation within the algebra. These $e$, $\nu$ fields are not necessarily frames because addition does not preserve the group subspace Sp(2,$\mathcal{Q}$) and geometrically, $e$ and $\nu$ are not sections of a principal fiber bundle but rather of an associated bundle with the Clifford algebra as fiber.

The source current $J$ in the theory is

$$J = \tilde{f}\kappa f = \overline{f}\kappa f \quad , \tag{5.8.4}$$

where $f$ is the frame section associated to the total field of the electron and the neutrino and $\kappa$ represents the orthonormal set. For the Sp(2,$\mathcal{Q}$) group the correlation reduces to conjugation.

Due to the properties that a neutral particle field should have (there are no strong electromagnetic or massive effects), we consider that the effect produced by $\nu$ should be small relative to the effect of $e$ . Hence we may assume that, in the composite system, $\nu$ is a perturbation of the order of the fine structure constant $\alpha$, the only physical constant in the theory,

$$f = e + \alpha\nu \quad . \tag{5.8.5}$$

Then the current becomes

$$J = \left(\overline{e} + \alpha\overline{\nu}\right)\kappa\left(e + \alpha\nu\right) = \overline{e}\kappa e + \alpha\left(\overline{e}\kappa\nu + \overline{\nu}\kappa e\right) + \alpha^2\overline{\nu}\kappa\nu \quad . \tag{5.8.6}$$

The intermediate terms may be considered as a perturbation of order $\alpha$ to a substratum electronic matter frame. The perturbation current may be written, by splitting $e$ into its even and odd parts and noticing that $\nu$ has only an even part,

$$J - \overline{e}\kappa e = \alpha\left[\left(\overline{\eta} + \overline{\xi}\overline{\kappa}^0\right)\kappa\nu_+ + \overline{\nu}_+\kappa\left(\eta + \kappa^0\xi\right)\right] \quad . \tag{5.8.7}$$

As usual in particle theory, we neglect gravitation, which is taken as an even SL(2,$\mathbb{C}$) connection. If we look for effects not imputable to gravitation, it is logical to center our attention on the odd part of the perturbation current as a candidate for the interaction current,

$$\alpha^{-}j^\mu = \alpha\left(\overline{\eta}\kappa^\mu\nu + \overline{\nu}\kappa^\mu\eta\right) \quad , \tag{5.8.8}$$

which has the structure of the Fermi weak current. It should be noted that the neutrino $\nu$ automatically associates itself, by Clifford addition, with the even part $\eta$ of the electron. This corresponds to the Weinberg-Salam association of the left handed components as a doublet, with the same Lorentz transformation properties.

If we apply perturbation theory to the field equations, we expand the potential $A$ in terms of the coupling constant $\alpha$. We have

$$J = J_0 + \alpha J_1 + \alpha^2 J_2 + \cdots \quad , \tag{5.8.9}$$

$$A = A_0 + \alpha A_1 + \alpha^2 A_2 + \cdots \quad . \tag{5.8.10}$$

The substratum equation and the first varied equation, which is second order in $\alpha$, have the following structure,

$$\left(D^*F\right)_E = 4\pi\alpha J_E \quad , \tag{5.8.11}$$

$$\delta\left(D^*F\right) = 4\pi\alpha\delta J \quad . \tag{5.8.12}$$

In the static limit, the $\delta J$ excitations reduce to the su(2)$_Q$+su(2)$_S$ subalgebra. The excitations are SU(2)$_Q$ representations and therefore their components are subject to the cuantization of their electromagnetic orientations, in accordance with section 4.6. We should express the term $J_1$ as a function of its odd component. In this manner, we let the first order terms be, as a function of the representation electrocone angle $\Theta$,

$$J_1 = j = {}^{-}j/\sin\Theta \quad , \tag{5.8.13}$$

$$A_1 = W \quad , \tag{5.8.14}$$

and we obtain for the variation, the linear equation

$$\alpha\left(d^*dW + LW\right) = 4\pi\alpha^2 j \quad , \tag{5.8.15}$$

where $L$ is a linear first order differential operator determined by the substratum. This equation may be solved in principle using its Green's function $\mathcal{G}$. The solution in terms of components with respect to a base $E$ in the algebra is

$$W_{\mu i} = 4\pi\alpha \int dx' \mathcal{G}^{\ j}_{\mu\nu}\left(x - x'\right) j^{\nu}_{j}\left(x'\right) \quad . \tag{5.8.16}$$

It is well known that the second variation of a lagrangian serves as lagrangian for the first varied Euler equations. Therefore, the $\Gamma J$ term in the covariant derivative present in the lagrangian provides an interaction coupling term that may be taken as part of the lagrangian for the process in discussion. When the $\Gamma J$ term is taken in energy units, considering that the lagrangian has an overall multiplier, it should lead to the interaction energy for the process. The second variation (or differential) in a Taylor expansion of the energy $U$, corresponds to the hessian of $U$.

The $AJ$ term gives the interaction lagrangian for the substratum, in energy units. For the perturbations, the interaction lagrangian is given by 1/2 the second variation or $\delta J.\delta\Gamma$,

$$\mathcal{L} = -\alpha^2 \frac{1}{4} \operatorname{tr} \frac{1}{2}\left(W_\mu j^\mu + j^\mu W_\mu\right) \quad . \tag{5.8.17}$$

It is clear that this interaction is carried by $\Gamma$ or $W$. Nevertheless, we wish to obtain a current-current interaction to compare with other theories at low energies. Substitution of eq. (5.8.16) in the last equation gives the corresponding action which, for clarity, we indicate by

$$\mathcal{W} = -2\pi\alpha^3 \frac{1}{4} \operatorname{tr} \int dx dx' \mathcal{G}^{\ j}_{i\mu\nu}\left(x - x'\right)\left[E^i E^l + E^l E^i\right] j^{\mu}_{j}\left(x\right) j^{\nu}_{l}\left(x'\right) \quad , \tag{5.8.18}$$

representing a current-current interaction hamiltonian, with a coupling constant derived from the fine structure constant. This new expression may be interpreted as a weak interaction Fermi lagrangian. The associated coupling constant is of the same order as the standard weak interaction coupling constant, up to terms in the Green's functions $\mathcal{G}$ .

# 5.9. Relation with Fermi's Theory.

The action, in terms of elements of the algebra and the trace, corresponds to the scalar product. For the case of the isotropic homogeneous constant substratum solution, eq. (1.3.2), the Greens function is a multiple of the unit matrix with respect to the algebra components. We assume here that for a class of solutions the Green's function have this property. Then for this class of solutions, the last equation may be written as

$$\mathcal{W} = -4\pi\alpha^3 \frac{1}{4} \operatorname{tr} \int dx dx' \mathcal{G}_{\mu\nu}\left(x - x'\right) j^{\mu}\left(x\right) j^{\nu}\left(x'\right) \quad , \tag{5.9.1}$$

where the $j$ are matrices.

The current defines an equivalent even current $j$ , which is the $\kappa^0$ component, by inserting $\kappa^0\kappa^0$ and commuting. Each $2\times2$ block component of this 4 dimensional real current matrix is an even matrix which may be represented as a complex number, by the known isomorphism between this complex algebra and a subalgebra of $2\times2$ matrices. Thus, we may use, instead of the generalized spinors forming the frame $f$, the standard bidimensional spinor pairs, $\eta_1$, $\eta_2$ and $\nu_1$, $\nu_2$ which form the $2\times2$ submatrices corresponding to the even part, respectively, of the electron and the neutrino. In this manner, if we use the standard quantum mechanics notation in Weyl's representation, *each matrix component* is of the form

$$\bar{j}^\mu = \frac{1}{2}\overline{\Psi}_e \gamma^\mu\left(1 + \gamma^5\right)\Psi_n = \eta^\dagger \overline{\sigma}^\mu \nu \quad , \tag{5.9.2}$$

which may be recognized as the standard nonhermitian Fermi weak interaction current for the electron neutrino system. We shall indicate this current by $j_F$ and we may determine that the 0-0 submatrix $j_0$ is the hermitian part of $j_F$.

We may evaluate the trace of the currents in the expression for $\mathcal{W}$. As discussed previously, within this theory, particles would be represented by excitations of frames. These fluctuations are matrices that correspond to a single representation of the subgroup in question. This means that for each pair of spinors of a spinor frame, only one is active for a particular fluctuation matrix. In other words, for a fluctuation, only one of the components of the current matrix is nonzero and we may omit the indices on the spinors and use complex number notation instead of $2\times2$ even matrices. The terms combine with their hermitian conjugates. It should be noticed that in going from the real $4\times4$ to the $2\times2$ complex matrix realizations the ¼ factor in front of the trace changes to 1/2 and due to the form of the excitation matrices introduces a factor of 1/2.

The resultant expression for $\mathcal{W}$ is

$$\mathcal{W} = -\frac{2\pi\alpha^3}{\sin^2\Theta} \int dx\,dx' \mathcal{G}_{\mu\nu}\left(x - x'\right) \bar{j}_0^{\dagger\mu}\left(x\right) \bar{j}_0^{\nu}\left(x'\right) \quad . \tag{5.9.3}$$

We now shall restrict the discussion to excitations around the constant complex substratum solution. Then the Green's function is determined from the fluctuation equation in section 4.5,

$$^*d\,{}^*d\delta A_{\delta}^{\alpha} + \left(\mathcal{M}/\sqrt{2}\right)^2 \delta A_{\delta}^{\alpha} + 2\left(\mathcal{M}/\sqrt{2}\right)^2 \delta A_{\rho}^{\hat{\rho}}\delta_{\delta}^{\alpha} = 4\pi\alpha\,\delta J_{\delta}^{\alpha} \quad ,$$  (5.9.4)

where $\mathcal{M}$ has a constant value, in the $4\times4$ representation. We may decouple the equations by assuming a zero value for the cross coupling term $\delta\eta^{\alpha}{}_{\alpha}$, and introduce the radial coordinate $r$ as was done in section 4.5. The result is a Yukawa equation, where we have defined the constant coefficient inside the parenthesis divided by $\alpha$ as the mass parameter $\mu$ in the corresponding representation of interest. In general, the Green's function has a Dirac time $\delta$ function that allows integration in $t$'. The spatial part of the Green's function should provide an equivalent range for the interaction. Since $\mu$ is a constant the equation reduces, for a point source, to the radial equation. Using the Green's function and further assumming that the currents vary slightly in the small region of integration, so that $j(x')$ approximately equals $j(x)$, we obtain an approximation for the hamiltonian contained in the geometric theory,

$$\mathcal{W}_w = -\frac{1}{2}\alpha\int dx\;{}^-j_0^{\dagger}(x)\bullet{}^-j_0(x)\int_0^{\infty}dr'r'e^{-\mu r'}\int_0^{4\pi}d^2\Omega' = \frac{-2\pi\alpha}{\mu^2\sin^2\Theta}\int dx\;{}^-j_0^{\dagger}(x)\bullet{}^-j_0(x).$$  (5.9.5)

The value of the $\Theta$ angle, which indicates the relation of the odd current with the total current, is determined by the component orientation cuantization in the corresponding current representation. Since the corrent $J$ is quadratic in terms of the frame $e$, its excitation correponds to a pair of su(2)$_Q$ fundamental excitations $\delta e$ and the current excitation $j$ is the representation with electromagnetic spin index $j=i=1$ in equation (4.6.5). We have then the corresponding electrocone angle which we shall indicate by $\Theta_I$ and

$$\sin\Theta_I = \sin\Theta_I^I = \frac{\left|^-j\right|}{|j|} = \frac{\left|^-j\right|}{\sqrt{\left|^-j\right|^2+\left|^+j\right|^2}} = \frac{\sqrt{i(i+1)}}{\sqrt{i(i+1)+1}} = \sqrt{\frac{2}{3}} \quad .$$  (5.9.6)

The expression of the currents in the integral may be rewritten recognizing the terms that equal the Fermi current $j_F$. We can evaluate the scalar $j_0.j_0$ in any coordinate system. We may choose a system where $j_F$ is along the timelike direction which has only a $J_F$ component. We may write then the product in terms of a phase $\varphi$. The angle $\varphi$ measures the degree of nonhermiticity of $j_F$. If $\varphi$ is zero $j_F$ is hermitic. Since $\varphi$ is not measured in standard experiments, we may expect its influence on $\mathcal{W}$ is through its average value $<\cos\varphi>$ and leave the expression in the form,

$$\mathcal{W}_w = \frac{-2\pi\alpha<\cos^2\varphi>}{\mu^2\sin^2\Theta_I}\int dx j_F^{\dagger}j_F \quad ,$$  (5.9.7)

which determines the lagrangian and the current $j_F$ assumed in Fermi's theory.

The value of the weak interaction constant $G_F$ is determined from results of experiments like the muon decay. As is it always done, in other weak interaction theories, we may fit the value of the constant in front of the integral to its experimental value by fixing values of $\mu$ and $\varphi$,

$$\frac{-2\pi\alpha<\cos^2\varphi>}{\mu^2\sin^2\Theta_I} = \frac{-G_F}{\sqrt{2}} \quad .$$  (5.9.8)

On the other hand, we now have at our disposal the possibility of theoretically calculating the value of $G_F$ from the knowledge of the constant substratum solution. The value of the mass $\mu$ is theoretically obtained in a geometric unit because of the used representation,

$$\mu = \frac{\mathcal{M}}{\sqrt{2}\alpha} \quad g^{-1} \quad .$$  (5.9.9)

This geometric unit g may be calibrated, in the SL(4,$\mathbb{R}$) representation induced by SL(2,$\mathbb{C}$), from the value of the proton mass $m_p$ in terms of a unit of mass as it was done in section 4.6.1. This is desirable because the expression for the proton and electron mass ratio agrees with the experimental value. The mass parameter for the proton, expressed in the geometric unit in the defining representation, is

$$m_p = \mathcal{M} \quad g^{-1} \quad ,$$  (5.9.10)

and we may calibrate the fundamental unit of length g by the experimental value of the proton mass. Because of similarities of the Fermi weak current with the electromagnetic current, which is hermitic, we may fix the $\varphi$ phase related factor equal to 1. The theoretical value of $G_F$, determined from the value of the proton mass, without any other experimental value, is

$$G_F = \frac{2\sqrt{2}\pi\alpha}{\mu^2 \sin^2\Theta_I} = \frac{2\sqrt{2}\pi\alpha}{\left(\frac{m_p}{\alpha\sqrt{2}}\right)^2 \left(\frac{\sqrt{2}}{\sqrt{3}}\right)^2} = 1.176668 \times 10^{-5} \quad \text{GeV}^{-2} \approx 1.16639 \times 10^{-5} \quad \text{GeV}^{-2} \ . \tag{5.9.11}$$

This value is subject to corrections of order $\alpha$ due to the approximations made.

The general perturbation for the interaction of the electron and the neutrino fields includes eqs. (5.9.2, 5.9.7, 5.9.11) which essentially are the current and lagrangian assumed in Fermi's theory [104, 105] of weak interactions [106, 107] of leptons. Fermi's theory is contained, as a low energy limit within the unified theory of potentials and frames, or physical geometry. If the full theory is accepted there are certainly new effects and implications, which should be determined without the simplifications made to display the relation of our theory to low energy weak interactions.

## 5.10. Results.

In general, the equation of motion is a geometric generalization of Dirac's equation which implies a generalized Pauli's equation. The geometric triple structure determines the triple interaction potential including the geometric interpretation of Weinberg's angle. The geometric Pauli equation of motion determines the anomalous bare magnetic moments of the proton, the electron and the neutron which agree with the experimental values. The "strong" electromagnetic part, without the help of any other force, generates nuclear range attractive potentials which are sufficiently strong to determine the values of the binding energies of the deuteron, the alpha particle, the neutron and other light nuclides, composed of protons and electrons. The strong *pep* links may be used to understand the structure of nuclei.

Fermi's theory of weak nuclear interactions and the constant $G_F$ are determined, as a low energy limit of the equations.

# 6. Particles and Interactions.

The group of the geometric space-time structure of special relativity is fundamental to the field theories of elementary particles, which are representations of this group. In contrast, the geometry of general relativity has not played such a fundamental role. Nevertheless, our geometric relativistic unified theory of gravitation and electromagnetism may have nontrivial applications for particle theory.

It is clear from the definition of excitation that a free particle is a representation of the structure group of the theory and consequently an algebraic element. A representation (and therefore a particle) is characterized by the eigenvalues of the Casimir operators. The states of a representation (particle) are characterized by the eigenvalues of a Cartan subspace basis operators. This provides a set of algebraic quantum numbers to the excitation. Of course, we must somehow choose the respective representations associated to these particles. It has been indicated that the physical particles are representations of the holonomy group of the connection, a subgroup of SL(2,$\mathcal{Q}$), induced from the subgroup SL(2,$\mathbb{C}$) and realized as functions on the coset spaces.

## 6.1. Geometric Classification of the Potential.

In order to classify the interactions we look for dynamical holonomy groups $H$ of the associated potential or connection. It is clear that $H$ must be a subgroup of the structure group of the theory, SL(2,$\mathcal{Q}$). For physical reasons we want that the interactions be associated to a dynamic evolution of the matter sources. The dynamical effects are produced in the theory by the action of the group, in particular accelerations should be produced by generators equivalent to the Lorentz boosts as seen by an observer. Therefore, we require that the boost generators, $\kappa_0\kappa_\alpha$, be present in a potential identifiable with a dynamical interaction as seen by the observer associated to the Minkowski subspace generated vectorially by $\kappa_\alpha$. Because of the nature of the source current,

$$J^\mu = \tilde{e}\kappa^{\hat{\alpha}}u_{\hat{\alpha}}^\mu e \ , \tag{6.1.1}$$

which corresponds to the adjoint action of the group on the algebra, it should be clear that all generators of the form $\kappa_{[\alpha}\kappa_{\beta]}$ of SL(2,$\mathcal{Q}$) should be present in the dynamical holonomy group of a physical interaction potential.

If we designate the even subgroup, generated by $\kappa_{[\alpha}\kappa_{\beta]}$, by $L$, the previous discussion means that

$$L \subseteq H \subseteq G \ . \tag{6.1.2}$$

Furthermore, $H$ must be simple. The different possibilities may be obtained from the knowledge of the subgroups of $G$. as discussed in chapter 2.

The possible simple subgroups are as follows:

1. The 10 dimensional subgroup $P$ generated by $\kappa_\alpha$, $\kappa_{[\mu}\kappa_{\nu]}$. This group is isomorphic to the groups generated by $\kappa_{[\alpha}\kappa_\beta\kappa_{\gamma]}$, $\kappa_{[\alpha\kappa\beta]}$ and by $\kappa_{[a}\kappa_b\kappa_{c]}$, $\kappa_{[a}\kappa_{b]}$, $\kappa_5$ and in fact to any subgroup generated by a linear combination of these three generators;

2. The 6 dimensional subgroup $L$, corresponding to the even generators of the algebra, $\kappa_{[\alpha}\kappa_{\beta]}$. This group is isomorphic to the subgroups generated by $\kappa_a$, $\kappa_{[a}\kappa_{b]}$, and by $\kappa_0\kappa_{[a}\kappa_{b]}$, $\kappa_{[a}\kappa_{b]}$ and in fact to any subgroup generated by a linear combination of these three generators;

2. The group $G$ itself.

The $P$ subgroup is Sp(2,$\mathcal{Q}$), as may be verified by explicitly showing that the generators satisfy the simplectic requirement [108]. This group is known to be homomorphic to SO(2,2), a De Sitter group. The $L$ subgroup is Sp(2,$\mathbb{C}$), isomorphic to SL(2,$\mathbb{C}$). In addition there are only two simple compact subgroups of $G$, nondynamical, generated by $\kappa_{[a}\kappa_{b]}$ and $\kappa_0$, $\kappa_5$, $\kappa_1\kappa_2\kappa_3$, apart from the unidimensional subgroups.

Then we have only three possible dynamical holonomy groups: $L$, $P$, or $G$. For each group we have an equivalence class of potentials and a possible physical interaction within the theory. The group chain $L \subset P \subset G$ has a symmetry because there is no unique way of identifying $L \subset G$ and $L \subset P$. The coset $G/L$ represents how many equivalent $L$ subgroups are in $G$. There is also an equivalence relation $R$ among the noncompact generators of $G$, all of them equivalent to a boost generator or space-time external symmetry. The subspace obtained as the quotient of $G/L$ by this relation is the internal symmetry group of $L \subset G$,

$$\frac{G/L}{R} = SU(2) \quad . \tag{6.1.3}$$

Similarly the coset $P/L$ gives, as the internal symmetry of $L \subset P$, the group

$$\frac{P/L}{R} = U(1) \quad . \tag{6.1.4}$$

The total internal symmetry of the chain $L \subset P \subset G$ is the product of the two groups SU(2)⊗U(1) which coincides with the symmetry group of the weak interactions. There is no geometrical reason to identify the structure group of the theory with the symmetry group because they are different geometrical concepts.

Other holonomy groups are not dynamical, in the sense that they do not produce a geometrical accelerating action on matter, as determined by an observer boost. This is the case of holonomy subgroups in the non simple chain $L \subset$ U(1)⊗$L \subset$ SU(2)⊗$L \subset G$ which may represent electromagnetism but do not provide, by their direct action, a geometric dynamics (force) on charged matter. The dynamics requires a separate Lorenz force postulate.

# 6.2. Algebraic Structure of Particles.

In general, the equation of motion for matter,

$$2\kappa^\mu \nabla_\mu e + \kappa^{\hat{\alpha}} \nabla_\mu u^\mu_{\hat{\alpha}} e = 0 \quad , \tag{6.2.1}$$

applies to the three classes of dynamical frames, according to the three dynamical holonomy groups. The three linearized equations, together with the nonlinear field equation should have solutions related to a substratum state solution. Then we may associate an excitation to each class of frames around the substratum solution. These excitations are elements of the Lie algebra of the structure group of the theory.

We shall now review and extend the discussion presented in sections 2.2 and 2.3. The maximally commuting subspace of the Lie algebra sl(4,$\mathbb{R}$) generated by the chosen regular element, [109] is a tridimensional Cartan subalgebra, which is spanned by the generators:

$$X_1 = \kappa_1\kappa_2 \quad , \tag{6.2.2}$$

$$X_2 = \kappa_0\kappa_1\kappa_2\kappa_3 \quad , \tag{6.2.3}$$

$$X_3 = \kappa_0\kappa_3 \quad . \tag{6.2.4}$$

It is clear that $X_1$, and $X_2$ are compact generators and therefore have imaginary eigenvalues. They should be associated, respectively, to the $z$-component of angular momentum and the electric charge. Both of them may be diagonalized simultaneously in terms of their imaginary eigenvalues.

The 4 members of the fundamental representation form a tetrahedron in the tridimensional Cartan space as described in chapter 2. They represent the combination of the two spin states and the two charge states of the associated particle, which are:

| charge | spin | flux | | |
|--------|------|------|--|--|
| $-1$ | $+1$ | $-1$ | negative charge with spin up | $f_-^\uparrow$ |
| $-1$ | $-1$ | $+1$ | negative charge with spin down | $f_-^\downarrow$ |
| $+1$ | $+1$ | $+1$ | positive charge with spin up | $f_+^\uparrow$ |
| $+1$ | $-1$ | $-1$ | positive charge  with spin down | $f_+^\downarrow$ |

The fundamental representation $f$ of SL(2,$\mathcal{Q}$), indicated  by $f_+^\uparrow$, $f_+^\downarrow$, $f_-^\uparrow$, $f_-^\downarrow$ groups together two excitation states of positive charge with two excitation states of negative charge. The presence of opposite charges in a representation forces us to make a clarification. To avoid confusion we should restrict the term *charge conjugation* to indicate the original Dirac operation to relate states of opposite charge.

There is also another fundamental representation $\tilde{f}$ dual to $f$ and of the same dimensions, with all signs reversed, which is inequivalent to the original one and represented by the inverted tetrahedron in the Cartan subspace. One of the two representations is arbitrarily assigned to represent the physical excitation. The mathematics of the representation algebra indicates that the state may also be described as composed of 3 members of its dual representation. In turn, the dual representation may also be similarly described as composed of 3 members of its dual, which is the original representation. This mutual decomposition is reminiscent of the idea of "nuclear democracy" proposed in the 1960's [110] but restricted to dual representations. To avoid confusion we restrict the use of the term *duality* to relate these inequivalent representations.

In standard particle language the antiparticle is the dual particle which implies that it also is the charge conjugate. Nevertheless, this assignment of a physical particle-antiparticle pair to the fundamental representation and its inequivalent dual is not a mathematical implication of standard particle theory, it is only a physical assumption. Since our fundamental representation includes opposite charges it is not appropriate to consider the dual representation as antiparticles. We may, as well, simply say that the antiparticle is determined by the algebra conjugation. The dual particle is just a necessary dual mathematical structure.

We have then for a particular state $p$ of the fundamental representation,

$$p^\uparrow \equiv f_+^\uparrow = \begin{pmatrix} \tilde{f}_+^\uparrow & \tilde{f}_+^\downarrow & \tilde{f}_-^\uparrow \end{pmatrix} \equiv \begin{pmatrix} q_+^\uparrow & q_+^\downarrow & q_-^\uparrow \end{pmatrix} \ , \tag{6.2.5}$$

in terms of the dual states $q$. Similarly for a particular $q$ state,

$$q_+^\uparrow = \begin{pmatrix} \tilde{q}_+^\uparrow & \tilde{q}_+^\downarrow & \tilde{q}_-^\downarrow \end{pmatrix} = \begin{pmatrix} p_+^\uparrow & p_+^\uparrow & p_-^\downarrow \end{pmatrix} \ . \tag{6.2.6}$$

It does not follow that the $q$ necessarily are states of a different physical excitation, only that the $q$ form a dual triplet mathematical representation of the $p$ representing the *same* excitation. This allows a different physical interpretation for these mathematical constructions. It should be noted that all $p$, $q$ have electric charge equal to the geometric unit, electron charge $\pm e$. Since these excitations have particle properties, there is a dual mathematical representation of the physical excitation (particle). We may raise the following question? What happens if we physically identify $p$ with the proton, which mathematically  may be expressed as *3 q*, interpreted as quarks? In our theory there is no need to assign fractional charges to quarks? In accordance with the "restricted nuclear democracy" we may assume that the quark states, in turn, may be mathematically expressed as 3 protons $p$, which may justify the large experimental mass of these unstable dual states.

The fundamental representation of Sp(2,$\mathcal{Q}$) which may be displayed as a square in a two dimensional Cartan space is $e_-^\uparrow$, $e_-^\downarrow$, $e_-^\uparrow$, $e_+^\downarrow$. Its dual Sp(2,$\mathcal{Q}$) representation, obtained by reversing all signs is mathematically the same $e_+^\downarrow$, $e_+^\uparrow$ $e_-^\downarrow$, $e_-^\uparrow$. Similarly the SL(2,$\mathbb{C}$) representation $\nu^\uparrow$, $\nu^\downarrow$ which may be displayed in a one dimensional Cartan space is mathematically its own SL(2,$\mathbb{C}$) dual. Therefore, the *only one* of the three excitations with a mathematically inequivalent dual structure is $p$. Since the Sp(2,$\mathcal{Q}$) and SL(2,$\mathbb{C}$) subgroups may be imbedded in SL(2,$\mathcal{Q}$), the corresponding Cartan spaces of Sp(2,$\mathcal{Q}$) and SL(2,$\mathbb{C}$) may also be imbedded in the Cartan subspace of SL(2,$\mathcal{Q}$). The plane subspace $Q=-1$, including dual states, has charges of one definite sign and is a representation of Sp(2,$\mathcal{Q}$). Another plane, $Q=1$, containing opposite charges is another representation of Sp(2,$\mathcal{Q}$) that completes the SL(2,$\mathcal{Q}$) Cartan subspace.

The space SL(4,$\mathbb{R}$)/SU(2)$\otimes$SU(2) is a nonadimensional riemannian space of the noncompact type. There are 9 boost generators $B^m{}_a$. The rotation SU(2), indicated by $S$, contained in in the maximal compact SO(4) acts on the $m$ index and the electromagnetic SU(2), indicated by $Q$, contained in SO(4) acts on the $a$ index,

$$Q_b^a B_a^n S_n^m = B_b'^m \ . \tag{6.2.7}$$

Similarly the subspace Sp(4,$\mathbb{R}$)/SU(2)$\otimes$U(1) is a hexadimensional riemannian space of the noncompact type. The comple-mentary subspace within SL(4,ℝ)/SU(2)⊗SU(2) is 3-dimensional. There is a triple infinity of these subspaces within the total space, reflecting the triple infinity of groups $P$ in $G$, depending on the choice of an electromagnetic generator among the three possible ones in SU(2).

In Sp(4,ℝ)/SU(2)⊗U(1), with a *fixed* electromagnetic generator, we have a vector in the odd sector, representing a momentum $k$ associated to an excitation $e$. Since we have this situation in SL(4,ℝ)/SU(2)⊗SU(2) for each generator in SU(2), we have, in effect, 3 momenta, $k_i$, that *may* characterize an excitation $p$. We must consider excitations characterized by 3 momenta, $k_i$, which may be interpreted as three subexcitations. This effect was used in the previous chapter to calculate quantum corrections to the proton and neutron magnetic moments.

Three P-subexcitations are necessary to characterize the protonic excitation $p$, but not those of $e$ nor those of $\nu$. Mathematically we have a system characterized by 3 trimomenta, $k_i$, that may be scattered into another system of 3 trimomenta, $k_j$'. A scattering analysis of excitations must include all momenta in some $\delta$ functions that appear in the scattering results. Experimentally this is a collision with a system of 3 pointlike scattering centers and is interpreted as a collision with partons inside the proton. In general it is clear that a pointlike proton excitation is not predicted by our theory. In particular, $e$ scattering from $p$ can be replaced by a sum of incoherent scatterings from P-excitations. Then group invariance implies that that the collision cross section can be expressed as two functions of the energy and momentum transfer and, for deep scattering, Bjorken's scaling law is obtained.

# 6.3. Interpretation as Particles and Interactions.

We have found properties of the geometric excitations which are particlelike. We take the position that this is no coincidence but indicates a geometrical structure for physics. The source current $J$ depends on a frame. To each holonomy group we may associate a class of frames thus giving three classes of matter.

As already discussed in section 5.8, the corresponding $L$-frame represents a zero bare mass, neutral, spin ½, left handed pointlike geometric excitation, which obeys eq. (5.8.1). It has the particle properties of the neutrino.

For the $P$-potential, the corresponding $P$-frame represents a massive, negative charge -1, spin ½, geometric pointlike excitation with bare magnetic $g$-factor -2, which obeys the Dirac equation. It has the properties of the electron.

For the $G$-potential, the corresponding $G$-frame represents a massive, charge +1, spin ½, 3-point geometric excitation with bare magnetic $g$-factor 2(2.780), which obeys the Dirac equation with a bare mass of 1836.12 times the bare mass of the previous excitation. It has the properties of the proton.

In this manner, we have associated to each of the three holonomy groups, one of the only three known stable particles.

For the $L$-potential it is not difficult to recognize that the interaction is gravitation, from the discussion in previous sections and the work of Carmeli [111]. Similarly, it also was shown in previous sections that electromagnetism (without dynamics) is associated to one of the SU(2) generators and that the physical Fermi weak interaction is related to the odd sector of a $P$-potential.

We propose here that the $P/L$ generators may be interpreted as weak electric nuclear interactions and the $G/P$ as strong magnetic nuclear interactions. Then the three dynamic holonomy classes of potentials may correspond to three classes of interactions as follows:

1. The $L$-potential describes gravitational interaction;
2. The $P$-potential describes coupled gravitational and electroweak interactions;
3. The $G$-potential describes coupled gravitational, electroweak and strong interactions.

The $L$-frames obey equations that may be obtained from the general equations of motion when the frame $e$ has only the even part $e_+$. From the field equation it is seen that a $P$-frame generates a $P$-potential and that a $G$-frame generates a $G$-potential.

From this classification it follows, in agreement with the physical interactions that:

1. All (matter) frames self interact gravitationally;
2. $L$-frames self interact only gravitationally (uncharged matter);
3. $G$-frames (hadrons) are the only frames that self interact strongly (hadronic matter);
4. $G$-frames self interact through all three interactions;
5. $G$-frames (hadronic matter) and $P$-frames (leptonic matter) self interact electroweakly and gravitationally;
6. $P$-frames self interact gravitationallly an electroweakly but not strongly (leptonic matter).

# 6.4. Topological Structure of Particles.

The framework of this theory is compatible with a phenomenological classification of particles in a manner similar to what is normally done with the standard symmetry groups.

First we should notice that the theory suggests naturally three stable ground particles ($\nu$, $e$, $p$). In fact, if we consider the possibility of different levels of excited states, each particle may generate a class of unstable particles or resonances.

In particular, since the equations for each of the three particle classes are the same, differing only in the subgroup that applies, it may be expected that there is some relation among corresponding levels of excitations for each class, forming families.

The ratio of the mass of the hadron in the ground level family (proton) to the mass of the lepton in the ground level family (electron) has been calculated from the ratio of volumes of coset spaces based on the existence of a inert nonzero constant solution for the substratum potential. This inert solution is the representative of a class of equivalent solutions generated by the action of the structure group.

Now consider only topological properties, independent of the potential, of the space of complete solutions (substratum plus excitation solutions). Scattering processes of excitations around a given substratum should naturally display these properties. An incoming scattering solution is a jet bundle local section, over a world tube in the space-time base manifold, that describes the evolution of the solution in terms a timelike parameter $\tau$ from past infinity to some finite time $t$. Similarly, an outgoing solution is a local section from time $t$ to future infinity. The local sections in the bundle represent classes of solutions relative to local observers. Scattering solutions at infinity are asymptotically free excitation solutions around a substratum. The substrata (incoming and outgoing) are equivalent to each other and to the constant inert solution if we choose observer frames adapted to the substrata.

The equations are of hyperbolic type and we should provide initial conditions, at past infinity $\tau = -\infty$, for the solutions over an initial tridimensional spatial hypersurface $\mathcal{I}_-$. We require that any incoming solution over the past infinity hypersurface $\mathcal{I}_-$ reduce to a free excitation around the substratum solution over the bidimensional subspace $\mathcal{I}_-(\infty)$ at spatial infinity $r = \infty$. Since the incoming solution are equivalent to the inert substratum solution at spatial infinity $\mathcal{I}_-(\infty)$, we may treat this spatial infinity as a single point, thus realizing a single point compactification of $\mathcal{I}_-$, so the initial hypersurface $\mathcal{I}_-$ is homeomorphic to $S^3$. All incoming solutions on $\mathcal{I}_-$ are classified by the functions over $S^3$. The same requirements may be applied to the outgoing future solutions and, in fact, to any solution along an intermediate tridimensional hypersurface $\mathcal{I}$, a section of the world tube. Thus, the final hypersurface at future infinity $\mathcal{I}_+$ is also homeomorphic to $S^3$. The incoming and outgoing substratum local sections over $\mathcal{I}_-$ and $\mathcal{I}_+$ must be pasted together in some common region around the present $t$, by the transition functions of the bundle. The scattering interaction is represented by the group action of the transition functions at $\tau = 0$. All generators of the group produce a transformation to a different, but equivalent under the group, expression for the solution. If the holonomy group of the solution is not the whole group, there is a reference frame that reduces the structure group to the particular holonomy subgroup. But in general for arbitrary observers, there are solutions formally generated by $SL(2,\mathcal{Q}) = SL(4,\mathbb{R})$. Since this transition region, the "equator" $\times R$, has the topology of $S^3 \times R$, the transition functions $\varphi$ define a mapping, at the $\tau = 0$ hypersurface,

$$\varphi : S^3 \to SL(2,\mathcal{Q}) \quad , \tag{6.4.1}$$

which is classified by the third homotopy group [112] of the structure group $SL(2,\mathcal{Q})$ or the respective holonomy subgroup. There are some solutions not deformable to the inert solution by a homeomorphism because $\varphi$ represents the twisting of local pieces of the bundle when glued together.

The homotopy group of $SL(2,\mathcal{Q})$ is $Z \otimes Z$ [9]. Similarly the homotopy group for both $Sp(2,\mathcal{Q})$ and $SL(2,\mathbb{C})$ is $Z$. The scattering solutions are characterized by topological quantum integer numbers $n$, for the three groups, and $n'$ only for group $G$, called winding or wrapping numbers. In all cases the scattering solutions are characterized by one topological winding number $n$ and in particular the hadronic scattering solutions have an additional topological winding number $n'$. This result implies that there are solutions $\omega_n$, $e_n$ that are not homotopically equivalent to $\omega_0$, $e_0$.

All $p$, $e$, $\nu$ excitation solutions with a given $n$ may be associated among themselves because of the isomorphism of the homotopy groups $Z$. This is an equivalence relation. Two $p$, $e$, $\nu$ excitation with the same $n$ are in the same topological class determined by the substratum. Each topological class, characterized by the topological quantum number $n$ defines a physical class, a family of particles with the same winding number, which is respected by transitions the same as the algebraic quantum numbers $s$, $q$, $f$.

For example, the association of a solution of each type to any value of $n$ may be part of a general scheme of relations labeled by $n$ as follows:

$$\begin{aligned}
&l_0 = e, \quad l_1 = \mu, \quad l_2 = \tau, \quad \cdots \\
&\nu_0 = \nu_e \quad \nu_1 = \nu_\mu \quad \nu_2 = \nu_l \quad \cdots \\
&h_0 = p_e \quad h_1 = p_\mu \quad h_2 \quad \cdots \quad ,
\end{aligned} \tag{6.4.2}$$

among the excitation levels of the electron (proper leptons), the excitation levels of the neutrino (other neutrinos) and the excitation levels of the proton (hadrons).

In any case, the possibility exists that there are excited states which may be interpreted as a particle composed of $p$, $e$ and $\nu$. As a matter of fact Barut [113] suggested that the muon should be considered as an electromagnetic excitation of the electron. Although the details should be different, because Barut's approach is only electromagnetic in nature and ours is a unified interaction, we may conjecture also that the muon is an excited electron state with a leptonic winding number $n = 1$. Similarly the $\tau$ would be an excitation with winding number $n = 2$

# 6.5. Geometric Excitation Masses.

If the leptons are topological excitations it should be possible to calculate their mass ratios [114] using the method given in chapter 4. The integrand $J\bullet\Gamma$ in eq. (4.1.10), which corresponds to a fundamental representation, is a local substratum constant equal for all values of $n$. The number of possible states of a fundamental representation only depends on the integration volume. Physically, when integrating over the $C_R$ subspace of De Sitter space $C$, corresponding to inequivalent observer states by an $L$ transformation, we have counted the number of $L$ equivalence classes of states (points in the mass hyperboloid are in the same relativistic equivalence class as the rest state of the representation) in local form (only counting $n=0$ local states). Taking in consideration the global characteristics of the topological excitation that would represent heavy leptons, the count should be larger, over all $L$ equivalence classes of excitation states with wrapping $n$. What is the number of states for each value of $n$?

Since we are working with a manifold with atlas, the integration over the symmetric space $K$ is realized on the atlas charts. On the principal fiber bundle $(E,M,G)$, the transition function preserves the projection and acts, as a group element, over the fiber which is the bundle structure group $G$ [9]. For two neighborhoods $U$, $V$ in the base manifold $M$, which corresponds to space-time, we have the following maps: homeomorphisms $h$ from the bundle to the model space, transition functions $\varphi$ among the charts and partial identity $i$ over the points $m\in M$.

The transition function $\varphi$ acts as an element of $G$ over the incoming scattering solution, defined on a hypersurface homeomorphic to $S^3$, to produce the outgoing scattering solution. Physically $\varphi$ represents a collision interaction. The active representation is the interaction of two excitations observed by a single observer. The passive representation is the observation of a single excitation by different observers. The nontrivial transition functions of class $[\varphi_n]$, where the index $n$ represents the set of wrapping numbers $(n, n')$, are classified by the mappings. Consider $G$ as a principal fiber bundle $(G,K,G_+)$. The induced representations, assigned to points $m\in M$, are sections of an associated bundle to $G$ that has Lorentz group representations $\mathcal{D}[L]$ as fiber and is denoted by $(D,K,\mathcal{D}[L],L)$ as shown in chapter 4. Under a change of charts, representing a change of observers, the transition function acts on the bundle $D$ (fig. 2) on the section $J\bullet\Gamma$ over $K$, which we shall indicate by $f(m)$.

The transition functions belong to the different classes determined by the third homotopy group of $G$. The class $[\varphi_n]$ may be expressed, using the homotopic product [115] in terms of the class $[\varphi_1]$. This class $[\varphi_n]$ may be considered generated by the product of $n$ independent elements of class $[\varphi_1]$ and one element of class $[\varphi_0]$. Observe that the homotopic mapping essentially determines a wrapping of subspaces $s(m)\subset G$, wrappings homeomorphic to spheres $S^3$, inside $(G,K,G_+)$. Each of the generating classes $\varphi$ is globally associated to a generating wrapped subspace $s$ in $(G,K,G_+)$, additional and necessary, linked to the original inert subspace by a transition function. In this manner the class $[\varphi_n]$ is associated to the trivial wrapping $[s_0]$ and to $(n, n')$ additional wrappings $[s_1]$, defined by the expression for the coordinates $u$ of point $m\in M$.

The excitation is not completely described by a single subspace that would correspond to a single observer $s(m)$ in the passive representation. For global reasons we must accept the presence of as many images of generating subspaces in the nontrivial chart as there are wrappings. Nevertheless some of the wrappings may be equivalent under the subgroup of interest. For the trivial subbundle $G$ there is only one independent wrapping, the trivial $(n=0)$ wrapping $s_0$, because the subspaces $s_n$ are equivalent under a $G$ transformation. For other subbundles, a tridimensional su(2) subalgebra acts on the complete group $G$ fiber bundle, generating transformations of $P$ and $L$ subbundles which are not equivalent under the corresponding subgroup. It is possible to generate no more than three wrappings $s_i(m)$ in geometrically independent $P_i$ or $L_i$ subbundles in $G$. In other words any other wrapping may be obtained by some combination of three transformations in $P_i$ or $L_i$. For this reason the number of independent wrappings $s_i$ in $P$ or $L$ bundles is $n\leq 2$ and determines three, and only three, families of excitations.

## 6.5.1. Leptonic Masses.

The actions of the $n+1$ subspaces $s_i(m)\subset P_i$ $(0 \leq i \leq n)$ on a Lorentz representation over $C$, the section $f\in\mathcal{D}$ of fiber bundle $(D,C,\mathcal{D}[L],L)$, map the fiber $\pi^{-1}$ over point $c$ onto fibers $\pi^{-1}$ over $s_i c$ by linear transformations $l_i$ [116], that determine a set of $n+1$ images of $f$, independent wrapped sections $f_i$ in the nontrivial chart, one for each independent wrapping. The Lorentz representations $f_i(s_i c)$ may be independent of any other $f_o(s_a c)$. Therefore, the physically possible states of a wrapped excitation of class $[\varphi_n]$ correspond 1 to 1 to all $L$ equivalence classes of states determined in each one of these $n+1$ section images $f_i$. Using the passive representation, on each image there are momentum coordinates $k_i^{\,\mu}$ relative to corresponding local observers $i$, states physically independent among themselves. The integration should be done over all independent variables $k_0^{\,\mu},k_1^{\,\mu},...k_n^{\,\mu}$, that is over a set of spaces, equivalent to the $C_R$ space defined in chapter 4, contained in a product of $n+1$ De Sitter spaces $C$.

As we indicated before for the substrate solutions, the integrand $J\bullet\Gamma$, corresponding to a fundamental representation, is a local constant equal for all values of $n$. The bare masses are determined by the volumes of integration. Let us denote the integration space that supports $f_i(m)$ by $C^*$, equal to the product of $n+1$ copies of the original $C_R$ space, corresponding to the $n+1$ observed wrappings,

$$C^* = \left(C_R\right)^{n+1} \quad . \tag{6.5.1}$$

For the trivial case $(n=0)$ the calculation is strictly local, it is not necessary to consider wrapped charts and the integration over $C_R$ may be done locally on any chart of the trivial class. The subspace $s_0$ is determined by a local observer.

For wrapped states ($n{\neq}0$) we are forced to use the nontrivial transition functions $[\varphi_n]$, with nonzero wrapping $n{\neq}0$, that relates a wrapped chart of class $[n]$ with the trivial chart. Therefore it is necessary to consider simultaneously (globally) these pairs of chart classes associated to the incoming and outgoing scattering solutions respectively. Physically it is necessary to consider all $L$-equivalence classes of possible states in accordance with observers related to nontrivial charts obtained from the trivial chart. Therefore, for the class $[n{\neq}0]$ it is necessary to integrate over all possible charts with "globally wrapped" sections, produced by the action of transition functions. In this case the transition functions may be any element of the group $P$. We restrict the integration to those non trivial charts that are inequivalent under a relativity transformation $L$. The number of possible charts corresponds to the volume of the $P/L$ subspace $C_R$ of $L$-equivalence classes.

The linear topological excitations are generated by the linear infinitesimal action of the differential $\varphi_*$ of the transition function $\varphi$ that represents the $\tau{=}0$ interaction. When restricting the transition functions to be elements of the subgroup $P$,

$$V \xrightarrow{\ h_U\ } \mathcal{A}(V) = V \times P \quad , \tag{6.5.2}$$

we should only consider the effect of the differential mapping $\varphi_*$. The odd subspace in the $P$ algebra, as differential mapping, generates the coset $C$. The only generator inequivalent to an $L$ "boost" is the compact u(1) generator. The differential action produces a smaller space of $L$-equivalence classes. If we denote the compact subgroup U(1)$\otimes$SU(2) of $P$ by $H$ we obtain the noncompact coset $B{\supset}C$

$$B = \frac{P}{H} \quad . \tag{6.5.3}$$

The group $P$ may be expressed as a principal fiber bundle $(P,B,H)$ over $B$. If we inject $L$ in $P$, the image of the rotation subgroup in $L$ should be the SU(2) subgroup in $H$ but the image of the "boost" sector in $L$ is not uniquely defined in $B$. The group $L$ acts on $P$ preserving its image in $P$. The u(1) subalgebra of $H$ acts on $B$, as a translation within $P$, mapping the $C$ subspace onto another subspace $C'$ in $B$ which is $P$-equivalent to $C$. Nevertheless $C'$ is not $L$-equivalent to $C$. Therefore, the possible $L$-inequivalent states correspond to the translation action by the U(1) compact subgroup not related [115] with the rotation SU(2). The number of $L$-equivalence classes of these possible charts, over which we should integrate, is determined by the volume of this electromagnetic U(1) group. The total wrapped space $C^T$ for an $n$-excitation ($n{\neq}0$) is the space formed by all possible translated, not relativistically equivalent, $C^*$ spaces

$$C^T = U(1) \times C^* \quad . \tag{6.5.4}$$

The number of states (different $k$ values) is, therefore, proportional to the volume of this total $C^T$ space, which may be calculated,

$$V\left(C^T\right) = V\left(U(1)\right) \times \left(V\left(C_R\right)^{n+1}\right) \qquad n \neq 0 \quad . \tag{6.5.5}$$

The bare mass of the trivial excitation ($n{=}0$) which is the electron, as previously indicated, is proportional to the volume of $C_R$

$$V\left(C_R\right) = \frac{16\pi}{3} \quad . \tag{6.5.6}$$

The bare masses corresponding to the $n$-excitations ($n{\neq}0$) are proportional to the volumes

$$V\left(C^T\right) = 4\pi \left(\frac{16\pi}{3}\right)^{n+1} \qquad 0 < n \leq 2 \quad , \tag{6.5.7}$$

that may be expressed in terms of the electron bare mass $m_0$,

$$\frac{m_n}{m_0} = \left(\frac{16\pi}{3}\right)^n 4\pi \qquad 0 < n \leq 2 \quad . \tag{6.5.8}$$

For excitations with wrappings 1 and 2 we have

$$\frac{m_2}{m_1} = V\left(C_R\right) = \frac{16\pi}{3} = 16.75516 \approx \frac{m_\tau}{m_\mu} = 16.818 \quad , \tag{6.5.9}$$

$$m_I = m_0 \left( \frac{16\pi}{3} \right) 4\pi = 0.5109989 \times 210.5516 = 107.5916 \text{ Mev } \approx m_\mu + O(\alpha) \quad , \qquad (6.5.10)$$

$$m_2 = m_0 \left( \frac{16\pi}{3} \right)^2 4\pi = 0.5109989 \times 3527.825 = 1802.7 \text{ Mev } \approx m_\tau + O(\alpha) \quad , \qquad (6.5.11)$$

that correspond to the bare masses of the $\mu$ and $\tau$ leptons, due to the electromagnetic U(1) interaction. Due to the electron U(1) electromagnetic interaction in the topological excitation, we should apply higher order energy corrections to $m_I$. For the subbundle $L$ we can make a similar calculation. The bare masses of the neutrinos for the 3 families are zero because the volume of the $L/L$ coset space is zero, as indicated in section 4.3. We shall refer to these topological excitations, respectively, as $T_n P$-excitations and $T_n L$-excitations.

## 6.5.2. Mesonic Masses.

The chart transformations under general $G$ transition functions translate, within $G$, the $C_R$ region of integration. The only possible additional $L$-inequivalent states correspond to the action of compact group sectors. In order to find the different possibilities we consider the two related chains of symmetric spaces in $G$, not related to $L$

$$
\begin{array}{ccccccc}
G & \supset & K_R & \supset & C_R & \supset & I \\
\cup & & \cup & & \cup & & \\
SU(2) & \supset & SU(2)/U(1) & \supset & U(1) & \supset & I \quad .
\end{array}
\qquad (6.5.12)
$$

The compact symmetric subspaces $C_R$ and $K_R$, respectively, contain one and two U(1) electromagnetic subgroups of the three equivalent U(1) subgroups contained in the electromagnetic SU(2)$_Q$. We note that one U(1) is common to both subspaces. We are interested in extending the translation within $G$, beyond the U(1) transition function, by adjoining compact sectors. The physical interpretation is that these excitations are subjected to additional nonlinear interactions that increase the energy and therefore their masses. In the same manner as an observable relativistic motion increases the rest mass to a kinetic mass, an observable relativistic interaction increases the free mass to a dynamic mass. This relativistic effect, included in the mass definition, is realized by the action of the transition functions at $\tau=0$.

For $n\neq0$ topological excitations the transition functions may be extended from U(1) to $C_R$. The physical interpretation is that these topological excitations are subjected to the additional electroweak nonlinear interactions of a leptonic $P$-system. We say that the excitations are "masked" by the nonlinear electroweak interactions. The geometric mass expressions for the class [$n\neq0$] masked excitations should be multiplied by the volume of the corresponding complete subspace of $L$-equivalence classes $V(C_R)$ rather than by its subspace $V$(U(1)). The weak interaction energy acquired by the $n$-excitations corresponds to the product of the values from equations (6.5.10) and (6.5.11) by the ratio of the volumes,

$$\frac{m_I'}{m_I} = \frac{V(C_R)}{V(U(1))} = \frac{4}{3} \approx \frac{m_\pi}{m_\mu} = 1.320957 \quad . \qquad (6.5.13)$$

If we use the experimental values for the lepton masses, we obtain the masked geometric masses,

$$m_I' = 140.8778 \text{ Mev } \approx m_\pi + O(\alpha) \;, \qquad (6.5.14)$$

$$m_2' = 2369 \text{ Mev } \approx m_f + O(\alpha) \;. \qquad (6.5.15)$$

We shall refer to these $C_R$ masked (electroweakly interacting) topological excitations as $T_n P_C$-excitations. The $T_n P_C$-excitations may be considered components of mesons. In particular, a masked muon $\mu'$ or $T_I P_C$-excitation, joined to a low energy $T_I L$-excitation, has the geometric mass and other properties of the pion $\pi$.

For $n\neq0$ topological excitations the transition functions may be further extended beyond $C_R$ to include the sector of $K_R$ corresponding to both U(1) subgroups in $K_R$ which is an SU(2)/U(1) compact sector that may be identified with the sphere S$^2$. The physical interpretation is that these excitations are subjected to the additional strong S$^2$ electromagnetic interaction of a hadronic $G$-system. The parametrization of group spaces and their symmetric cosets is, to a certain extent, arbitrary. They map different points in the linear Lie algebra to the same group operation. Since both equivalent U(1) subgroups in S$^2$ have the same significance because of the symmetry of S$^2$, they both should contribute equally to the invariant volume of S$^2$ and it is convenient to choose a parametrization that explicitly displays this fact. A parametrization that accomplishes this is

$$SU(2) = \exp\left(\alpha_+ J_+\right)\exp\left(\alpha_3 J_3\right)\exp\left(\alpha_- J_-\right) = \Sigma_+ U(I)\Sigma_- \quad, \tag{6.5.16}$$

$$V\left(SU(2)\right) = 16\pi^2 = V\left(\Sigma_+\right)V\left(U(I)\right)V\left(\Sigma_-\right) = 4\pi V^2\left(\Sigma_+\right) \quad. \tag{6.5.17}$$

We adjoin the group subspace $\Sigma \supset K_R$, defined by the expression in equation (6.5.16), to subspace $C_R$. Because of the extended integration in $\Sigma$, the $n$-excitations acquire a strong interaction energy corresponding to the ratio of the volumes

$$\frac{m_I''}{m_I'} = \frac{V\left(\Sigma C_R\right)}{V\left(C_R\right)} = V(\Sigma) = \left(4\pi\right)^{\frac{1}{2}} = 3.5449 \approx \frac{m_K}{m_\pi} = 3.5371 \quad. \tag{6.5.18}$$

We obtain the geometric masses for $n=I$ and $n=2$,

$$m_I'' = m_\pi V(\Sigma) = m_\pi \left(4\pi\right)^{\frac{1}{2}} = 494.76 \quad \text{Mev} \ = m_K + O(\alpha) \quad. \tag{6.5.19}$$

$$m_2'' = m_2' V(\Sigma) = \left(\frac{16\pi}{3}\right)m_I'\left(4\pi\right)^{\frac{1}{2}} \approx \left(\frac{m_\tau}{m_\mu}\right)m_K = 8303 \quad \text{Mev} \quad. \tag{6.5.20}$$

These excitations are not contained in the $P$ sector defined by the $n=0$ wrapping, but rather in a combination of inequivalent $P$ sectors inside $G$. Therefore, strictly, they do not have a proper constituent topological $P$-excitation. We shall refer to these $\Sigma$ masked (strongly interacting) topological excitations as $T_n P_\Sigma$-excitations. In particular, a masked muon $\mu''$ or $T_I P_\Sigma$-excitation, joined to a low energy $T_I L$-excitation by the strong interaction, has the geometric mass and other properties of the kaon $K$.

The physical interpretation of these masked leptons suggests that the corresponding geometric excitations may be combined to form lepton pairs that may be considered particles. In accordance with this interpretation, the combinations under nuclear interactions are only possible if there is, at least, one masked lepton. Take the $n=I$ topological leptons and construct a doublet $l$ of a $^c SU(2)$ combinatory group associating a masked muon and a stable lepton. We should indicate that the conjugation in the geometric algebra, which is equivalent to the dual operation in sp(4,$\mathbb{R}$), is not the dual operation in $^c$su(2). Both $l$ and its conjugate are $^c$su(2) fundamental representations **2**. The product is

$$\mathbf{2} \otimes \mathbf{2} = \mathbf{3} \oplus \mathbf{1} \tag{6.5.21}$$

The first possibility is that $\mu$ is a weakly masked $\mu'$, part of the SU(2) leptonic system doublet $(\mu', \nu)$ characterized by the charge $Q$ and the muonic number $L_\mu$. We get the amplitude for the $\pi$ representation:

$$[\mu',\nu]\otimes[\overline{\mu}',\overline{\nu}] \equiv \left[\pi^-,\pi^0,\pi^+\right]\oplus x = \left[\overline{\nu}\mu', \left(\overline{\mu}'\mu'-\overline{\nu}\nu\right)\tfrac{1}{\sqrt{2}}, \overline{\mu}'\nu\right]\oplus\left(\left(\overline{\mu}'\mu'+\overline{\nu}\nu\right)\tfrac{1}{\sqrt{2}}\right) \quad. \tag{6.5.22}$$

The only other possibility is the more complex combination where the muon is a strongly masked $\mu''$. In addition to the coupling to $\nu$, since $\mu''$ has a strong $S^2$ electromagnetic interaction, $\mu''$ also couples strongly to the electron $e$, forming two related SU(2) hadronic systems. We substitute $\nu$ by $e$ as the leptonic doublet partner. We get the $K$ representation amplitude:

$$[\mu'',\nu]\otimes[\overline{\mu}'',\overline{\nu}] \equiv \left[K^-,K_L,K^+\right]\oplus x_L' = \left[\overline{\nu}\mu'', \left(\overline{\mu}''\mu''-\overline{\nu}\nu\right)\tfrac{1}{\sqrt{2}}, \overline{\mu}''\nu\right]\oplus\left(\left(\overline{\mu}''\mu''+\overline{\nu}\nu\right)\tfrac{1}{\sqrt{2}}\right) \quad, \tag{6.5.23}$$

$$[\mu'',e]\otimes[\overline{\mu}'',\overline{e}] \equiv \left[\overline{K}^0,K_s,K^0\right]\oplus x_S' = \left[\overline{e}\mu'', \left(\overline{\mu}''\mu''-\overline{e}e\right)\tfrac{1}{\sqrt{2}}, \overline{\mu}''e\right]\oplus\left(\left(\overline{\mu}''\mu''+\overline{e}e\right)\tfrac{1}{\sqrt{2}}\right) \quad. \tag{6.5.24}$$

The masses of these $\pi$ and $K$ geometric particles equal the masses of ground states belonging to the product representation of these representations. Their geometrical masses, essentially determined by the mass of its principal component or masked heavy lepton as indicated in equations (6.5.14) and (6.5.19), approximately correspond to the masses of all physical pions and kaons. We may also define

$$x' = \left(x_L' + x_S'\right)\tfrac{1}{\sqrt{2}} = \left(2\overline{\mu}''\mu'' + \overline{e}e + \overline{\nu}\nu\right)\tfrac{1}{2} \equiv \eta' \quad. \tag{6.5.25}$$

$$\left(x + x'\right)\tfrac{1}{\sqrt{2}} \equiv \eta \quad. \tag{6.5.26}$$

These definitions correspond to physical particles whose geometrical masses, essentially determined by the masked heavy lepton mass, are approximately,

$$m_{x'} \approx 990 \text{ Mev } + O(\alpha) \approx 957.8 \text{ Mev} = m_{\eta'} \tag{6.5.27}$$

and using the experimental value of $m_\eta$,

$$m_x \approx 549 \text{ Mev } + O(\alpha) \approx 547.3 \text{ Mev} = m_\eta . \tag{6.5.28}$$

These results suggest a higher approximate symmetry for the $\pi K$ combination, that may be contained in the pseudoscalar meson representation as indicated in figure 5. In other words, the scalar mesons may be considered lepton-antilepton pairs, as suggested by Barut [117].

In addition to the proper topological $P$-excitations just discussed, we may consider **subexcitations inside G**. For [n=0] $P$-subexcitations, considered in $G$, there is a set of $C$ subspaces that correspond to the infinite ways of choosing $P \subset G$, not to the unique $C$ defined for an electron $P \subset P$. These additional $C$ spaces are not $P$-equivalent and must be included in the integration. For a triple electromagnetic interaction there are as many $C$ spaces as U(1) subgroups inside SU(2), in other words, as points on the sphere SU(2)/U(1). The volume of integration should be multiplied by $4\pi$. The same thing happens with $L$-excitations. In other words, class [n=0] leptonic subexcitation masses $m_o$, present in a hadronic $G$-system, are masked by the nonlinear strong interactions. We shall refer to these $S^2$ masked (strongly interacting) subexcitations as $G_s$-subexcitations. The $G_s$-subexcitations geometric mass values corresponding to the fundamental leptonic geometric excitations are $4\pi$ times the previous leptonic geometric mass values,

$$m'''(e, \nu_e, \mu, \nu_\mu, \tau, \nu_\tau) = (0.00641, \, 0, \, 1.77, \, 0, \, 29.9, \, 0) \text{ Gev} . \tag{6.5.29}$$

The masked excitations are characterized by the same quantum numbers that characterize the leptonic excitation but the value of the masked geometric mass includes the energy increase due to the other interactions. If the masked excitations were to be ejected (injected) from (to) a $G$-system they would loose (gain) the energy due to the extra interactions, and they would exit (enter) with the bare geometric mass value, as standard free leptons. There are no observable free particles with the masked masses. Experimentally the masked mass would never be detected by long-range methods. These masked leptonic excitations behave as quarks.

In particular, the $G_s$-subexcitations correspond, one to one, to the leptons. These $G_s$-subexcitations determine a hexadimensional space. Consequently any other possible geometric $G_s$-subexcitation can be expressed as a linear superposition of these fundamental masked geometric subexcitations. Alternately we may interpret some of these subexcitations as quark states. The structure of masked leptonic subexcitations is geometrically equivalent to the physical quark structure. For example, a superposition of 2 low velocity masked electrons $e$ plus 1 masked low energy neutrino $\nu_e$ has a $2/3$ charge and a total $4.2$ Mev invariant mass and may be interpreted as a $u$ quark. Similarly a superposition of 2 masked low velocity muons $\mu$ plus 1 masked low energy neutrino $\nu_\mu$ has $2/3$ charge and $1.2$ Gev total mass. It should be noted that, if free quarks are unobservable, all experimental information about their masses comes from bound quark states (meson resonances) and must depend somehow on theoretical arguments about these states. There is no quark confinement problem.

As shown in section 6.2, a proton contains leptonic subexcitations. We may consider that there are $G_s$-subexcitations inside the hadrons. In this manner we may identify six flavors of geometric masked leptonic excitations inside all hadrons providing an equivalent quark flavor structure. There is also another triple structure associated to the dual representations, also shown in that section, which could in principle provide another quark structure. Nevertheless, we have said duality is just a necessary mathematical equivalent structure. Additionally, there is no spin-statistics problem with these geometric components and there is no need for another degree of freedom (color).

# 6.6. Barut's Model.

On the other hand, Barut has suggested [117, 118, 119, 120, 121] a process to construct particles in terms of the stable particles and the muon. Therefore, it appears convenient to constitute geometric excitations that represent particles starting from the geometric fundamental excitations $G$ (the proton), $P$ (the electron) and $L$ (the neutrino), together with the strongly interacting masked leptonic $T_n P_s$-excitations. In this section the symbols $\mu$, $\mu$', $\mu$'' and $\tau$ represent masked unstable leptons. On hadronic time scales a weakly masked muon is considered stable and may be represented here by its stable components $e \, \bar{\nu}_e \, \nu_\mu$.

In order to adapt our geometric model to Barut's model we establish that the quantum numbers of an excitation correspond to the quantum numbers of its stable components. In this manner, following Barut's ideas, we define the charge of an excitation as the net number of charge quanta of the component excitations,

$$Q = N_p - N_e . \tag{6.6.1}$$

Similarly, the barionic (atomic) number of an excitation is the number of constituent $G$-excitations, representatives of protons $p$,

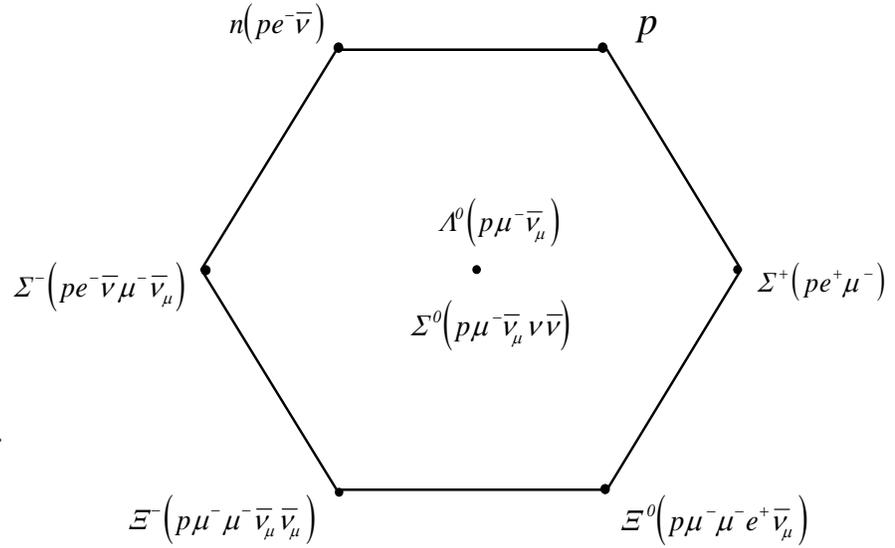

**Figure 4.**

Barionic

octet $\left(J^P = \frac{1}{2}^+\right)$.

$$A \equiv N_p \; , \tag{6.6.2}$$

and the leptonic number of an excitation is the number of constituent $P$-excitations and $L$-excitations, in $n$ classes according to their topological $T_n$ excited states, representatives of electrons $e$ and neutrinos $\nu$,

$$L \equiv N_e + N_\nu \; . \tag{6.6.3}$$

In a hadronic system of excitations, strangeness is the number of component muons $\mu$, masked by the strong interaction, capable of a $T_1 P_\Sigma$-excitation or $\mu$'',

$$-S \equiv N_\mu \; , \tag{6.6.4}$$

and, similarly, beauty is the number of strongly masked $\tau$ leptonic components capable of $T_2 P_\Sigma$-excitations,

$$B \equiv N_\tau \; , \tag{6.6.5}$$

Charm and truth are, respectively, the number of component $T_1 L$-excitations and $T_2 L$-excitations, representatives of neutrinos $\nu_\mu$ and $\nu_\tau$,

$$C \equiv N_{\nu(\mu)} \; , \tag{6.6.6}$$

$$T \equiv N_{\nu(\tau)} \; . \tag{6.6.7}$$

The isotopic spin depends on the number of stable constituent $G$-excitations, $P$-excitations and $L$-excitations representatives of $p$, $e$ and $\nu$ in the following manner,

$$2I_3 \equiv N_p - N_e + N_\nu \; . \tag{6.6.8}$$

From these definitions we may derive the Gell-Mann-Nakano-Nishijima formula,

$$Q = I_3 + \tfrac{1}{2}\left(A+S\right) \; , \tag{6.6.9}$$

and define strong hypercharge

$$Y \equiv A + S \; . \tag{6.6.10}$$

In additon to these Barut definitions, it is also possible to define weak hypercharge for a $P$-excitation in terms of the leptonic number and the rarity $r$ [9] (see eq. A.25 in the appendix),

$$y \equiv -\left(L + r\right) \; . \tag{6.6.11}$$

The hadronic states contain protons. In figure 4 we show the barionic octet $J^P = 1/2^+$ and in figure 6 the barionic decuplet $J^P = 3/2^+$. The meson states are bound states of two leptons $l\bar{l}$ as indicated in the mesonic octect $J^P = 0^-$ in figure 5.

Furthermore, the geometric theory allows a discussion of the approximate quantum interaction between two dressed

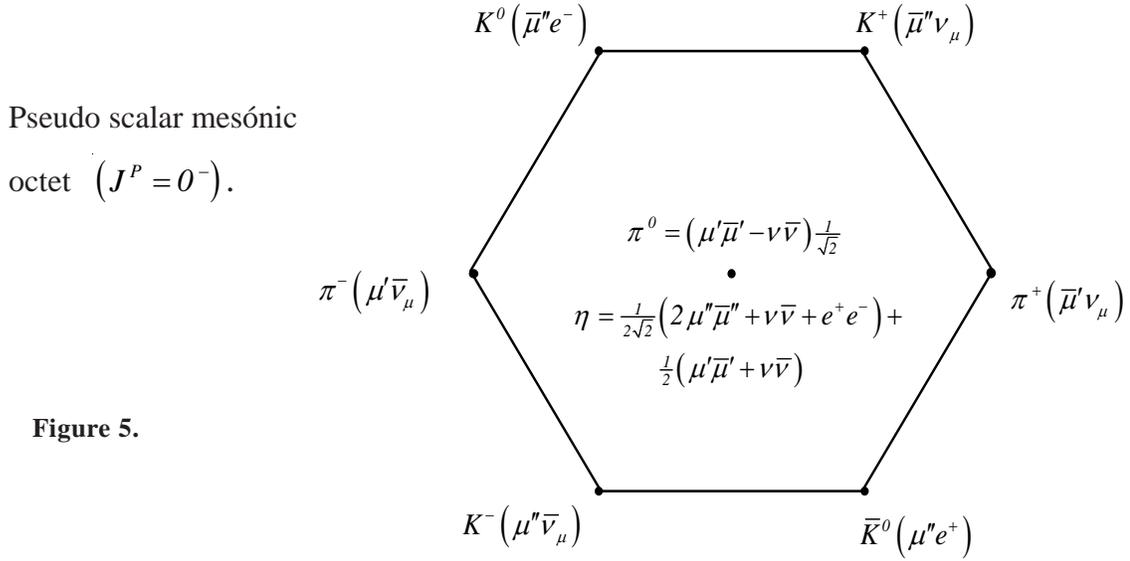

Pseudo scalar mesónic

octet $\left(J^P = 0^-\right)$.

**Figure 5.**

leptonic excitation states forming a system (quarkonium?). We speculate that the nonrelativistic effective potential should be similar to the one in QCD because there are similarities of the mathematics of the theories. If this were the case, (this has to be shown) the nonrelativistic effective potential would be [122]

$$V(r) = -\frac{a}{r} + kr \qquad\qquad (6.6.12)$$

and then we may fit the experimental $\Psi$ excited levels accordingly, as is done in QCD.

The full frame excitation has protonlike properties. Within the geometric theory the quark excitations are not the fundamental building blocks of matter. They are only subexcitations that are useful and necessary in describing a series of hadronic excitations.

# 6.7. Relation with Particle Theory.

The geometric triple structure determines various structures compatible with the quark description. On top of this geometry, as is done in the standard model over the geometry of special relativity, it is possible to add an approximate structure to help in the phenomenological understanding of the physical particles. It is known that heavy nuclei may be partially understood by using groups SU(N), U(N), O(N) to associate protons and neutrons in a dynamic symmetric or supersymmetric manner [123]. This is essentially the use of group theory to study the combinations of the two building blocks, protons and neutrons, assumed to form nuclei. In the same manner we can use certain groups to describe the combinations of the three fundamental geometrical building blocks introduced by the physical geometry, protons, electrons and neutrinos assumed to form other particles.

We are interested in combining $G$, $P$ and $L$ excitations. We attach $L$ excitations to $P$ excitations and then to $G$ excitations. As indicated in section 6.1, this combination may not be done uniquely, rather it depends on the identification of a subgroup $H$ with a subspace in the group fiber bundle space $G$. Any noncompact generator is equivalent to a boost or external symmetry, by the adjoint action of a compact generator. Thus, the compact generators generate an internal symmetry of the combination of excitations.

In particular, the identification of $L$ within $P$ is not unique. The group $L$ may be expressed as the principal bundle $(L,{}^3B,S)$, where the fiber $S$ is the $SU(2)_R$ associated to rotations and ${}^3B$ is the tridimensional boost symmetric space,

$$^3B = \frac{SL(2,\mathbb{C})}{SU(2)_R} \quad . \qquad\qquad (6.7.1)$$

The group $P$ may be expressed as the principal bundle $(P,{}^6B,S\otimes U(1))$ where ${}^6B$ is the hexadimensional double boost symmetric space,

$$^6B = \frac{Sp(4,\mathbb{R})}{SU(2)_R \otimes U(I)_Q} \quad . \qquad\qquad (6.7.2)$$

The choice of the boost sector of $L$ (even generators), inside $P$, depends on the action of the U(1) group in the fiber of $P$,

**Figure 6.**

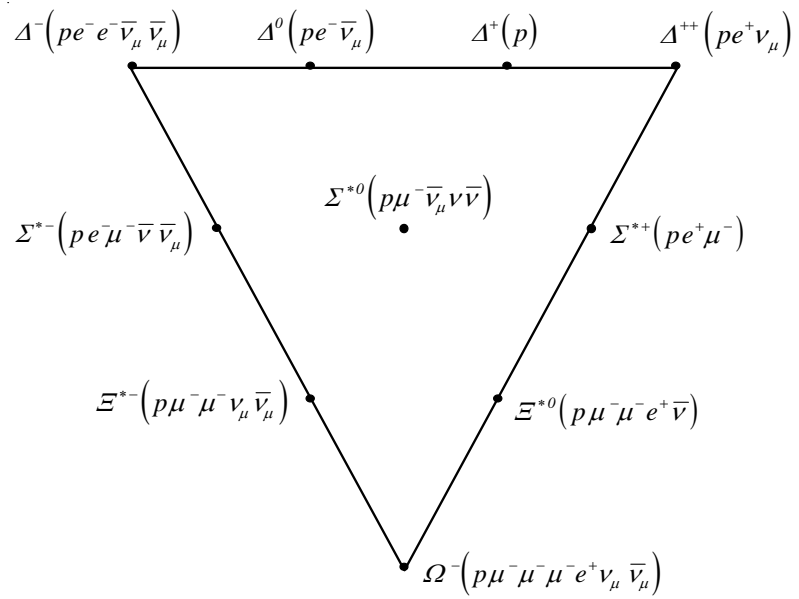

Barionic

decuplet $\left(J^P = \frac{3}{2}^+\right)$.

generated by $\kappa_0$, a rotation within the double boost sector in $P$.

$$\left[\kappa_0, \kappa_I\right] = 2\kappa_0\kappa_I \quad . \tag{6.7.3}$$

Mathematically this corresponds to the adjoint action of the only compact generator in the odd quotient $P/L$.

Similarly, the identification of $P$ inside $G$ is not unique either. The group $G$ may be expressed as the bundle $(G, {}^9B, S \otimes Q)$ where $Q$ is the $SU(2)_Q$ associated to charge and ${}^9B$ is the nonadimensional triple boost symmetric space,

$$^9B = \frac{SL(4, \mathbb{R})}{SU(2)_R \otimes SU(2)_Q} \quad . \tag{6.7.4}$$

The choice of the boost sectors of $L$ and $P$ sector inside $G$ depends on the action of the $SU(2)$ in the fiber of $G$, generated by the compact electromagnetic generators. Mathematically this corresponds to the action of all compact generators in the quotient $G/L$. The total internal symmetry of first identifying $L \subset P$ and then $(L \subset P) \subset G$ is the product of the two groups. Therefore, there is an internal symmetry action of a group equal to $SU(2) \otimes U(1)$, *but different to the subgroups of $G$,* on the identification of the subgroups in the chain $G \supset P \supset L$.

There is an induced symmetry on the combination of $L$ and $P$ excitations on $G$ excitations corresponding to this $SU(2) \otimes U(1)$ symmetry. The action of this combinatorial group may be interpreted as the determination of the possible combinations of $L$ and $P$ excitations to give flavor to states of $G$ and $P$ excitations and thus related to the weak interactions.

This $SU(2)_F$ group relates electron equivalent $P$-excitation states or neutrino equivalent $L$-excitation states. Clearly at sufficiently high energies the mass of any excitation kinematically appears very small and its effects on results are small deviations from those of a zero mass excitation that always corresponds to an excitation of the even subgroup or $L$-excitation. For this reason, at high energies, the even part (left handed part) of a $P$-excitation may be related to an even $L$-excitation by an $SU(2)$ transformation. Both corresponding physical leptonic excitations, the even part $e_+$ or $e_0$ and $\nu$, may be considered members of a doublet, labeled by rarity 0 or weak hypercharge -1, while the odd part $e_-$ or $e_1$ is a singlet labeled by rarity 1 or weak hypercharge -2. This weak interaction association of leptons into hypercharged states has the approximate symmetry $SU(2)_F$. In addition, there is an undetermined orientation of $L_1$ inside $G$ that depends on the action of $SU(2) \otimes U(1)$. Under this approach, the standard physical electromagnetic potential $A$, the potential of $U_+(1)$ in $SL_1(2, \mathbb{C})$, has an orientation angle within the chiral $SU(2) \otimes U(1)$ that should be related to the possible polar angles $\Theta$ in eq. (4.6.5).

Since our description may be made in terms of the *proton representation or, alternately, in terms of its dual quark representation* as indicated in section 6.2, we could obtain complementary dual realizations of the relations among different experimental results. These are really different perceptions or pictures of the same physical reality of matter. The proton or $G$-excitation has a three point structure: it behaves as if formed by quarks. If we use the quark representation to build other particles, we may consider that in $G$ the 3 quark states span a 2 dimensional Cartan subspace. There is a combinatorial unitary symmetry, $SU(3)$, related to this subspace of a type $A_3$ Cartan space, that may be interpreted as the color of the combinations of states of $G$ and thus related to strong interactions. States of $G$-excitations display an $SU(3)_C$ symmetry.

In this manner, the physical geometry determines an approximate combinatorial internal symmetry characterized by the combinatorial symmetry group $S_C$,

$$S_C = SU(3)_C \otimes SU(2)_F \otimes U(I) \quad . \tag{6.7.5}$$

We combine $L$-excitations, $P$-excitations and $G$-excitations using mathematical and physical considerations. Any subgroup $H{\in}G$ geometrically determines a fiber bundle $(G,G/H,H,U)$. The structure group $U$ of the bundle cannot be uniquely, mathematically, identified with a ***fixed*** $H$ subspace in $G$. It acts on each equivalent $H$ subspaces, while the fiber is transported throughout the $G/H$ space, and represents an equivalence class of subspaces. We may interpret $U$ as a ***different*** group isomorphic to each $H$. Mathematically, we require that the combinations of $G{\supset}P{\supset}L$ chain subgroups be realized as direct products. Nevertheless from the physical point of view we recognize that all chain subgroups are related. The compact $SU(2)_S$ spin subgroup in $L$ is interpreted as physical rotations. The other two compact subgroups in $P$ and $G$ are $U(1)$ and $SU(2)_Q$ and are interpreted as internal electromagnetic rotations. The complementary noncompact sectors are interpreted as generalized (internal) boosts. Physically we require that quantum numbers of product representations in the chain $G{\supset}P{\supset}L$ obey an addition rule. As indicated in previous sections, combinations of excitations appear to determine masses, magnetic moments, quantum numbers, etc. of physical particles. The neutron and the pion would correspond, respectively, to the fundamental combinations of the $G{\supset}P{\supset}L$ and $P{\supset}L$ chains.

The constructions discussed in this section should share certain features and results with the standard model while there would be differences in some other features. This actually represents a program to be carried out. It is true that the standard model [112] has led to success, but this does not imply there is no better way of reordering the physical without large number of empirical parameters. Other geometric models, like string theories [124] involve a large number of dimensions which are not clearly related to a fundamental underlying theoretical physical interaction. Due to the richness and vastness of particle physics experimental results, it is not clear how differences compare with experiments, in particular because the geometrical ideas and the group $SL(4,\mathbb{R})$ [125] may introduce a rearrangement of experimental results.

## 6.8. Results.

The interaction may be classified into three subclasses by the dynamical holonomy groups of the possible physical potentials. The three classes correspond to gravitation, electro-weak and strong physical interactions. Solutions to the three corresponding frame excitation equations representing matter, show algebraic and topological quantum numbers. The algebraic numbers correspond to electric charge, spin and magnetic flux. The topological numbers correspond to wrapping (winding) numbers of higher levels of excitation, defining three 2-member leptonic families. We are able to identify three classes of fundamental excitations, corresponding to the three stable physical particles: neutrino, electron and proton. The calculated masses agree with the experimental values.

The bare masses of the leptons in the three families are calculated as topological excitations of the electron in agreement with experimental values. The masses of these excitations increase under the action of a strong potential (relativity of energy) producing the meson masses, also in agreement with experiment. The geometry determines the geometric excitation mass spectrum, which for low masses, essentially agrees with the physical particle mass spectrum. The proton shows a triple structure that may be related to a quark structure. The combinations of the three fundamental geometric excitations (associated to the proton, the electron and the neutrino), forming other excitations, may be used to represent particles and show a symmetry under the group $SU(3){\otimes}SU(2){\otimes}U(1)$.

# 7. Gravitation and Geometry.

The standard geometrical approach to gravitation requires the construction of an energy momentum or stress energy tensor $T$ determined by the classical equations of macroscopic matter and fields. In general these matter fields are described by the classical fluid equations. It may be claimed that these stress energy tensors are not truly geometrical. Einstein [126] himself was unsatisfied by the nongeometrical character of $T$ and spent his later years looking for a satisfactory unified theory. Now we use the geometrical theory to obtain the Eintein equations with a geometric stress energy tensor.

It is also known that the value of the fine structure constant $\alpha$ is essentially equal to an algebraic expression in terms of $\pi$ and entire numbers that is related to the quotient of the volume of certain groups. Here we determine this expression using the invariant measure defined by the substratum solution of this geometrical theory.

## 7.1. Einstein's Equation.

The third equation (1.2.8) determined by a variational principle has the structure

$$\mathrm{tr}\Big[ 4F_{\bar{\rho}\nu}F^{\mu\nu} - u_{\bar{\rho}}^{\mu}F^{\kappa\lambda}F_{\kappa\lambda} \Big] = X \tag{7.1.1}$$

of a sum of terms, each of the form corresponding to the electromagnetic stress energy tensor. Using this equation we may define a tensor with the structure of stress energy. It is clear that this tensor would not represent the source term of the field equations. It represents the energy of the interaction potential and matter frame fields.

Eq. (1.4.3) indicates that the even part of the curvature in this equation represents the classical interaction fields of gravitation and electromagnetism. Due to this reason we shall separate the terms in the equations with respect to the even subalgebra and its coset space. In particular, the SL(2,C) curvature form corresponds to the Riemann curvature $R$ in standard spinor formulation. It obeys field equations which are different from Einstein's equations and represent a spinor gravitational formulation that resembles Yang's [18] theory restricted to its SO(3,1) subgroup. Yang's gravitational theory may be seen as a theory of a connection or potential in the principal bundle of linear frames $TM_m$ with structure group GL(4,R). In Yang's theory the group is taken to act on the tangent spaces to $M$ and therefore is different from the theory under discussion. The potential in Yang's theory is not necessarily compatible with a metric on the base space $M$, which leads to well known difficulties. Nevertheless, when Yang's theory is restricted to its SO(3,1) subgroup its metrical problems are eliminated. The well known homomorphism between this group and our even subgroup SL(2,C) establishes a relation between these two restrictions of the theories.

It may be claimed that the Einstein equation of the geometric unified theory is equation (1.2.8) rather than the field equation. When we consider the external field problem, that is, space-time regions where there is no matter, the gravitational part of the field equations for $J=0$ are similar to those of Yang's gravitational theory. All vacuum solutions of Einstein's equation are solutions of these equations. In particular the Schwarzchild metric is a solution and, therefore, the newtonian motion under a $1/r$ gravitational potential is obtained as a limit of the geodesic motion under the proposed equations. Nevertheless, there are additional spurious vacuum solutions for the field equation in Yang's theory, which are not solutions for Einstein's theory. Fairchild [127] has shown that setting $X$ equal to zero in equation (7.1.1) is sufficient to rule out these spurious vacuum solutions. The geometric equation when restricted to vacuum eliminates the solutions found by Pavelle [128, 129] and Thompson [130].

Nevertheless, the interior problem [131] provides a situation where there are essential differences between the unified physical geometry and Einstein's theory. For this problem where the source $J$ is nonzero, our theory is also essentially different from Yang's theory. Since in the physical geometry the metric and the so(3,1) potential remain compatible, the base space remains pseudoriemannian with torsion avoiding the difficulties discussed by Fairchild and others for Yang's theory.

The presence of a matter current term in the equation for the stress energy affords the possibility of getting a geometric stress energy tensor that could enter in a fully geometric Einstein equation for the metric, thus resembling Einstein's theory. The right hand side of the main equation (1.2.8) is interpreted as a stress energy tensor $^j\Theta$ related to the matter current source, in terms of the vector density $\iota$,

$$^j\Theta^\mu_{\hat\rho} = -\frac{\alpha}{4}\,\mathrm{tr}\left[e^{-1}\iota \circ u^{(\mu}\nabla_{\hat\rho)}e - \tfrac{1}{4}u^\mu_{\hat\rho}e^{-1}\iota \circ u^\lambda\nabla_\lambda e\right] \quad . \tag{7.1.2}$$

The left hand side terms with the familiar quadratic structure may be split into the even Lorentz curvature components and the odd coset components which define a coset field stress energy tensor $^c\Theta$,

$$^c\Theta_{\rho\mu} \equiv \frac{-1}{4\pi}\,^c g_{\hat a\hat b}\left[^c\Omega^{\hat a}_{\ \rho\nu}\,^c\Omega^{\hat b}_{\ \mu}{}^\nu - \frac{1}{4}g_{\rho\mu}\,^c\Omega^{\hat a\kappa\lambda}\,^c\Omega^{\hat b}_{\ \kappa\lambda}\right] \quad . \tag{7.1.3}$$

We may express the even Lorentz curvature components in terms of the Riemann tensor $R^a_{\ bmn}$, defined by the space-time metric, and an explicit dependence on the torsion, $\Sigma$,

$$^{L'}\Omega^\alpha_{\ \beta\kappa\lambda} = \left(^n\Omega^\alpha_{\ \beta\kappa\lambda} + R^\alpha_{\ \beta\kappa\lambda}\right) + Z^\alpha_{\ \beta\kappa\lambda} \equiv {}^nR^\alpha_{\ \beta\kappa\lambda} + Z^\alpha_{\ \beta\kappa\lambda} \quad , \tag{7.1.4}$$

where $^nR^\alpha_{\ \beta\mu\nu}$ is defined as a nonriemannian curvature including the Riemann tensor and

$$Z^\alpha_{\ \beta\kappa\lambda} = \nabla_\kappa\Sigma^\alpha_{\ \beta\lambda} - \nabla_\lambda\Sigma^\alpha_{\ \beta\kappa} + \Sigma^\alpha_{\ \gamma\kappa}\Sigma^\gamma_{\ \beta\lambda} - \Sigma^\alpha_{\ \gamma\lambda}\Sigma^\gamma_{\ \beta\kappa} \quad . \tag{7.1.5}$$

There are terms in eq. (1.2.8) that may be grouped to form a stress energy tensor associated to the torsion,

$$^t\Theta_{\rho\mu} \equiv \frac{g_{\rho\mu}}{32\pi}\left(Z^{\hat\alpha}_{\ \hat\beta}{}^{\kappa\lambda}Z^{\hat\beta}_{\ \hat\alpha\kappa\lambda} + Z^{\hat\alpha}_{\ \hat\beta}{}^{\kappa\lambda}\,^nR^{\hat\beta}_{\ \hat\alpha\kappa\lambda} + {}^nR^{\hat\alpha}_{\ \hat\beta}{}^{\kappa\lambda}Z^{\hat\beta}_{\ \hat\alpha\kappa\lambda}\right) - \frac{1}{8\pi}\left(Z^{\hat\alpha}_{\ \hat\beta\rho\nu}Z^{\hat\beta}_{\ \hat\alpha\mu}{}^\nu + Z^{\hat\alpha}_{\ \hat\beta\rho\nu}\,^nR^{\hat\beta}_{\ \hat\alpha\mu}{}^\nu + {}^nR^{\hat\alpha}_{\ \hat\beta\rho\nu}Z^{\hat\beta}_{\ \hat\alpha\mu}{}^\nu\right) . \tag{7.1.6}$$

Because of the nonlinearity of the equations there is a contribution of gravitation to its own source, defining a geometric energy-momentum tensor contribution,

$$^g\Theta_{\mu\rho} \equiv \frac{1}{8\pi}\left[\frac{^nR}{3}\left(g_{\rho\mu}\frac{R}{4} + {}^n\Omega_{\rho\mu} - g_{\rho\mu}\frac{^n\Omega}{4}\right) + 2\,^nC^\kappa_{\ \mu\lambda\rho}\,^nR^\lambda_{\ \kappa}\right] \tag{7.1.7}$$

where $C$ is the Weyl conformal tensor.

With these definitions the equation (1.2.8) may be identically written in terms of the Einstein tensor $G$ [9]

$$\frac{{}^nR}{3} G_{\rho\mu} + 8\pi \left( {}^g\Theta_{\rho\mu} + {}^t\Theta_{\rho\mu} + {}^c\Theta_{\rho\mu} \right) = 8\pi \, {}^j\Theta_{\rho\mu} \quad . \tag{7.1.8}$$

where ${}^nR$ is the nonriemannian curvature scalar. We have then a generalized Einstein equation with geometric stress energy tensors. Nevertheless, as in (1.2.8), the energy momentum tensor ${}^j\Theta$ of the matter current is equivalent to the *total* of geometric *field* energy momentum contributions including the Einstein tensor.

If ${}^nR$ is nonzero we may write formally this Einstein equation,

$$G_{\rho\mu} = 8\pi \frac{3}{{}^nR} \left( {}^j\Theta_{\rho\mu} - {}^c\Theta_{\rho\mu} - {}^g\Theta_{\rho\mu} - {}^t\Theta_{\rho\mu} \right) \equiv 8\pi \frac{3}{{}^nR} \Theta_{\rho\mu} \equiv 8\pi G T_{\rho\mu} \quad , \tag{7.1.9}$$

and relate two stress energy tensors: the geometrical $\Theta$ and the classical $T$. The geometric energy momentum tensor has terms which represent the energy and motion of matter and interaction potentials in a similar way to known physical situations. Nevertheless, it is possible that the new geometric terms included in the energy tensor $\Theta$ may be related to the so called dark matter and energy. In certain phenomenological macroscopic situations it is also possible that this tensor approaches only a combination of the tensors normally used in astrophysics. In any case the fundamental difference rests in the presence of ${}^nR$ in rather than the gravitational constant $G$ in the equation, caused by its quadratic structure.

It should be kept in mind that, as we said before, that this equation should be considered an energy momentum balance equation rather than the proper field equation. Nevertheless, the conservation of the tensor $G$ with respect to the induced Levi-Civita connection in the bundle $TM$ implies the conservation of a tensor defined by

$$\nabla_\rho \, G^{\rho\mu} = 8\pi \, \nabla_\rho \left( \frac{3\Theta^{\rho\mu}}{{}^nR} \right) = 0 \quad . \tag{7.1.10}$$

There should be compatibility of the resultant equations with those obtained from the conservation of the current $J$, which determine the equations of motion, [9]. If the stress energy tensor is decomposed in terms of a multipole expansion we find the usual equations of motion [7, 132]: geodesic equation for a monopole, equations for a spinning particle and other multipole equations of motion.

For the case of a pure metric gravitation theory there are zero coset fields and zero torsion. We have then

$$G_{\rho\mu} = 8\pi \frac{3}{R} \left( {}^j\Theta_{\rho\mu} - {}^g\Theta_{\rho\mu} \right) \equiv 8\pi \frac{3}{R} \, {}^m\Theta_{\rho\mu} \tag{7.1.11}$$

where we have defined the matter current energy momentum tensor ${}^m\Theta$.

When we consider the external field problem, that is, space-time regions where there is no matter current, the gravitational part of the field equations for $J=0$ are similar to those of Yang's gravitational theory. Only ${}^g\Theta$ remains and we get back the Stephenson-Yang equation for vacuum, All vacuum solutions of Einstein's equation are solutions of these equations. In particular the Schwarzschild metric is a solution and, therefore, the newtonian motion under a $1/r$ gravitational potential is obtained as a limit of the geodesic motion under the proposed equations. Nevertheless, in the interior field problem there are differences between the unified physical geometry and Einstein's theory.

## 7.1.1. Newtonian *G* and the Schwarzschild Mass.

It is usual to assume, as a newtonian approximation of the interior problem, that the characteristic parameters of a newtonian solution of Einstein's equation $v/c \ll 1$, $\varphi \ll 1$ are of the same order $\varepsilon^2$ [133] in a small dimensionless parameter $\varepsilon$. This small parameter $\varepsilon$ may be related to the orthonormal tetrad $u$, thus characterizing the propagation of gravitational disturbances. The geometric formulation of Newton's gravitation is the Newton-Cartan theory [134]. The newtonian limit of Einstein type theories of gravitation was previously discussed [9]. The main mathematical difficulty in obtaining the limit is that the geometric Newton gravitation does not provide a metric but relies on a nonriemannian affine connection, a tensor of valence (2,0) and rank 3 (singular metric) and a scalar time function. There [9] it is shown that, in the limit $\varepsilon \to 0$, the metric becomes singular in $\varepsilon$. Nevertheless the connection remains regular in the limit and defines a newtonian affine connection not related to a metric. Since we have taken the connection as the fundamental representation of gravitation, the gravitational limit may be defined appropriately. The corresponding newtonian curvature tensor is the limit of the Riemann tensor. The projection of this tensor on the tridimensional time hypersurfaces $t$ defines a Riemann tensor on them, which is not necessarily flat.

Nevertheless, standard newtonian assumptions on the stress energy tensor determine that the tridimensional Newton space is flat. In this case the nonvanishing components ${}^0\Gamma_{00}{}^a$ of the limit connection would give the only nonvanishing components of the curvature tensor determining Poisson's equation.

In the geometric theory, the newtonian limit $\varepsilon \to 0$ should be obtained from eq. (7.1.9) The curvature scalar $R$ may become singular in the limit due to the singularity in the metric. It is possible to make assumptions on the geometry to avoid this singularity, but it is also possible to let the geometry be determined by the stress energy tensor, making assumptions of regularity for the latter tensor. We actually may do the same to handle the possible $R$ singularity present in the Einstein tensor $G$ adjoining it to $\Theta$. Equivalently it is also possible to move the term from the left side to the right side. We also assume that the contravariant stress energy tensor remains regular in the limit. As shown before [9] the curvature tensor and the curved time hypersurfaces become flat in the limit, as in Einstein's theory. The field equation for the corresponding tridimensional Riemann tensor ${}^0\overline{R}$

$$
{}^0\overline{R}_{mn} = \lim_{\varepsilon \to 0} \overline{R}_{mn} = \lim_{\varepsilon \to 0} \kappa \left( g_{m\alpha} g_{n\beta} T^{\alpha\beta} - \frac{1}{2} g_{mn} g_{\alpha\beta} T^{\alpha\beta} \right) = \lim_{\varepsilon \to 0} \mathcal{O}\left( \varepsilon^2 \right) = 0 \tag{7.1.12}
$$

determines that ${}^0\overline{R}$ is zero in the limit. Similarly the other space components of the Riemann tensor $R$ also vanish and there is a time vector $t^\mu$ orthogonal to the space-like hypersurfaces.

When establishing this limit, for any connection solution $\Gamma$, we may use the connection defined in eq. (1.3.3) by subtracting the tensorial potential form $\Lambda$ corresponding to the inert substratum solution,

$$
\bar{\Gamma} \equiv \Gamma - \Lambda = \Gamma - \left( \frac{-\mathcal{M}}{4} J \right) . \tag{7.1.13}
$$

The even so(3.1) curvature component may be split into the substratum part ${}^s\Omega$, determined in terms of the new metric, the Riemann tensor of the new connection and a complementary nonriemannian part ${}^n\Omega$ determined only by the total odd connection [9],

Einstein's equation (7.1.9) may thus be written as

$$
G_{\rho\mu} = 8\pi \frac{3}{{}^nR} \Theta_{\rho\mu} = 8\pi \frac{3}{3\mathcal{M}^2 + {}^n\widehat{\Omega} + \widehat{R}} \Theta_{\rho\mu} \equiv 8\pi G T_{\rho\mu} \tag{7.1.14}
$$

The only nonriemannian curvature contribution due to the substratum is through the scalar ${}^nR$. All other contributions of the conformally flat curvature cancel themselves.

Since Newton's theory is a pure gravitational theory, we assume that the new connection, as we approach the limit, is strictly gravitational. It is necessary to make the previous assumptions of regularity of the matter stress energy tensor. When we take the newtonian limit of the generalized Einstein equation (7.1.9), the three dimensional space curvature vanishes [9]. Then the only surviving component of the Riemann tensor satisfies the equation

$$
\widehat{R}_{00} = \frac{8\pi}{2} \lim_{\varepsilon \to 0} \left( \frac{3\Theta_{\hat{0}\hat{0}}}{{}^nR} \right) = 4\pi \lim_{\varepsilon \to 0} \left( \frac{3\Theta_{\hat{0}\hat{0}}}{{}^s\Omega + {}^n\widehat{\Omega} + \widehat{R}} \right) . \tag{7.1.15}
$$

If in the newtonian limit we simultaneously approach the inert substratum solution, $R$ and ${}^n\Omega$ are negligible, in the limit, with respect to the constant ${}^s\Omega$. Taking in consideration the limit of ${}^nR$ the field equation in the newtonian limit may then be written in the following form:

$$
\partial_a \partial^a \varphi = 4\pi \lim_{\varepsilon \to 0} \left( \frac{\Theta_{\hat{0}\hat{0}}}{\mathcal{M}^2} \right) = \frac{4\pi}{\mathcal{M}^2} \lim_{\varepsilon \to 0} \Theta_{00} = 4\pi G \rho \tag{7.1.16}
$$

and the gravitational field equation becomes Poisson's equation.

It is possible to equate $\mathcal{M}^{-2}$ to the value of the gravitational constant $G$ in eq. (7.1.16) and $m$ to the value of a fundamental particle inertial mass without contradictions. Physically we consider that the inert substratum provides two mass scales conceptually related by the coupling constant $\alpha$: in the defining matrix representation it determines the gravitational parameter $G$ that characterizes a macroscopic mass scale and in the induced representations it determines a fundamental particle mass $m$ that characterizes a microscopic mass scale.

We also have shown [9] that for a pure gravitational relativistic spherical solution, the Schwarzschild mass $\mathcal{M}$ is the integral of the energy density using this ${}^nR/3$ coupling,

$$
\int_0^r dr\, 4\pi r^2 \left( \frac{3\Theta_{\hat{0}\hat{0}}}{{}^nR} \right) = \int_0^r \frac{dr}{2} \frac{d}{dr} \left( r\left( 1 - e^{-2\varphi} \right) \right) = \frac{r}{2}\left( 1 - e^{-2\varphi} \right) \equiv G\mathcal{M} . \tag{7.1.17}
$$

This mass $\mathcal{M}$ determines the Schwarzschild geometry geodesics and the particle motion without explicit knowledge of a parameter $G$. In principle, similar relations should exist for other solutions. The density expressed in equation (7.1.16) is in agreement with the definition of this total mass $\mathcal{M}$ In other words the general density of matter, corresponds to the source of Poisson's equation in Newton´s theory.

Therefore, the inert substratum interaction energy density is related to mass in different contexts. In particular the energy determines the following: 1- the Dirac particle mass parameter $m$, in corresponding representations; 2- the Schwarzschild metric gravitational mass parameter $\mathcal{M}$; 3- the gravitational parameter $G$. The concept of macroscopic active gravitational mass is consistent with the concept of microscopic inertial mass.

# 7.2. Dark Matter Effects.

The replacement of the inverse of the newtonian gravitational constant $G$ by the scalar field parameter ${}^{n}R/3$ modifies the Einstein equation by a factor that may be interpreted as an "apparent" increment (or decrement) of an effective energy momentum tensor. The Einstein tensor may be written

$$G_{\rho\mu} = \frac{8\pi\Theta_{\rho\mu}}{\mathcal{M}^2 + \left({}^{n}\widehat{\Omega} + \widehat{R}\right)/3} = 8\pi G \frac{\Theta_{\rho\mu}}{1 + G\left({}^{n}\widehat{\Omega} + \widehat{R}\right)/3} \quad . \tag{7.2.1}$$

In general solutions, as shown in equation (7.1.9), the even curvature scalar ${}^{n}R$ plays the role of a newtonian gravitational coupling. In solutions near the substratum solution the scalar variable ${}^{n}R/3$ should have an inverse value near the constant $G$, which would be the exact value in the newtonian limit.This variation may be responsible for the detection of dark matter and energy. In particular, the Friedmann cosmological metric

$$d\tau^2 = dt^2 - a^2\left(t\right)\left(d\chi^2 + \Sigma\left(d\theta^2 + \sin^2\theta d\phi^2\right)\right) \quad \begin{cases} k=1 & \Sigma = \sin\chi \\ k=0 & \Sigma = \chi \\ k=-1 & \Sigma = \sinh\chi \end{cases} \tag{7.2.2}$$

would correspond to an equation for the Einstein tensor which may be written as follows, using a well known result [135] for its scalar curvature $R$ and neglecting $\Omega$,

$$G_{\rho\mu} = 8\pi G \frac{\Theta_{\rho\mu}}{1 - 2G\left(\dfrac{\dot{a}^2 + k}{a^2} + \dfrac{\ddot{a}}{a}\right)} \quad . \tag{7.2.3}$$

Of course, the new solutions for the radius $a$ of the universe, using the different known energy momentum tensors, will differ from the presently known solutions. This geometric equation affects the theoretical motion of galaxies and may also be interpreted as the existence of an apparent effective stress energy tensor which includes contributions from "dark matter or dark energy",

$$T_{\rho\mu} = \frac{\Theta_{\rho\mu}}{1 + GR/3} = \frac{\Theta_{\rho\mu}}{1 - 2G\left(\dfrac{\dot{a}^2 + k}{a^2} + \dfrac{\ddot{a}}{a}\right)} \quad . \tag{7.2.4}$$

In the newtonian limit the geometric and classic densities, $\Theta_{\hat{0}\hat{0}}$ and $\rho$ are related by the inert substratum constant $\mathcal{M}$ which determines a gravitational constant $G$. The densities determine the active gravitational mass. In general solutions, as shown in equation (7.1.3), the even curvature scalar ${}^{n}R$ plays the role of a coupling function.

For example, assume that a body of mass $M$ produces a Friedmann metric with zero $k$ and no acceleration. Using eqs. (7.1.9) or (7.2.4) the newtonian velocity of a particle rotating under this field is predicted by the physical geometry to be

$$v^2 = \frac{MG}{r\left(1 - 2G\left(\dot{a}/a\right)^2\right)} = \frac{v_0^2}{1 - 2G\left(\dot{a}/a\right)^2} \quad , \tag{7.2.5}$$

where $v_0$ is the velocity of the particle if the Friemann metric velocity $\dot{a}$ were to be zero. This pure geometric effect that increases the apparent rotational velocity may be blamed on "dark matter or energy".

# 7.3. The Alpha Constant.

In section 7.1 we indicated tha the newtonian constant $G$ is related by $\mathcal{M}^{-2}$ to the only fundamental constant in the geometric theory which is the coupling constant $\alpha$ as discussed in section 3.3. Geometrically it is also known that this alpha constant or fine structure constant is essentially equal to an algebraic expression in terms of $\pi$ and entire numbers that is related to the quotient of the volume of certain groups [136, 137]. Here we determine this expression using the invariant measure defined by the substratum solution. Starting from the structure group $G$ of the unified geometric theory and using the geometric fact that space-time is a hyperbolic manifold modeled on the symmetric space $G/G_+$ [57, 138] as discussed in chapter 4, the invariant measure on this symmetric space $K$ is obtained from groups associated to the structure group $G$. The $\alpha$ expression (invariant measure) is transported to space-time using the fact that the fiber of the tangent space to space-time is the image of a minkowskian subspace of the geometric algebra.

The current $*J$ is a 3-form on space time $M$ valued in the Clifford algebra $A$. It is constructed starting from a vector field on a symmetric space $K$. This space is $G/G_+$ where $G$ is the simple group whose action produces the automorphisms of $A$ and $G_+$ is the even subgroup, relative to the orthonormal base of the algebra. The vector field is the image, under a Clifford injection $\kappa$ of a vector field on space-time $M$. This injection allows us to define $*J$ as the pullback form of a 3-form on $K$. The integration of this 3-form on a tridimensional boundary of a region $R$ in $K$ is equivalent to the integration of the pullback 3-form on a tridimensional boundary of the image of the region $R$ in the space time $M$. The forms are defined by the existence of a geometric invariant measure on $G/G_+$. The constant coefficient of this invariant measure may be calculated in the particular case where the fiber bundle is flat and the field equation reduces to the linear equation equivalent to electromagnetism. This relation defines a geometric interpretation for the coupling constant of the geometric unified theory: "The coupling constant is the constant coefficient of $*J$ introduced by the invariant measure on the symmetric space $G/G_+$".

## 7.3.1. Symmetric Space K.

As indicated before, the group $G$ is $SL(2,\mathcal{Q})$ and the even subgroup $G_+$ is $SL_1(2,\mathbb{C})$. The symmetric space $K$ is a noncompact real form of the complex symmetric space corresponding to the complex extension of the noncompact SU(2,2) and its quotients [9]. The corresponding series of symmetric spaces coincides with the series characterized by the homomorphic group SO(4,2). In particular we can identify the quotients with the same character, +4, in order to write the series of spaces in the following form,

$$R \equiv \frac{SO(4,2)}{SO(4)\times SO(2)} \approx \frac{SL(4,\mathbb{R})}{SL(2,\mathbb{C})\times SO(2)} \cong K \cong \cdots \approx \frac{SO(6)}{SO(4)\times SO(2)} \ . \tag{7.3.1}$$

These quotients include the two symmetric spaces of interest: the noncompact riemannian hermitian $R$ and the noncompact pseudoriemannian nonhermitian $K$ symmetric spaces. Since some of these groups and quotients are noncompact we shall use the normalized invariant measure $\mu_N$ calculated from a known measure, as usually done when working with noncompact groups. For compact groups the integral of the invariant measure over the group parameter space gives the group volume. In general, the normalized measure gives only the functional structure of the volume element, in other words, the invariant measure up to a multiplicative constant.

## 7.3.2. Wyler's Measure on the K Space.

In principle we may obtain the common coefficient of the measures of the symmetric spaces in the series using any corresponding group in the series. The more transparent calculation is using the group SO(5,2) as Wyler did. In what follows, we indicate the result of this calculation of the value of the constant coefficient of the measure on $Q^5$, Silov's boundary of $D^5$. This measure is obtained by constructing Poisson's measure invariant under general complex coordinate transformations on $D^5$ by the group SO(5,2) of analytic mappings of $D^5$ onto itself.

Under this construction the coefficient of Poisson's kernel over $Q^5$, which is the constant coefficient of the normalized measure $\mu_{N_0}$, defines the coefficient of the measure over $D^4$,

$$\alpha_g = \frac{\left(V\left(D^5\right)\right)^{\frac{1}{4}}}{V\left(Q^5\right)} \frac{V[I(4,2)]}{V[I(5,2)]} = \frac{\left(V\left(D^5\right)\right)^{\frac{1}{4}}\times V\left(SO(4)\right)}{V\left(Q^5\right)\times V\left(SO(5)\right)} \quad . \tag{7.3.2}$$

The indicated volumes are known [139]. The volume of the polydisc is

$$V\left(D^n\right) = \frac{\pi^n}{2^{n-1}n!} \tag{7.3.3}$$

and the volume of Silov's boundary is the inverse of the coefficient in the Poisson kernel,

$$V\left(Q^n\right) = \frac{2\pi^{\frac{n}{2}+1}}{\Gamma\left(\frac{n}{2}\right)} \quad . \tag{7.3.4}$$

In particular we have,

$$V\left(D^5\right) = \frac{\pi^5}{2^4 \times 5!} \quad , \tag{7.3.5}$$

$$V\left(Q^5\right) = \frac{2^3\pi^3}{3} \quad . \tag{7.3.6}$$

Substitution of theses expressions in equation (7.3.2) gives Wyler's coefficient of the induced invariant measure,

$$\alpha_g\left[Q^4\right] = \frac{3^2}{2^6\pi^5}\left(\frac{\pi^5}{2^4 \times 5!}\right)^{\frac{1}{4}} \quad . \tag{7.3.7}$$

## 7.3.3. Value of the Geometric Coefficient.

The use of a normalized measure makes sense and is necessary on a noncompact subspace of $M$. In order to calculate the coeficient of the invariant measure on this noncompact space we must restrict the measure $\mu_{Ng}$ to this subspace by integrating over a complementary compact subspace. The latter space should be the characteristic compact subspace of the harmonic functions on $M$. There is a contribution from the integration on this space to the value of the geometric coefficient on the complementary noncompact space. Since $\mu_{Ng}$ is defined on the cotangent space at a point $*TM_m$ and only depends on the properties of the symmetric fiber $K$ at $m$, this contribution should be independent of any solution of the field equation. Therefore, without loss of generality we may assume spherical symmetry and staticity conditions that allow a decomposition of space-time $M^4$ and its forms in two orthogonal subspaces, spatial compact spheres $S^2$ and the complementary noncompact space-time $M^2$ with their forms. The space $M^2$ locally has the null-cone structure of a relativistic bidimesional Minkowski space.

In order to find the complete coefficient we are free to restrict the problem to the special particular case of pure electromagnetism on a flat space-time $M$. All solutions of this restricted problem may be found as a sum of fundamental solutions that correspond to the Green's function for the electromagnetic field. The Green's function determines the field of a point source which always corresponds to a spherically symmetric static field relative to an observer at rest with the source. The spherically symmetric harmonic potential solutions are determined, using Poisson's integral formula, by its value on a boundary sphere. We see that the characteristic boundary space, where integration should be performed, is the sphere $S^2$ in $R^3$. Geometrically, in accordance with our theory, the form $\mu_{Ng}$ should be integrated over the image of a sphere, determined by the Clifford mapping $\kappa$, on the space $K$. Therefore, the characteristic space is $\kappa(S^2)$. Since $S^2$ is compact the measure provided by the volume element $\mu$ does not require normalization. The measure $\mu_{Ng}$ integrated over the characteristic space gives the volume of $\kappa(S^2)$ which is twice the volume of $S^2$ due to the 2-1 homomorphism between standard spinors and vectors determined by its homomorphic groups SU(2) and SO(3).

Hence the complete geometric coefficient $\alpha$ is obtained by multiplying eq. (7.3.7), which is the value of Wyler's coefficient calculated in the previous section, by $8\pi$,

$$\alpha = \frac{2^3\pi 3^2}{2^6\pi^5}\left(\frac{\pi^5}{2^4 \times 5!}\right)^{\frac{1}{4}} = \frac{9}{16\pi^3}\left(\frac{\pi}{120}\right)^{\frac{1}{4}} = \frac{1}{137.03608245} \approx \alpha_e \quad , \tag{7.3.8}$$

which is equal to the experimental physical value of the alpha constant $\alpha_e$.

# 7.4. Results.

We have obtained a generalized Einstein equation with geometric stress energy tensors as the source. In vacuum the known gravitational solutions with a cosmological constant are obtained, including the Schwarzschild solution.

The interior Schwarzchild problem defines the Schwarzchild mass. There is a nonlinear particular internal solution which may represent inertial properties. It is a hyperbolic symmetric space determuned by the field equation and a purely geometric energy momentum tensor. The constant curvature parameter $\mathcal{M}$ (geometric energy) of this hyperbolic solution, in the newtonian limit, defines Newton's gravitational constant $G$.

If the nonriemannian connection fields contribute to the scalar curvature parameter of a general solution, the parameter $G$ would be variable diminishing with the field intensity. This effect may be interpreted as the presence of dark matter or energy.

The only independent constant in the theory, the fine structure alpha coupling constant, is also determined geometrically from the volume of spaces related to the structure group.

# 8. Quantum Fields.

We may raise the following question: Can we also obtain quantum field theory from this geometry without recurring to classical or quantum mechanics? This question is in line with the ideal aspiration indicated by Schrödinger [140].

In quantum field theory the fields are considered operators and form an algebra. It appears possible that this algebra arises from geometrical properties of sections in the principal bundle, which are properties related to group elements. Here we discuss this question, which is interesting since the group is associated to a Clifford algebra with an anticommuting algebraic structure.

The postulates of Newton's classical mechanics [141] are based on the concept of point particles and the free motion along straight lines of euclidian geometry. A classical field is usually seen as a mechanical system where classical mechanics is applied. With the advent of general relativity and gauge theories, there should be a recognition that the geometry of the physical world is not as simple as that provided by classical greek geometry. The idea of a physical geometry determined by the evolution or flow of mass and energy [142, 143] is attractive as a criterion for setting the fundamental laws of nature.

It was natural when quantum mechanics was born [144, 145] to base its development on classical mechanics and euclidian geometry. Postulates disclosed the difficulties with the simultaneous measurement of position and velocity of a point particle. Another approach, with hindsight, may be the realization that a point is not an appropriate geometrical element on which to superpose physical postulates. The concepts of fields and currents are close to modern geometrical ideas and points in the direction of establishing the physical postulates on a general geometry, away from prejudices introduced by classical geometry and classical mechanics.

In consequence, we discuss a generalization of field theory, directly to geometry, by-passing the intermediate step of mechanics. Why should we introduce these intermediate concepts which history of physics proved need revision in relativistic and quantum mechanics? Mechanics may be seen, rather than as a fundamental theory, as a simplification when the evolution of matter can be approximated as the motion of a point particle.

If we take into consideration the geometrical structure of the principal fiber bundle $E$ and the affine fiber bundle $W$, related to the algebraic structure of their fibers, a process of variations of the equations of the theory leads to an interpretation of the extended Jacobi fields as quantum operators. It is possible to define a bracket operation which becomes the commutator for the Jacobi fields associated to the connection or potential and becomes the anticommutator or commutator for those associated to the frame. This bracket operation leads the quantization relations of quantum field theory for bosonic interaction fields and fermionic matter fields.

We discuss, in particular, whether the theory gives quantum electrodynamics (QED) including its probabilistic interpretation. A basic feature of the proposed theory is nonlinearity. A solution cannot be obtained by the addition of two or more solutions and therefore it is not possible to build exact solutions from small subsystems. Nevertheless, it is possible to study its local linearized equations which represent excitations that evolve approximately under the influence of effects inherited from the nonlinear equation.

If we represent particles as excitations, then an interaction between two particles corresponds to the interaction of two of these geometric excitations, generated by these geometric Jacobi operators. The geometric coupling is between the connection or potential and the frame. The interaction among electrons (frame excitations) is mediated by photons (potential excitations).

## 8.1. Linearization of Fields.

Both the frame and the potential are determined by the equations, in terms of an orthonormal frame $u$ and an orthonormal set $\kappa$ that generates the geometric algebra [146] associated to space-time.

The technique necessary for attacking this problem is known as global nonlinear analysis [147]. If we introduce the bundle of connections $W$, a connection may be taken as a section of this affine bundle [9, 56]. When we consider an equation relating the connection or potential and the frame, we are dealing with differentiable manifolds of sections of the fiber bundles $E$, $W$ and nonlinear differential operators which define differentiable maps between these manifolds of sections.

Generally, it is accepted that the process of quantization requires the existence of a classical mechanical theory which is quantized by some fundamental rules. Instead, we think of the physically related geometry giving rise directly to fields, currents and in an approximate way to both classical and quantum theories.

We make the conjecture that the process of field quantization is the technique of replacing the nonlinear problem, just indicated, by a linear problem obtained by variations of the nonlinear maps, reducing Banach differentiable manifolds of sections to Banach linear spaces and the nonlinear differentiable maps between manifolds of sections, to linear maps between Banach spaces.

From a geometrical point of view, this means working at the tangent space of these section manifolds at some particular "point" (section). In order to give a rigorous definition to the concept of excitations we use the geometric version of the calculus of variations. The variation of a section leads to the geometric concept of the "jet" of a section. This term may physically be interpreted, in a loose manner, as a variation or excitation of potential and frame currents.

Certain manifolds, convenient to work with bundle sections, are the jet bundles of order $k$, indicated by $J^k E$, which are, essentially, manifolds of sections equal at a point modulus derivatives [9, 147] of higher order $k+1$. Of interest is the manifold of solutions $\gamma$, which is the submanifold of all the sections that satisfy a given nonlinear differential equation. Sections obtained from a solution by the action of the structure group are equivalent solutions. The quotient of $\gamma$ by this equivalence relation are the physical solutions.

If we have a variational problem, its critical sections (solutions) may be characterized geometrically as follows: A section is critical if and only if the Euler-Lagrange form is zero on the 1-jet prolongation [9] $js$ of the section $s$,

$$\left( D\Lambda + f\eta \right)_{js} = 0 \quad . \tag{8.1.1}$$

The set of all the critical sections forms the differentiable manifold $\gamma$ of solutions of the variational problem. On this manifold there are vector fields whose flow generates the space of solutions. These fields are sections of the tangent space $T\gamma$. Instead of studying $\gamma$ it is possible to study the linear spaces $T\gamma_s$. We may introduce in $T\gamma$ a vector $\Psi$ representing a solution of the associated varied equations. A "quantum operator" field may be related to a Jacobi field on the bundles in consideration.

Really a solution to the equations corresponds to a mapping between two such manifolds of solutions: one $\gamma_E$, corresponding to the bundle $E$, representing a matter frame solution and the other $\gamma_W$ corresponding to the bundle $W$, representing the interaction connection or potential solution.

## 8.2. Frame Solutions.

For the manifold of frame solutions $\gamma_E$ [56] we shall consider its tangent $T\gamma$. We define a Jacobi vector field as a section $V_s$, of $s*T^*E$, induced by the section $s$ from the vertical sub-bundle of $TE$, such that the corresponding 1-jet prolongation satisfies

$$\mathcal{L}_{jV} \left( D\Lambda + f\eta \right) = 0 \quad . \tag{8.2.1}$$

Here $\mathcal{L}$ indicates the Lie derivative. The space of all the Jacobi vector fields forms the tangent space of $\gamma$ at a given section, denoted by $T\gamma_s$. The last equation is a linearization of the nonlinear field eq. (8.1.1) which is eq.(1.2.2) in another form. A Jacobi field may be seen as the tangent vector to a differentiable curve of solutions $s_s$ of the nonlinear equations at a given solution $s$. When we have such a curve, we may consider it to be the integral curve of an extended Jacobi vector field $V$ which takes the values $V_s$, at each point of the curve.

The fiber of the bundle $E$ is the group $SL(2,\mathcal{Q})$. The Lie algebra $A$ of this group is enveloped by a geometric Clifford algebra with the natural product of matrices of the orthonormal subset of the Clifford algebra. The elements of $A$ may be expressed as fourth degree polynomials in terms of the orthonormal subset of the Clifford algebra.

When variations are taken, as usual, in terms of the fiber coordinates, the algebraic properties of the fiber are not brought to full use in the theory. In order to extract the additional information contained in the fiber which, apart from being a manifold, is related to algebra $A$, we note that in many cases we have to work with the fiber representative by the bundle homeomorphisms. We are dealing then with elements of $G$.

The vertical spaces of $s*T^*E$ are homeomorphic to the vertical spaces of $TG$. This means that the fiber of $s*T^*E$ is $A$, taken as a vector space. The association of the algebraic structure of $A$ to the vertical spaces of $s*T^*E$ depends on the image of the section $s$ in $G$ given by the principal bundle homeomorphisms.

When an observer frame is chosen, the local homeomorphism is fixed and the algebraic structure of $A$ may be assigned to the vertical spaces. A variation of the observer $h$ produces an effect equivalent to a variation of the section $s$, and should be taken into consideration. The total physical variation is due to a variation of the section and/or a variation of the observer in the composition mapping $hos$, defined on a local chart $U$. A variation of $hos$ is represented by a variation vector, a generalized Jacobi vector, valued in the Clifford algebra.

We are working with a double algebraic structure, one related to the algebra $A$, of the fiber of $TE$ and the other to the Lie algebra of vector fields $\mathcal{F}$ on the jet bundle $JE$. This allows us to assign a canonical algebraic structure to the Jacobi fields, which should be defined in terms of these natural structures in each of the related spaces.

For instance, if we use both algebraic structures and calculate the following Lie bracket of vectors,

$$\left[ x\kappa^0 \partial_\mu, y\kappa^1 \partial_\nu \right] \quad , \tag{8.2.2}$$

where $\mu$, $\nu$ are coordinates on the jet bundles and $x$, $y$ the components, the result is not a vector because of the anticommuting properties of the orthonormal set. Nevertheless, if we calculate the anticommutator, the result is a vector.

In particular, it is also known [148] that the Clifford algebra, taken as a graded vector space, is isomorphic to the exterior algebra of the associated tangent space, which in our case is *TM. There are two canonical products, called the exterior and the interior products, that may be defined between any two monomials in the algebra using the Clifford product. For any element $\kappa_\alpha$ of the orthonormal set the product of a monomial $a$ of degree $p$ by $\kappa_\alpha$ gives a mapping that may be decomposed in the exterior product $\kappa_\alpha \wedge a$ and the interior product $\kappa_\alpha \vee a$.

Due to the associative property of the Clifford product we may extend this decomposition to products of monomials. We may define the grade of the product $\alpha\beta$, indicated by $\mathrm{gr}(\alpha\beta)$, as the number of interior products in the Clifford product. The grade is equal to the number of common elements in the monomials.

For example, $(\kappa_0 \kappa_1)(\kappa_2 \kappa_0 \kappa_3)$ is of grade one. This decomposition may be applied to the product of any two elements of $A$. It is clear that the maximum grade is the number of orthonormal elements and that the product of grade zero is the exterior product.

The use of the exterior product instead of the Clifford product turns the algebra $A$ into a Grassmann algebra isomorphic to the exterior algebra of differential forms on the tangent spaces. This fact leads us to look for gradation of the Lie algebra structure [149].

The structure group of the theory was chosen as the simple group of inner automorphisms of the Clifford algebra. In this sense, the group, its Lie algebra and their respective products arise from the Clifford algebra. In fact the Lie bracket is equal to the commutator of Clifford products. To avoid an unnecessary factor of 2 we may define a Lie product as the antisymmetrized product (1/2 of the commutator).

We may consider that the elements of $A$ also satisfy the postulates to form a ring, in the same manner as the complex numbers may be considered as an algebra over $\mathbb{R}$ or as a field. In this manner, the elements of $A$ play the role of generalized numbers, the Clifford numbers. The candidates for the ring product are naturally the geometric Clifford product, the exterior Grassmann product, the interior product and the Lie product. As mentioned, all these products arise from the Clifford product. In fact, for any two monomials $\alpha, \beta \in A$, all the other products $\alpha \cdot \beta$ are either zero or equal to the Clifford product $\alpha\beta$. The geometric Clifford product is more general and fundamental, and the others may be obtained as restrictions of the Clifford product. Besides, the Lie product is not associative and may need generalizations of the algebraic structures. On the other hand, the fiber bundle of frames, $E$, is a principal bundle and its vertical tangent bundle $TE$ has for fiber a Lie algebra structure inherited from the group, it would be more natural that the chosen product be closed in the algebra so that the result be also valued in the Lie algebra. For the moment, we take the more general geometric Clifford product as the product of the ring structure of $A$ and we shall specialize to other products when needed.

It is convenient, then, to work with the universal enveloping associative algebra of the Lie algebra of vector fields $\mathcal{F}$. In this manner we have an associative product defined and we may represent the brackets as commutators of the elements. The vertical vectors on $JE$ may be taken as a module over the ring $A$. Using the canonical algebraic structure we may define a bracket operation on the elements of the product $A \times \mathcal{F}$. For any monomial $\alpha, \beta, \gamma \in A$ and $V, W, Z \in \mathcal{F}$, define

$$\{\alpha V, \beta W\} = \alpha \bullet \beta \times [V, W] = \gamma\, Z \quad, \tag{8.2.3}$$

turning the module into an algebra $\mathcal{A}$. This bracket gives the commutator or anticommutator depending on the arguments

$$\{\alpha V, \beta W\} = -(-1)^{|\alpha\beta|} \{\beta W, \alpha V\} \quad, \tag{8.2.4}$$

where $|\alpha\beta|$ is the gradation of the product $\alpha\beta$ or the bracket, equal to the number of permutations given by

$$|\alpha\beta| = |\alpha||\beta| - \mathrm{gr}(\alpha\beta) \quad, \tag{8.2.5}$$

in terms of the degree $|\alpha|$ of the monomials in $A$ and the grade of the product $\alpha\beta$.

The previous formulas are valid in general if we restrict the Clifford product to the exterior or the interior products taking the proper number of permutations for each case. For example in the case of Grassmann product, $\mathrm{gr}(\alpha\beta)$ is always zero. If we restrict to the Lie product, the bracket is always symmetric in eq. (8.2.3) and in the graded Jacobi identity.

# 8.3. Potential Solutions.

For the manifold of potential solutions $\gamma_w$, [56] we may proceed similarly, taking in consideration that the bundle of connections $W$ has a different geometrical structure than the bundle of frames $E$. A connection on the principal bundle may be defined by an splitting of a short sequence of vector bundles [150]. The connection or potential may be identified with a section of the bundle of connections $\pi: W \to M$ [9]. It may be seen that each point $w_m$ of the vertical space $W_m$ of the fiber bundle $W$ corresponds to a vector space complement of $AdE$ in $T_G E$. It is known [146] that the space of linear complements of a vector subspace in a vector space has a natural affine structure. Therefore the fiber of the bundle $W$ is an affine space with linear part $L(T_G E \,/\, AdE, AdE) \approx L(TM, AdE)$.

We define Jacobi vector fields $V_s$, as sections of $s^*T^*W$ and associate quantum operators to the prolongation of the extended Jacobi vectors. A difference arises, in this case, because the algebraic structure of the fiber of $W$, which is not a principal bundle, differs from that of the fiber of $E$. A connection or potential in $E$ is defined by giving a section in $W$. This defines horizontal vector subspaces in $TE$. The fiber of the manifold of connections $W$ is the space of complements of the vertical space in $T_eE_e$. The fiber of $W$ is an affine space and we say that $W$ is an affine bundle.

The algebraic structure of the tangent space to the fiber of $W$ is then isomorphic to $TP_p$ and consequently isomorphic to $V$ with the usual operation of addition of vectors. Correspondingly the vertical vectors on $JW$ form a vector space over a commutative ring and the bracket defined in the previous section reduces to the Lie bracket, since no sign permutation arises under commutation.

# 8.4. Bracket as Derivation.

In both cases, frame and potential solutions, we have that the bracket is defined using the natural geometric product related to the algebraic structure of the fiber of the corresponding bundles $E$ and $W$.

It is clear that the bracket is a derivation,

$$\{\Xi, \Phi\Psi\} = \{\Xi, \Phi\}\Psi + (-1)^{|\Xi\Phi|}\Phi\{\Xi, \Psi\} \quad , \tag{8.4.1}$$

and we have, therefore, a generalized Lie derivative with respect to the prolongation of Jacobi fields given by

$$\mathcal{D}_{jV}jW = \{jV, jW\} \quad , \tag{8.4.2}$$

where the bracket is the anticommutator or the commutator depending on the number of permutations for frames and is always the commutator for connection or potentials. Physically this means that matter fields should be fermionic and that interaction fields should be bosonic. The quantum operators in quantum field theory may be identified with the prolongation of the extended Jacobi vector fields.

The meaning of these brackets, eq. (8.4.2), is equivalent to the postulate of quantum theory that gives the change in some quantum operator field $\Phi$ produced by the transformation generated by some other operator $\Psi$. In the present context, this equation is not a separate postulate, but rather it is just the result of taking the $\mathcal{D}$ derivative with respect to a direction tangent to a curve in the manifold of solutions $\gamma$ and it is due to the geometry of the bundles.

The $\mathcal{D}$ derivative may be generalized to tensorial forms valued over the ring $A$. These derivatives represent the variation of sections along some direction in $T\gamma$, which corresponds to a generator of some transformation on the jet bundle along a vertical direction.

The complete geometrical formulation of this physical variational problem may be carried by constructing the extension of the vertical vector space $T^eE_e$ to a module over the ring $A$. In a manner similar to the complexification of a real vector space we consider the dual $(T^eE_e)^*$ and define corresponding mappings. These mappings are elements of a right $A$-module which we designate by $^AT^eE_e$. If we form the union of these spaces over the manifold $E$ we get a fiber bundle $^AT^eE_e$ over $E$.

Further, we may construct the bundles $^AT^eJE_e$ over $JE$ and $s*s*T^eE$ over $M$. The standard geometric version of variations may be generalized to these physical variations by substituting appropriately these bundles for the bundles $T^eE$, $T^eJ$ and $s*s*T^eE$ respectively, using the $\mathcal{D}$ derivative and keeping track of the noncommutative products [9, 150].

# 8.5. Geometric Theory of Quantum Fields.

The operator bracket leads to the quantization relations of quantum field theory using Schwinger's action principle. The generator of a variation $F$ may be written in the jet bundle formalism by the appropriate term [9]. It may be seen that the elements entering in the expression are tensorial forms on the jet bundle that inherit the algebraic properties of the fiber and therefore have the properties of operators.

The generating function $F$ determines the variations as indicated in eq. (8.4.2). We shall write the generator $F$ in the equivalent standard expression [151] used in quantum theory obtaining

$$\delta\Psi(x) = \{\Psi, F\} = \left\{\psi(x), \int_\sigma \Pi^\mu \delta\Psi d\sigma_\mu\right\} \quad , \tag{8.5.1}$$

for the field operator $\Psi$ variations. In this expression we recognize that the elements are quantum operators, as defined above, which obey the defined bracket operation. If we express the bracket in this relation as in eq. (8.4.1), we get

$$\delta\Psi(x) = \int_\sigma \{\Psi(x), \Pi^\mu(y)\}\delta\Psi(y)d\sigma_\mu \pm \int_\sigma \Pi^\mu(y)\{\Psi(x), \delta\Psi(y)\}d\sigma_\mu \ .$$

(8.5.2)

It is clear that the resultant commutators for the potential operators and anticommutators or commutators for the frame operators are none other than the quantization relations in quantum field theory for boson fields and fermion fields.

The last equation implies that

$$\left\{ \Psi\left(x\right),\delta\Psi\left(y\right)\right\} = 0 \quad , \tag{8.5.3}$$

$$\left\{ \Psi\left(x\right),\Pi^{\mu}\left(y\right)\right\} = -\delta^{\mu}\left(x,y\right) \quad , \tag{8.5.4}$$

where the bracket is interpreted as the commutator or anticommutator depending on the gradation of the bracket according to the type of field.

From this formulation, the geometrical reasons for the existence of both bosonic and fermionic fields also become clear. Another advantage of this geometric formulation is that the operator nature of the fields may be explicitly given in terms of the tangent vector operators on the space of solutions and the Clifford algebra matrices.

A tangent vector may be considered as an infinitesimal action on functions on a manifold. In particular for $\gamma$, the manifold of sections that are solutions, the quantum operator may be considered to act on a solution section, called the substratum or background section, producing solution perturbations, called quantum excitations. The substratum or background section takes the place of the vacuum in conventional quantum field theory.

If the manifold of sections is taken as a Hilbert manifold its tangent spaces are Hilbert linear spaces. This means that the quantum operators, which act on tangent spaces, would act on Hilbert spaces as usually assumed in quantum theory.

If the Hilbert manifolds in question admit local harmonic sections we may introduce fundamental harmonic excitations. Any arbitrary excitation may be linearly decomposed in terms of these harmonic excitations. The eqs (8.5.3, 8.5.4) determine commutation rules for the harmonic excitation amplitudes which correspond to the creation, annihilation and particle number operators of harmonic field excitations at each point, with the same properties of those of the quantum harmonic oscillator. The energy of the harmonic field excitations may be defined in the usual manner leading to the concept of a quantum particle.

Nevertheless, if this geometric model is taken seriously, field (second) quantization reduces to a technique for calculating field excitations or perturbations to exact solutions of theoretical geometrical equations. There may be other techniques for calculating these perturbations. In fact, it is known that quantum results may be obtained without quantization of the electromagnetic field. See, for example, an alternate technique to QED [152] where the self-field is taken as fundamental.

In general relativity, the self-field reaction terms do not appear as separate features in the nonlinear equations but they appear in the linearized equations obtained using a perturbative technique. Once the geometrical nonphenomenological structure of the source term $J$ is known, the exact equations of motion for the fields describing matter, including particles, would be the conservation law for $J$. This relation, being an integrability condition on the field equations, includes all self-reaction terms of the matter on itself. *There should be no worries about infinities produced by self reaction terms.* A physical system would be represented by matter fields and interaction fields which are solutions to the set of simultaneous equations. When a perturbation is performed on the equations, for example to obtain linearity of the equations, the splitting of the equations into equations of different order brings in the concepts of field produced by the source, force produced by the field and therefore self-reaction terms. We should not look at self-reaction as a fundamental feature but, rather, as an indication of the need to use nonlinear equations.

## 8.5.1. Product of Jacobi Operators in QED.

The product $\alpha\beta$ in the ring $A$ has been taken, in general, as the Clifford product [9]. Since the fiber bundle of frames $E$ is a principal bundle and its tangent bundle $TE$ has for fiber a Lie algebra structure inherited from the group, it is necessary that the chosen product be closed in the algebra so that the result be also valued in the Lie algebra. Geometrically we should specialize the ring product, as indicated before, to be the Lie product. Then the product $\alpha\beta$ is zero whenever its gradation is even. The bracket survives only when the commutator $\alpha\beta$ is nonzero because its gradation is odd and corresponds to the anticommutator. With this product, the bracket defined before is the anticommutator for the matter Jacobi operator fields.

The fiber bundle of connections, $W$, is an affine bundle and the ring product associated to the fiber of $TW$ is commutative. Therefore, the bracket is the commutator for the connection or potential Jacobi operator fields.

As discussed before, the Jacobi vectors $V_s$ represent fluctuations of sections $s$ of the fiber bundle $E$. The corresponding jet prolongation $_jV$ of an extension $V$ of $V_s$ is a vector field on $JE$ which acts as an operator on functions on $JE$. A Jacobi field, as a vertical vector on the section $s(M)$ may be considered as a displacement of $s$. Similarly, its jet prolongation may be considered as a displacement of $^js$, in other words a variation,

$$\frac{d^{j}s}{d\lambda} = {}_jV\left(^js\right) \quad . \tag{8.5.5}$$

This linear action of the Jacobi fields on the substratum sections allow us to associate them with quantum theory objects.

We may consider the inert substratum section as the vacuum state $|0\rangle$ of the physical system and the Jacobi fields as physical linear operators $\Psi$ that have a natural geometric action on the states.

A given physical matter section may be expressed in terms of reference frame sections. This reference frames are also physical systems. Hence they may evolve under the action of the same group that acts on the matter sections. This represents the known equivalence of the active and passive views of evolution. We may consider the physical system evolving in a fixed frame or, equivalently, we may consider the system fixed in an evolving reference frame.

We may choose the reference frame so that the substratum section does not evolve. In this condition the substratum section remains fixed and the Jacobi operators obey the first order linearized equations of motion. (Heisenberg picture).

The Jacobi vector fields and their jet prolongations transform under the adjoint representation of the group that transforms the sections. We may use these transformations to a new reference frame where the Jacobi operators do not evolve. In this condition, the time dependence of the Jacobi operators may be eliminated and the substratum sections obey equations of motion, indicating that the state is time dependent (Schrödinger picture).

# 8.6. Geometric Electrodynamics.

## 8.6.1. Free Particles and Currents.

Now we discuss, in particular, whether the theory gives quantum electrodynamics (QED) including its probabilistic interpretation [153]. It should be clear that an exact solution of QED is not possible with this linearized technique. The reason for this is the presence of the nonlinearity of the self interaction in the equations. We must consider only approximate solutions. The interaction between frame excitations is mediated by potential excitations. In QED the interaction is imposed on special fields called "free fields", the electronic field and the radiation field. Here we have to discuss which frame and potential sections correspond to free fields. A frame excitation representing a "free" electron must be chosen. Similarly, we must choose a potential excitation representing a "free" photon.

The equation of motion

$$\kappa^\mu \nabla_\mu e = 0 \ , \tag{8.6.1}$$

includes self interaction terms. It is not possible to set the substratum connection to zero because it will automatically eliminate the self energy and the mass parameter in accordance with our theory. We understand as "free" a frame excitation with the correct mass parameter determined by the substratum. The geometric approximation is then to assume that all self interaction effects, to first order, are due to the single mass parameter. The fluctuation equation is

$$\kappa^\mu \partial_\mu e = me + \text{ interac. } . \tag{8.6.2}$$

It is natural to consider a free frame field as one that satisfies the previous equation without the interaction term. A solution to this free equation provides a frame excitation section that can be used with our technique. The linearized fluctuation equations, with the interaction term, determine the evolution of the potential and frame excitation fields.

The field equation,

$$D^*A = 4\pi\alpha \ ^*J \ , \tag{8.6.3}$$

when the potential has only a $\kappa^0$ electromagnetic component reduces to pure electrodynamics,

$$d^*dA = 4\pi\alpha \ ^*j \ . \tag{8.6.4}$$

It is natural to consider a free potential field as one that satisfies the previous equation with $j=0$. A solution to the field equation provides a substratum potential section that can be used with the proposed technique.

In order to get the standard QED from the geometric theory, we have to reduce the structure group to one of the 3 U(1) subgroups in the SU(2) electromagnetic sector. The corresponding u(1) component of the fluctuation of the generalized connection is the electromagnetic potential $A_\mu$. Similarly, an electromagnetic fluctuation of the matter section determines a fluctuation of the generalized current. This fluctuation must be generated by one of the three electromagnetic generators, for example $\kappa^5$. The standard QED electric current is the electromagnetic sector component of this current fluctuation,

$$\delta j = \tfrac{1}{4}\text{tr}\left(\kappa^1\kappa^2\kappa^3\delta J\right) = \tfrac{1}{4}\text{tr}\left(\kappa^1\kappa^2\kappa^3 \tfrac{1}{2}\left[J,\kappa^5\right]\right) = \tfrac{1}{4}\text{tr}\left(\kappa^5\kappa^1\kappa^2\kappa^3 J\right) = \tfrac{1}{4}\text{tr}\left(\overline{\kappa}^0 J\right) \ . \tag{8.6.5}$$

It was already shown that the standard electric current in quantum theory is related to the $\kappa^0$ component of the generalized current,

$$\tfrac{1}{4}\text{tr}\,\overline{\kappa}^0 J^\mu = \tfrac{1}{4}\text{tr}\left(\overline{\kappa}^0\tilde{e}\kappa^\mu e\right) = \overline{\Psi}\gamma^\mu\Psi \ , \tag{8.6.6}$$

which is the electric current for a particle with a charge equal to one quantum in the geometric units.

## 8.6.2. The Interaction Hamiltonian.

We may choose a reference frame so that both substratum sections, which include the corresponding self interactions of the two "free" systems, also include an observer dependent free motion for the two excitations. The two substratum sections are not solutions to the interacting system, they are free background solutions. This situation corresponds to an interaction picture in QED. By choosing an appropriate reference frame, the substratum sections (states) have the "free motion", including self interaction effects and the excitation Jacobi operators represent the dynamics of the interaction of the two systems, excluding the free self interactions.

The only invariant effect of the self interaction in the free motion of the fields is the mass parameter $m$. The remaining effects of the total physical interaction, associated with the fluctuation fields, correspond to an effective net interaction energy that may be written as

$$\tilde{e}\kappa^{\mu}eA_{\mu} - mI = H \quad . \tag{8.6.7}$$

The first part of $H$ is the total physical interaction. The meaning of the second part is that massive self energy is not included in the fluctuation and this method (or QED), by construction, *is not adequate for a calculation of bare mass*.

It is known that the lagrangian for the first variation of the Lagrange equations is the second variation of the lagrangian. For both free fields, the corresponding second variation of the lagrangian gives the free Maxwell-Dirac field lagrangian. Since the current variation has only a $\kappa^0$ component given by eq.(8.6.6) and the potential variation has $A_{\mu}$ for $\kappa^0$ component, the second variation or interaction hamiltonian gives, in terms of these interaction field Jacobi operators $\delta\Gamma$, $\delta J$,

$$H \equiv \delta^2 L = \tfrac{1}{4}\mathrm{tr}\left(\delta A \delta J\right) = \tfrac{1}{4}\mathrm{tr}\left(\kappa^0\kappa^0\right)\bar{\Psi}\gamma^{\mu}\Psi A_{\mu} = -\bar{\Psi}\gamma^{\mu}\Psi A_{\mu} \quad , \tag{8.6.8}$$

which is the standard interaction hamiltonian in QED.

We have obtained a framework of linear operators (the prolongations of Jacobi fields) acting on sections forming a Banach space (states) where the operators obey the bracket commutation relations, commutators for $A$ and anticommutators for $\Psi$. Furthermore, the geometric lagrangian of the theory reduces to a lagrangian in terms of operators which is the standard lagrangian for QED.

From this point we can proceed using the standard techniques, language and notation of QED, equivalent to the geometric techniques, for any calculation we wish to carry in this approximation of the geometric theory. The result of the calculation should be interpreted physically, in accordance with the geometric ideas. This is our next task, to obtain a statistical interpretation to fluctuations in the geometrical theory.

## 8.6.3. Statistical Interpretation.

The physical significance of the local sections in the nonlinear geometrical theory is the representation of the influence of the total universe of matter and interactions. The geometry of the theory, including the notion of excitations is determined by the large number of excitation sources in the universe. Within this geometric theory and its interpretation, it is not appropriate to consider the sections to be associated with a single particle or excitation. Rather, they are associated to the global nonlinear geometric effects of bulk matter and radiation.

This is not the customary situation in which physical theories are set. It is usual to postulate fundamental laws between elementary microscopic objects (particles). In order to translate the global universal geometry to the usual microscopic physics, since we associate particles to geometric excitations, we shall establish what we call a "many excitations microscopic regime", distinguished from the situation described in the previous paragraph.

Statistics enter in our approach in a manner different than the usual one, where fundamental microscopic physical laws are postulated between idealized elementary particles and statistical analysis enters because of the difficulties that arise when combining particles to form complex systems. Here, the fundamental laws are postulated geometrically among all matter and radiation in the system (the physical universe) and statistical analysis arises because of the difficulties and approximations in splitting the nonlinear system into elementary microscopic linear fluctuation subsystems. As a consequence of this holistic situation, the results associated to fundamental excitations must have a natural (classical) statistical character corresponding to that of quantum theory.

We consider that we may work in two different regimes of the geometrical theory. One is the geometrical holophysical regime, not representing particles, where there are exact nonlinear equations between the local frame sections, representing matter, and a potential section representing the interaction. The other is the many-excitations microscopic regime where we have linearized approximate equations between the variations of the frame and potential sections, representing particles and fields.

In the many-excitations regime, the nonlinear local effects of the interaction remain hidden in a substratum solution and are replaced by linear approximate local effects among the subsystems, seen as a collection of excitations on the substratum.

The number of excitations is naturally very large and the cross interactions among them preclude an exact treatment for a single excitation. Instead it is necessary to treat the excitations as one among a large ensemble and use classical statistical theory.

The geometric excitations form statistical ensembles of population density $n_i$. It is not possible to follow the evolution of any single one because of the previous arguments. It is absolutely necessary to use statistics in describing the evolution of the excitations. The situation is similar to that of chemical reactions or radiation, where atoms or particles are statistically created or annihilated.

There are adequate classical statistical techniques for describing these processes. The mathematical methods used in physical chemistry may be applied to the geometric excitations. In particular we may use the theory of irreversible thermodynamics [154] to calculate the rates of reaction between different geometric excitations. The process is described by the flux density $\mathcal{J}$ characterizing the flow of $n$ excitations between two systems or reaction rate.

It is also necessary to introduce a driving force function $\mathcal{F}$, which is called affinity and represents differences of thermodynamic intensive parameters. The use of these classical concepts is derived from fundamental statistical analysis. In case of equilibrium between two different subsystems, both the affinity and the flux are zero.

The identification of the affinity is done by considering the rate of production of entropy $s$. The affinity associated to any given excitation is given in terms is the energy $u$, the temperature $T$ and is the excitation potential $\mu$, similar to the chemical potential used to determine the statistics of chemical reactions.

The statistical flux is a function of the affinity and we see that the statistics of reaction of geometric excitations should depend on the classical geometric energy of the excitations. An excitation is an extensive parameter $x$ that has associated an intensive parameter $F$ thermodynamically related with the change of energy. At a point, the excitation is a function of time that may always be decomposed in harmonic functions. In general, the energy of harmonic oscillations depends on the excitation amplitude. Thus, *the probability of occurrence of a single reaction event*, implicit in the reaction rate determined by the excitation potential, *depends on the excitation amplitude*. This is the significance of the probabilistic interpretation of quantum fields.

In some cases, for linear markoffian systems (systems whose future is determined by their present and not their past), the flux is proportional to the affinity and the calculation of reaction rates is simplified, indicating explicitly the dependence of the flux on the difference of excitation potentials.

The radiation excitation must be a massless, long range, geometric connection potential which is a representation of the group $G=SL(2,\mathcal{Q})$ induced from the subgroup $P=Sp(2,\mathcal{Q})$. As indicated in section 2.1, this representation is characterized by the quantum number helicity determined by a representation of the isotropy subgroup $SO(2,1)$. The corresponding fundamental irreducible representation, which is the photon, must carry a quantum of angular momentum. The presence of helicity means that a photon is a harmonic oscillation of some frequency $\nu$. Then the Jacobi vector operator $\Psi$ may be decomposed in its irreducible photons using a Fourier series decomposition.

It should be clear that the *geometric* commutation or anticommutation relations of the Jacobi vector operators $\Psi$ imply similar relations for the fundamental oscillator amplitudes (Fourier components) $a$. Consequently the energy of these geometric oscillators is quantized, with the same results of the quantum theory. These amplitudes become creation or annihilation operators $a$. Under these conditions the number operator,

$$N = a^{\dagger}a \ , \tag{8.6.9}$$

has discrete eigenvalues $n$ which indicate the number of photons and determine the energy of the oscillators. The energy quanta are the eigenvalues

$$\varepsilon = \left(n + \frac{1}{2}\right)\nu \ , \tag{8.6.10}$$

in geometric units where Planck's constant $\hbar$ equals 1. The energy of the excitations has quanta of value equal to the frequency $\nu$.

In order to illustrate this statistical character of the linearized regime, we discussed [9] the wave corpuscular duality of light and matter within the concepts of the geometric theory. There are excitations that correspond to multiple particles and lead to Schrödinger's "entangled states", typically demonstrated in Young's double interference experiment as shown in figure 7. In the last decades there has been a revolution in the experimental preparation of multiparticle entanglements [155, 156, 157, 158, 159]. There are no exact equations for physical situations with a single excitation. The outcome of microscopic transition experiments depends on the excitation potentials of the systems. Any real physical situation in the microscopic linear regime of the theory requires the use of statistics for the total system.

A physical particle (f.e. a photon) is an excitation of the corresponding substratum field. In geometric terms, a Jacobi vector $\Psi$, associated to a variation of a section $e$ or $\omega$, represents the particle. As indicated in the previous section, it is not possible to follow the evolution of a single excitation $\Psi$. The classical statistical approach is to treat the radiation as a thermodynamic reservoir of excitations. This approach was used in blackbody radiation [160, 161, 162] and was the origin of Planck's quantum theory.

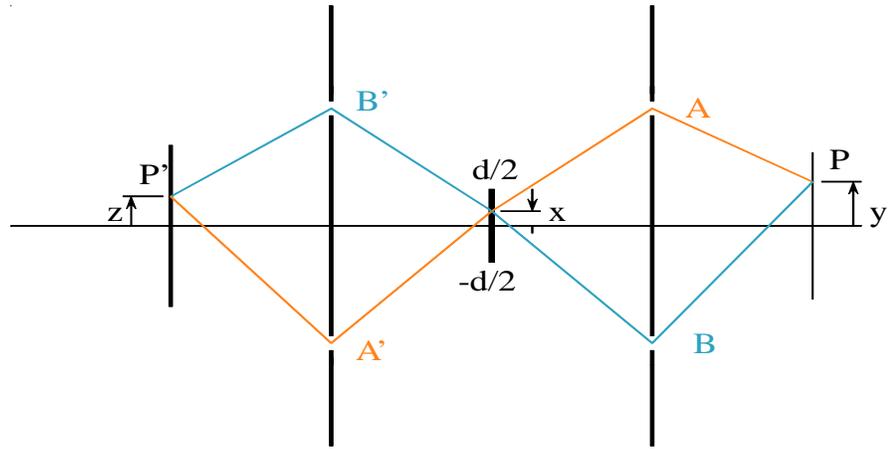

Young's double interferometer.

Figure 7

The amplitude of the Jacobi excitation, which is a product representation of the two excitations, is

$$\Psi \propto \frac{a}{d} \int_{-d/2}^{+d/2} dx \cos\left[\frac{\pi\theta}{\lambda}(x+y)\right] \cos\left[\frac{\pi\theta}{\lambda}(x+z)\right].$$ (8.6.11)

If $d$ is much smaller than $\lambda/\theta$ the integral gives the product of two Young's patterns as it should. If, on the contrary, $d$ is much larger than $\lambda/\theta$, this integral gives,

$$\Psi \propto \frac{a}{2} \cos\left(\frac{\pi\theta}{\lambda}(z-y)\right).$$ (8.6.12)

and the excitation potential, which gives the probability for the simultaneous detection of particles at P and P', is

$$\mu \propto \left[\cos\left(\frac{\pi\theta}{\lambda}(z-y)\right)\right]^2.$$ (8.6.13)

There are regions of high and low probabilities because of the interference of the excitation amplitudes.

In recent years, similar ideas lead to introduction of stochastic quantization [163, 164], which has been shown to be equivalent to path integral quantization. Our approach is different, for example we do not introduce evolution along a fictitious time direction as done in stochastic quantization. We rely on the existence of a global nonlinear geometry which makes a practical necessity the statistical treatment of the linearized equations describing the evolution of microscopic subsystems.

The requirement of statistics necessarily leads to the probability interpretation of quantum theory. If, within a particular experimental setup, we can physically distinguish between two excitation states, there is no room to apply statistics and there follows the absence of the interference pattern. This is the content of Feynman's statement about experimentally indistinguishable alternatives [165]. Fundamentally the quantum statistics are the classical statistics of section excitations in this physical geometry. The objections raised by Einstein [166, 167] to the probabilistic interpretation are resolved automatically because these statistics enter due to the lack of detailed knowledge of the states of many excitations.

## 8.7. Results.

The process of field quantization is the technique of replacing the nonlinear geometric problem by a linear problem obtained by variations of the nonlinear maps. The properties of the linear variations determine a bracket operation which becomes the commutator for the varied potential fields and also becomes the anticommutator or commutator for the varied matter fields. This bracket operation leads the quantization relations of quantum field theory for bosonic interaction field and fermionic matter fields. The substratum plays the role of the quantum vacuum.

Quantum electrodynamics is obtained in the corresponding limit. The nonlinear geometric equations apply to the total universe of matter and radiation. If we work with excitations, this implies the need to use statistical theory when considering the evolution of microscopic subsystems. Use of classical statistics, in particular techniques of irreversible thermodynamics, determines that the probability of absorption or emission of a geometric excitation is a function of the classical energy density. Emission and absorption of geometric excitations imply discrete changes of certain physical variables, but with a probability determined by its wave energy density. The geometric properties of the excitations are the germ of quantum physics.

Quantum Field Theory (QFT) should be used to calculate the perturbation corrections to the geometric bare values of the microscopic physical magnitudes. In this manner the proper results of particular experiments are determined. As an example, higher order corrections for the magnetic moments were calculated in section 5.2.

# Appendix

Introduce the real anticommuting matrices,

$$\rho = \begin{bmatrix} 0 & 1 \\ 1 & 0 \end{bmatrix} \quad , \tag{A.1}$$

$$\sigma = \begin{bmatrix} 1 & 0 \\ 0 & -1 \end{bmatrix} \quad . \tag{A.2}$$

These anticommuting elements form an orthonormal subset which may also be designated by,

$$\rho_i = \{\rho, \quad \sigma\} \quad , \tag{A.3}$$

and generates two additional matrices

$$\varepsilon = \begin{bmatrix} 0 & 1 \\ -1 & 0 \end{bmatrix} \quad , \tag{A.4}$$

$$e = \begin{bmatrix} 1 & 0 \\ 0 & 1 \end{bmatrix} \quad . \tag{A.5}$$

which together are a base for the $R_{20}$ geometric algebra formed by the real 2×2 matrices,

$$\begin{bmatrix} \rho & \sigma & \varepsilon & e \end{bmatrix} = \begin{bmatrix} \rho_1 & \rho_2 & \rho_1\rho_2 & \rho_1\rho_1 \end{bmatrix} \quad . \tag{A.6}$$

The generators $E$ of the $A(3,1)$ algebra $R_{31}$ may be expressed as 2×2 matrices over the $\mathcal{Q}$ ring. Explicitly in this representation, the orthonormal subset of the algebra $A$ is, up to equivalence under automorphisms,

$$\kappa_0 = \begin{bmatrix} 0 & \rho \\ -\rho & 0 \end{bmatrix} = \begin{bmatrix} 0 & 0 & 0 & 1 \\ 0 & 0 & 1 & 0 \\ 0 & -1 & 0 & 0 \\ -1 & 0 & 0 & 0 \end{bmatrix} \quad , \tag{A.7}$$

$$\kappa_1 = \begin{bmatrix} -\rho & 0 \\ 0 & \rho \end{bmatrix} = \begin{bmatrix} 0 & -1 & 0 & 0 \\ -1 & 0 & 0 & 0 \\ 0 & 0 & 0 & 1 \\ 0 & 0 & 1 & 0 \end{bmatrix} \quad , \tag{A.8}$$

$$\kappa_2 = \begin{bmatrix} \sigma & 0 \\ 0 & \sigma \end{bmatrix} = \begin{bmatrix} 1 & 0 & 0 & 0 \\ 0 & -1 & 0 & 0 \\ 0 & 0 & 1 & 0 \\ 0 & 0 & 0 & -1 \end{bmatrix} \quad , \tag{A.9}$$

$$\kappa_3 = \begin{bmatrix} 0 & \rho \\ \rho & 0 \end{bmatrix} = \begin{bmatrix} 0 & 0 & 0 & 1 \\ 0 & 0 & 1 & 0 \\ 0 & 1 & 0 & 0 \\ 1 & 0 & 0 & 0 \end{bmatrix} \qquad . \tag{A.10}$$

This orthonormal subset $\kappa_\mu$ generates the rest of the base,

$$\kappa_0 \kappa_1 = \begin{bmatrix} 0 & 0 & 1 & 0 \\ 0 & 0 & 0 & 1 \\ 1 & 0 & 0 & 0 \\ 0 & 1 & 0 & 0 \end{bmatrix} = \begin{bmatrix} 0 & I \\ I & 0 \end{bmatrix} = \sigma_1 \qquad , \tag{A.11}$$

$$\kappa_0 \kappa_2 = \begin{bmatrix} 0 & 0 & 0 & -1 \\ 0 & 0 & 1 & 0 \\ 0 & 1 & 0 & 0 \\ -1 & 0 & 0 & 0 \end{bmatrix} = \begin{bmatrix} 0 & -\varepsilon \\ \varepsilon & 0 \end{bmatrix} = \sigma_2 \qquad , \tag{A.12}$$

$$\kappa_0 \kappa_3 = \begin{bmatrix} 1 & 0 & 0 & 0 \\ 0 & 1 & 0 & 0 \\ 0 & 0 & -1 & 0 \\ 0 & 0 & 0 & -1 \end{bmatrix} = \begin{bmatrix} I & 0 \\ 0 & -I \end{bmatrix} = \sigma_3 \qquad , \tag{A.13}$$

$$\kappa_1 \kappa_2 = \begin{bmatrix} 0 & 1 & 0 & 0 \\ -1 & 0 & 0 & 0 \\ 0 & 0 & 0 & -1 \\ 0 & 0 & 1 & 0 \end{bmatrix} = \begin{bmatrix} \varepsilon & 0 \\ 0 & -\varepsilon \end{bmatrix} = i\sigma_3 \qquad , \tag{A.14}$$

$$\kappa_2 \kappa_3 = \begin{bmatrix} 0 & 0 & 0 & 1 \\ 0 & 0 & -1 & 0 \\ 0 & 1 & 0 & 0 \\ -1 & 0 & 0 & 0 \end{bmatrix} = \begin{bmatrix} 0 & \varepsilon \\ \varepsilon & 0 \end{bmatrix} = i\sigma_1 \qquad , \tag{A.15}$$

$$\kappa_3 \kappa_1 = \begin{bmatrix} 0 & 0 & 1 & 0 \\ 0 & 0 & 0 & 1 \\ -1 & 0 & 0 & 0 \\ 0 & -1 & 0 & 0 \end{bmatrix} = \begin{bmatrix} 0 & I \\ -I & 0 \end{bmatrix} = i\sigma_2 \qquad , \tag{A.16}$$

$$\kappa_1\kappa_2\kappa_3 = \begin{bmatrix} 0 & 0 & 1 & 0 \\ 0 & 0 & 0 & -1 \\ -1 & 0 & 0 & 0 \\ 0 & 1 & 0 & 0 \end{bmatrix} = \begin{bmatrix} 0 & \sigma \\ -\sigma & 0 \end{bmatrix} \quad , \tag{A.17}$$

$$\hat{a} = \kappa_5^\dagger a \kappa_5 \quad , \tag{A.24}$$

induces a direct sum decomposition of $A$ into its even subalgebra $A_+$ and the complementary odd part $A_-$. Associated to the eigenvalues of the main involution, we may introduce a quantum number called rarity $r$, defined by the eigenvalues of the odd projection operator,

$$\mathrm{rar}(a) = \frac{a - \hat{a}}{2} \quad . \tag{A.25}$$

Any element of the algebra may be written in terms of its even and odd parts as

$$a = a_+ + \kappa^0 a_- \tag{A.26}$$

and may be represented by a matrix of twice dimensions with even components,

$$a \rightarrow \begin{bmatrix} a_+ & -\bar{a}_-^\dagger \\ a_- & \bar{a}_+^\dagger \end{bmatrix} \quad . \tag{A.27}$$

Using this technique we may represent the various geometric objects as follows:

$$e \rightarrow \begin{bmatrix} \eta & -\bar{\xi}^\dagger \\ \xi & \bar{\eta}^\dagger \end{bmatrix} \quad , \tag{A.28}$$

$$\Gamma \rightarrow \begin{bmatrix} \Gamma_+ & -\bar{\Gamma}_-^\dagger \\ \Gamma_- & \bar{\Gamma}_+^\dagger \end{bmatrix} \tag{A.29}$$

and, since

$$\kappa^\mu = \kappa^0 \kappa^{0\dagger} \kappa^\mu = \kappa^0 \bar{\sigma}^\mu \quad , \tag{A.30}$$

$$\kappa^\mu \rightarrow \begin{bmatrix} 0 & -\sigma^\mu \\ \bar{\sigma}^\mu & 0 \end{bmatrix} \quad . \tag{A.31}$$

It is possible to give another representation to this geometric algebra $A$ [9 appendix A2]. Define the fields $\mathbb{L}$ and $\mathbb{Q}$ of matrices isomorphic to the quaternion field, with respective bases,

$$\lambda_\alpha = \left\{ I, \kappa_1\kappa_2, \kappa_2\kappa_3, \kappa_3\kappa_1 \right\} \quad , \tag{A.32}$$

$$q_\alpha = \left\{ I, \kappa_0, \kappa_1\kappa_2\kappa_3, \kappa_0\kappa_1\kappa_2\kappa_3 \right\} \quad . \tag{A.33}$$

The orthonormal subset $\kappa_\alpha$ may be expressed in terms of the ring

$$\kappa_\alpha = \left\{ q_1\lambda_0, \quad q_2\lambda_1, \quad q_2\lambda_2, \quad q_2\lambda_3 \right\} \quad . \tag{A.34}$$

This tensor product of matrices is homomorphic to the set of 4×4 matrices generated by the orthonormal subset with the standard matrix product. The homomorphism is 2 to 1 due to the direct product in the definition. Its base spans the geometric Clifford algebra $A$.

# Table of Contents

# INDEX